%% file: ms.tex
\documentclass[times,3p,twocolumn,final]{elsarticle}

\usepackage{framed,multirow}

\usepackage{amssymb}
\usepackage{latexsym}

\usepackage{amsmath}
\usepackage{caption}
\usepackage{subcaption}
\usepackage{booktabs}
\usepackage{pgfplots}
\usepackage{tikz}
\usepackage[binary-units=true]{siunitx}
\usepackage[export]{adjustbox}
\usepackage[inline]{enumitem}

\usepgfplotslibrary{groupplots}
\pgfplotsset{compat=1.7}

\usetikzlibrary{arrows.meta,spy,backgrounds}
\tikzset{font=\scriptsize}

\DeclareMathOperator*{\argmin}{arg\,min}

\usepackage[hidelinks]{hyperref}

\begin{document}

\begin{frontmatter}

\title{Learning Neural Implicit Representations with Surface Signal Parameterizations}

\author[1]{Yanran Guan\corref{cor1}}
\cortext[cor1]{Corresponding author}
\emailauthor{yanran.guan@carleton.ca}{Y. Guan}
    
\author[2,3]{Andrei Chubarau}
\author[3]{Ruby Rao}
\author[2]{Derek Nowrouzezahrai}

\address[1]{Carleton University}
\address[2]{McGill University}
\address[3]{Huawei Technologies Canada}

\begin{abstract}
\input{abstract}
\end{abstract}


\end{frontmatter}

\input{introduction}
\input{related}
\input{method}
\input{results}
\input{conclusion}
\input{acknowledgments}

\bibliographystyle{cag-num-names}
\bibliography{references}

\end{document}

%% file: abstract.tex
Neural implicit surface representations have recently emerged as popular alternative to explicit 3D object encodings, such as polygonal meshes, tabulated points, or voxels. While significant work has improved the geometric fidelity of these representations, much less attention is given to their final appearance. Traditional explicit object representations commonly couple the 3D shape data with auxiliary surface-mapped image data, such as diffuse color textures and fine-scale geometric details in normal maps that typically require a mapping of the 3D surface onto a plane, i.e., a surface parameterization; 
implicit representations, on the other hand, cannot be easily textured due to lack of configurable surface parameterization. 
Inspired by this digital content authoring methodology, we design a neural network architecture that implicitly encodes the underlying surface parameterization suitable for appearance data. 
As such, our model remains compatible with existing mesh-based digital content with appearance data. Motivated by recent work that overfits compact networks to individual 3D objects, we present a new weight-encoded neural implicit representation that extends the capability of neural implicit surfaces to enable various common and important applications of texture mapping. Our method outperforms reasonable baselines and state-of-the-art alternatives.

%% file: introduction.tex
\section{Introduction}\label{sec:introduction}

The 3D surface of an object can be encoded \emph{implicitly} as the zero isocontour of a 3D scalar field, such as a distance field~\cite{park2019deepsdf,chibane2020neural}. Neural networks have become an alluring and powerful tool for parameterizing these fields, and these \emph{neural implicit representations} are capable of encoding a variety of 3D shapes. One such recent approach looks to efficiently encode a \emph{single 3D object} as the weights of a small, overfit neural network~\cite{davies2020overfit}, unlike the original approaches that seek to learn a latent embedding of \emph{many} 3D shapes~\cite{park2019deepsdf}. In most cases, however, these neural implicit representations focus exclusively on the underlying scalar distance field that encodes the surface geometry, ignoring auxiliary appearance data commonly co-authored during the digital content creation process, such as diffuse colors and fine-scale geometric deviations. 

\input{figures/teaser}

On the other hand, those works that do consider the neural representation of object or scene appearance tend to implicitly define the appearance properties, such as with continuous radiance fields~\cite{mildenhall2020nerf}, in a manner that does not disentangle them from objects' geometry. Here, it is difficult to decouple the auxiliary appearance data from its underlying geometry, e.g., to edit it separately from the object data --- as is standard in 3D digital content creation pipelines.

We instead seek to learn neural representations that are suitable for surface geometry \emph{and} appearance data, without entangling the two. Inspired by traditional explicit 3D object representations, i.e., meshes, we rely on surface parameterizations~\cite{sheffer2006mesh} to treat auxiliary appearance data in a manner that naturally disentangles it from the underlying geometry. These parameterizations --- also called UV maps --- map 3D points on the surface onto a 2D chart. During visualization, we can invert this mapping to project 2D image data onto the 3D surfaces: a process called \emph{texture mapping}. 

Our main idea is to treat the surface parameterization as an implicit bijective function, which maps 3D input locations onto unique 2D UV coordinates, and we parameterize this function with a compact neural network. We jointly learn an overfit neural implicit surface network with our appearance mapping network for texture mapping. Figure~\ref{fig:teaser} illustrates a collection of 3D objects with their textures applied using our learned surface parameterization. The main challenge here is that the neural surface parameterization must reason about the ``unwrapping'' of 3D surfaces onto the 2D UV texture domain. 
In practice, object textures comprise a discontinuous collection of piecewise smooth charts for each mesh segment, known as a texture atlas.

\input{figures/overview}

The discontinuous, multimodal nature of these charts makes them challenging to learn, and we propose a simple and effective learning strategy --- which relies on applying a spatial decomposition when conditioning the input of the network --- to tackle this difficulty. The decomposition components are defined for the input 3D locations based on the parameterization charts, each of which corresponds to a segment of the surface. We then apply a weight-encoded neural implicit representation to the surface parameterization, overfitting a feed-forward multilayer perceptron (MLP) to individual objects and their UV maps. 

Our model can be applied directly to neural implicit surfaces and enables texture mapping on these surfaces, as illustrated and explained in Figure~\ref{fig:overview}. In this paper, we apply our model jointly with overfit representations of neural implicit surfaces (e.g.,~\cite{davies2020overfit}) with only modest overhead.


%% file: figures/teaser.tex
\begin{figure}[t!]
    \centering
    \includegraphics[width=\linewidth,trim={0 100 0 920},clip]{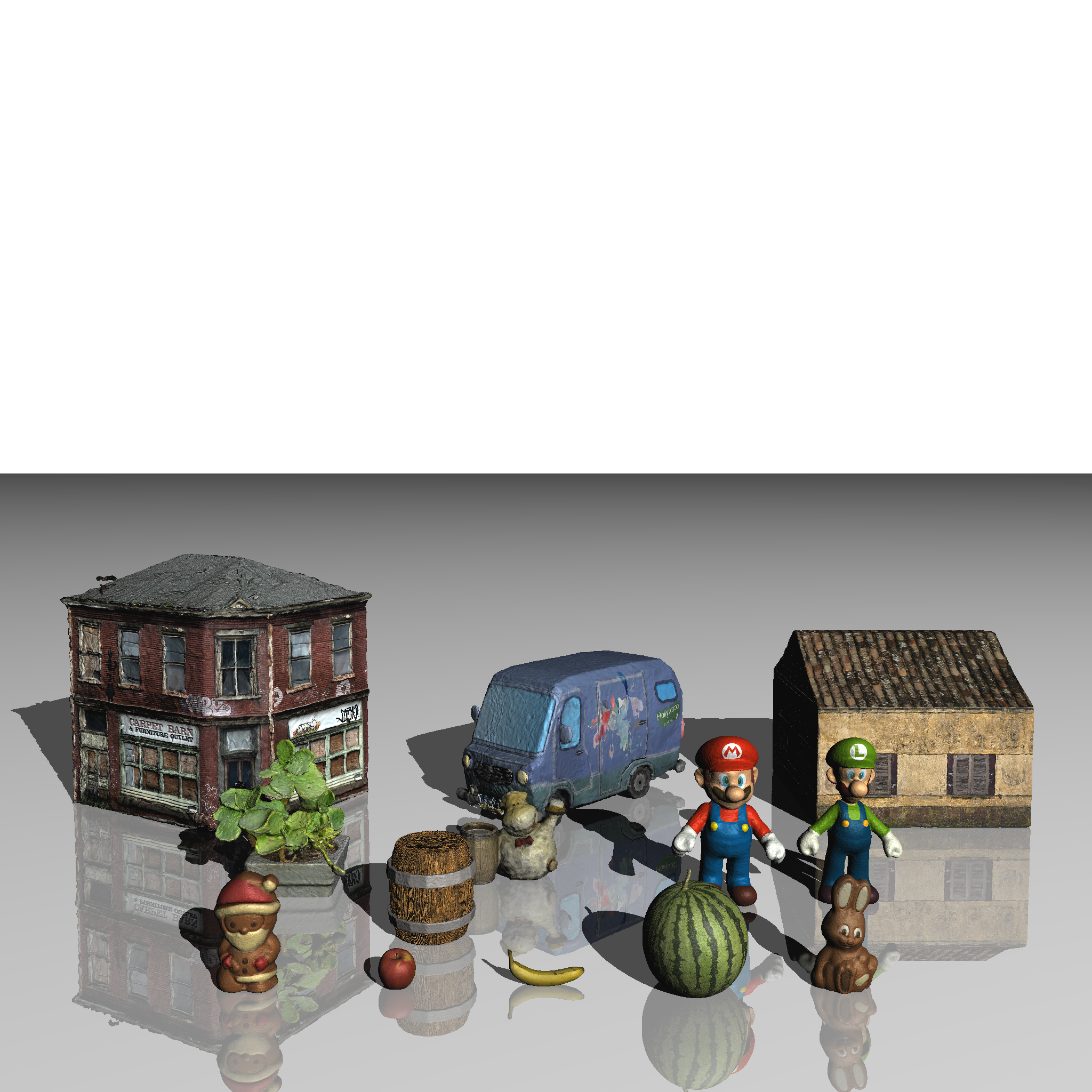}
    \caption{A collection of textured objects, rendered using a sphere tracer~\cite{hart1996sphere}, where the geometry is represented by the neural implicit surface encoded using the model of Davies et al.~\cite{davies2020overfit} and the texture is applied through our neural surface parameterization.}
    \label{fig:teaser}
\end{figure}

%% file: figures/overview.tex
\begin{figure}
    \centering
    \begin{tikzpicture}
        \node at (3.3,0.8) {\includegraphics[width=0.12\linewidth]{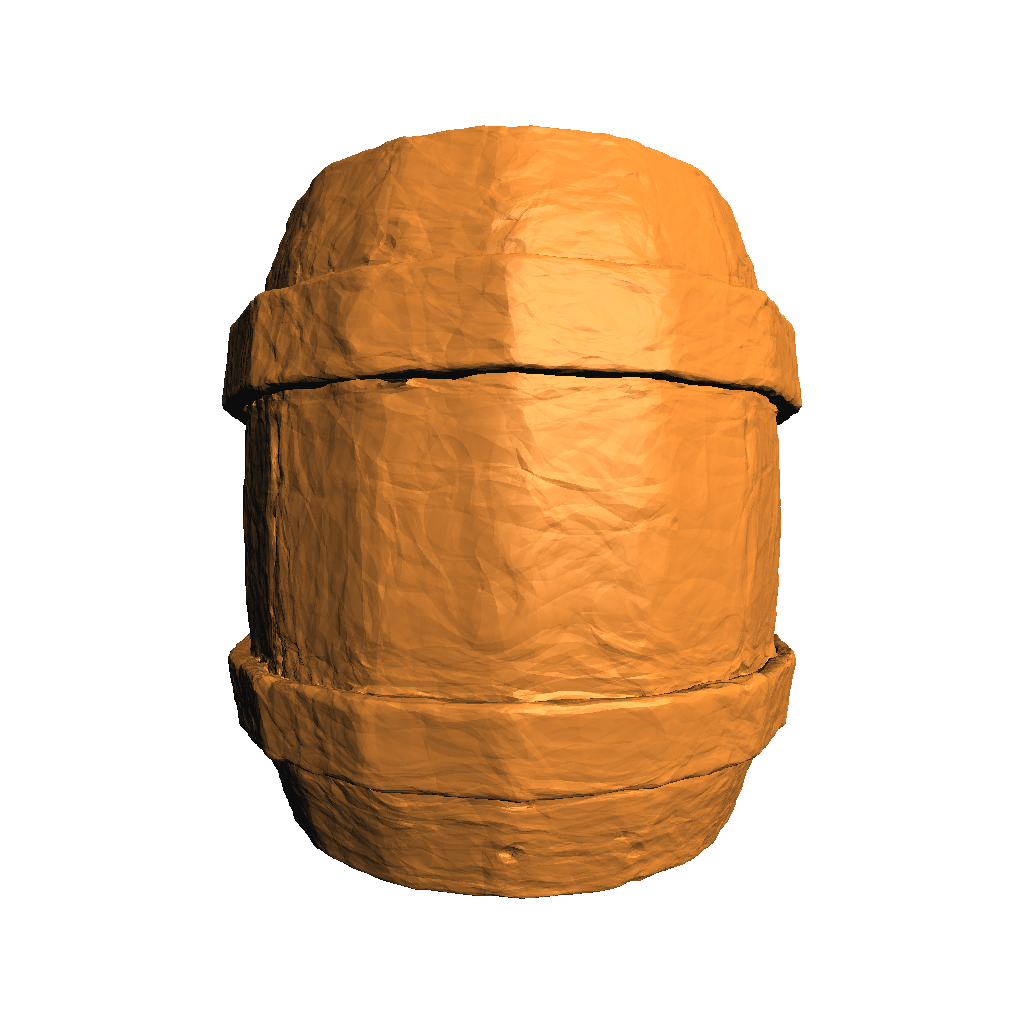}};
        \node at (3.3,-0.8) {\includegraphics[width=0.12\linewidth,trim={250 100 250 100},clip]{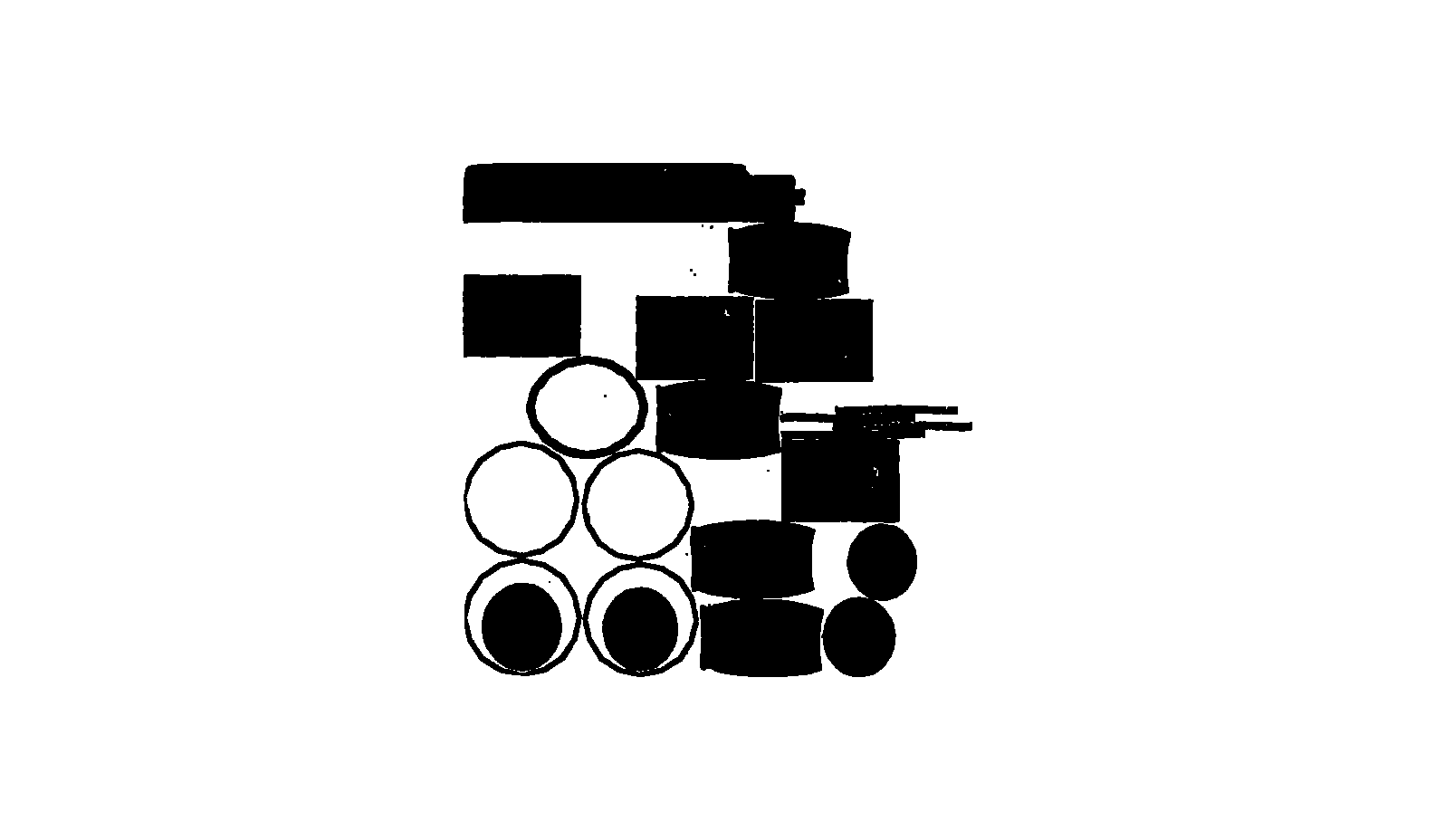}};
        \node at (5.4,-0.4) {\includegraphics[width=0.16\linewidth,trim={120 120 120 120},clip]{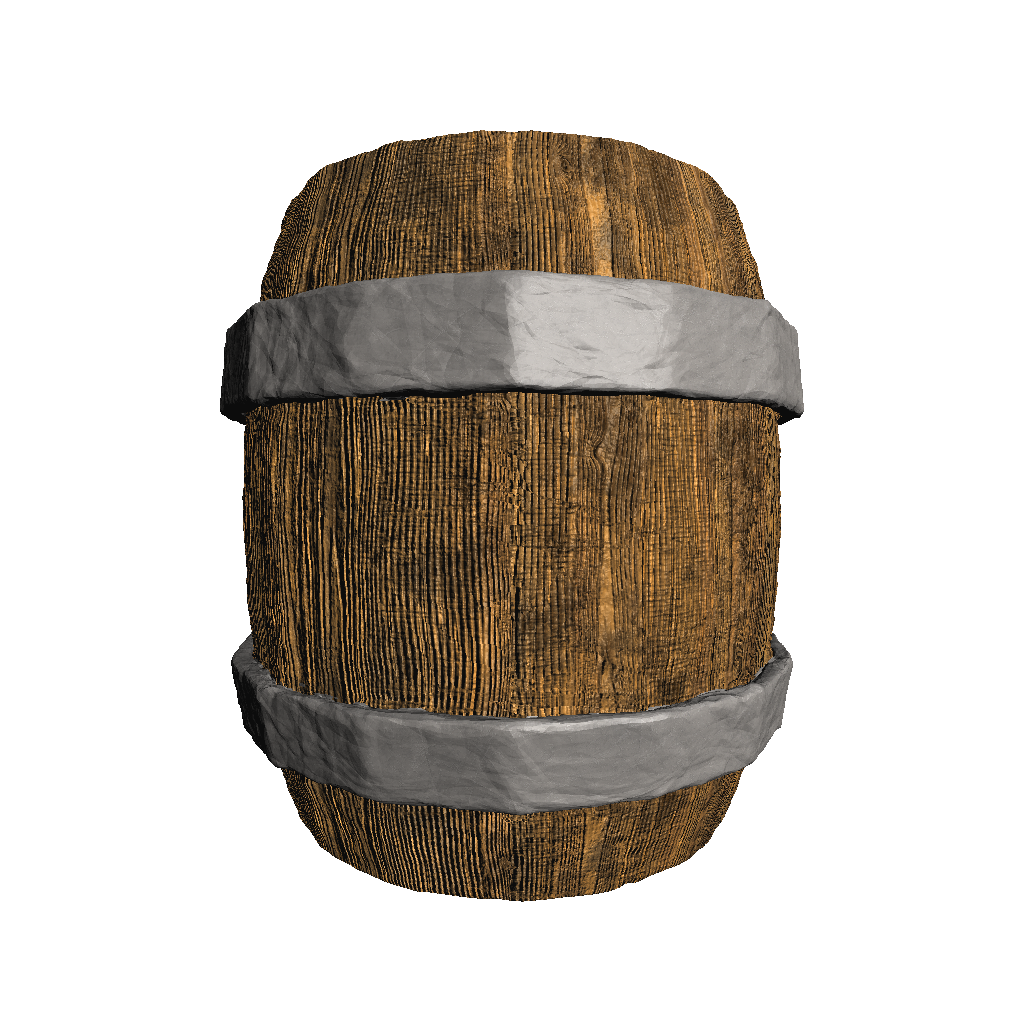}};
        \node at (5.4,-1.3) {Rendered result};
        \node at (0,0.8) {\includegraphics[width=0.12\linewidth,trim={340 100 340 100},clip]{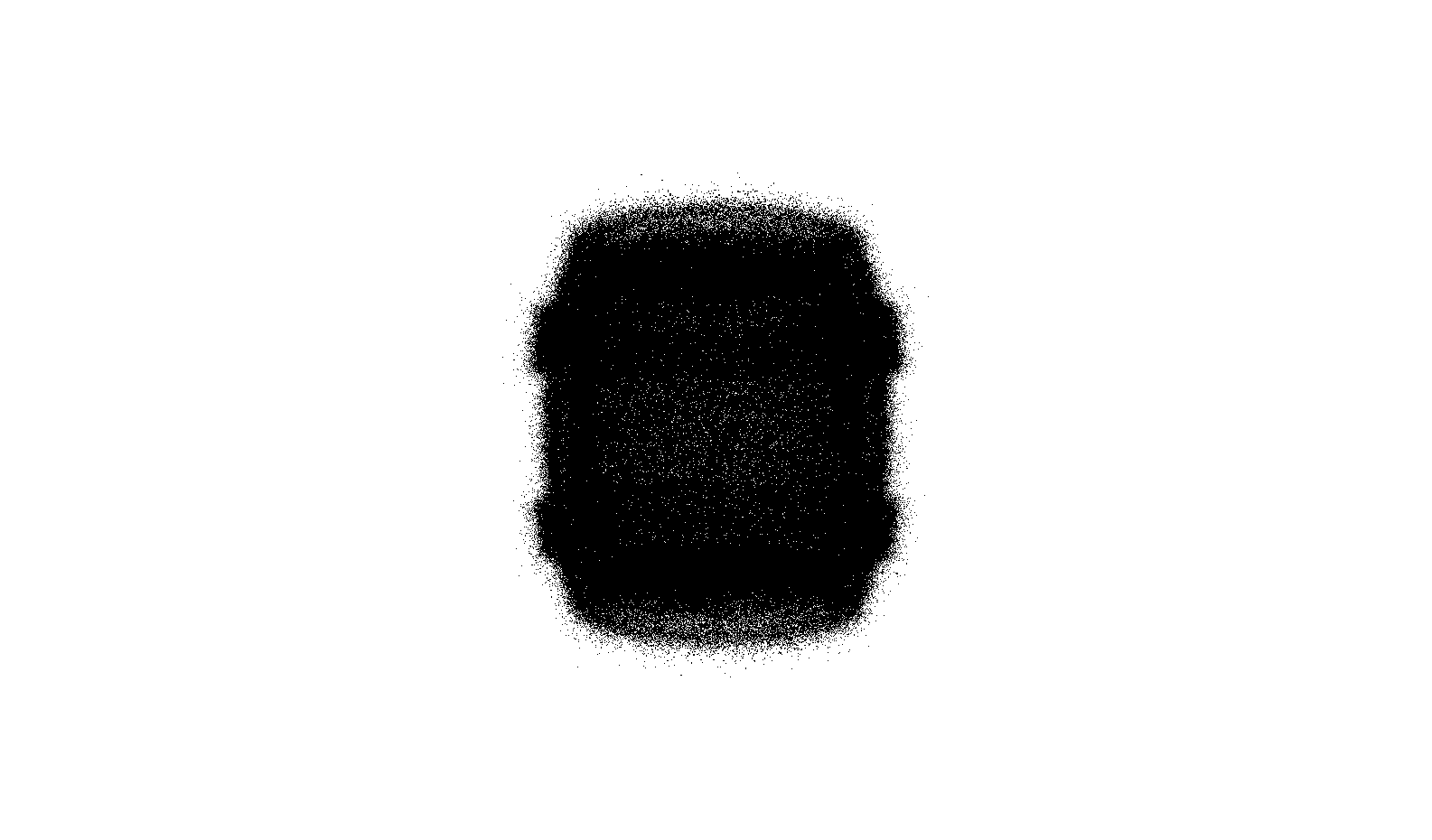}};
        \node at (0,0.2) {Sample points};
        \draw[red!60,-{Triangle[width=18pt,length=5pt]},line width=10pt] (0.55,0.8) -- (0.85,0.8);
        \draw[red!60,-{Triangle[width=18pt,length=5pt]},line width=10pt] (0,0) -- (0,-0.3);
        \node at (0,-0.8) {\includegraphics[width=0.12\linewidth,trim={340 100 340 100},clip]{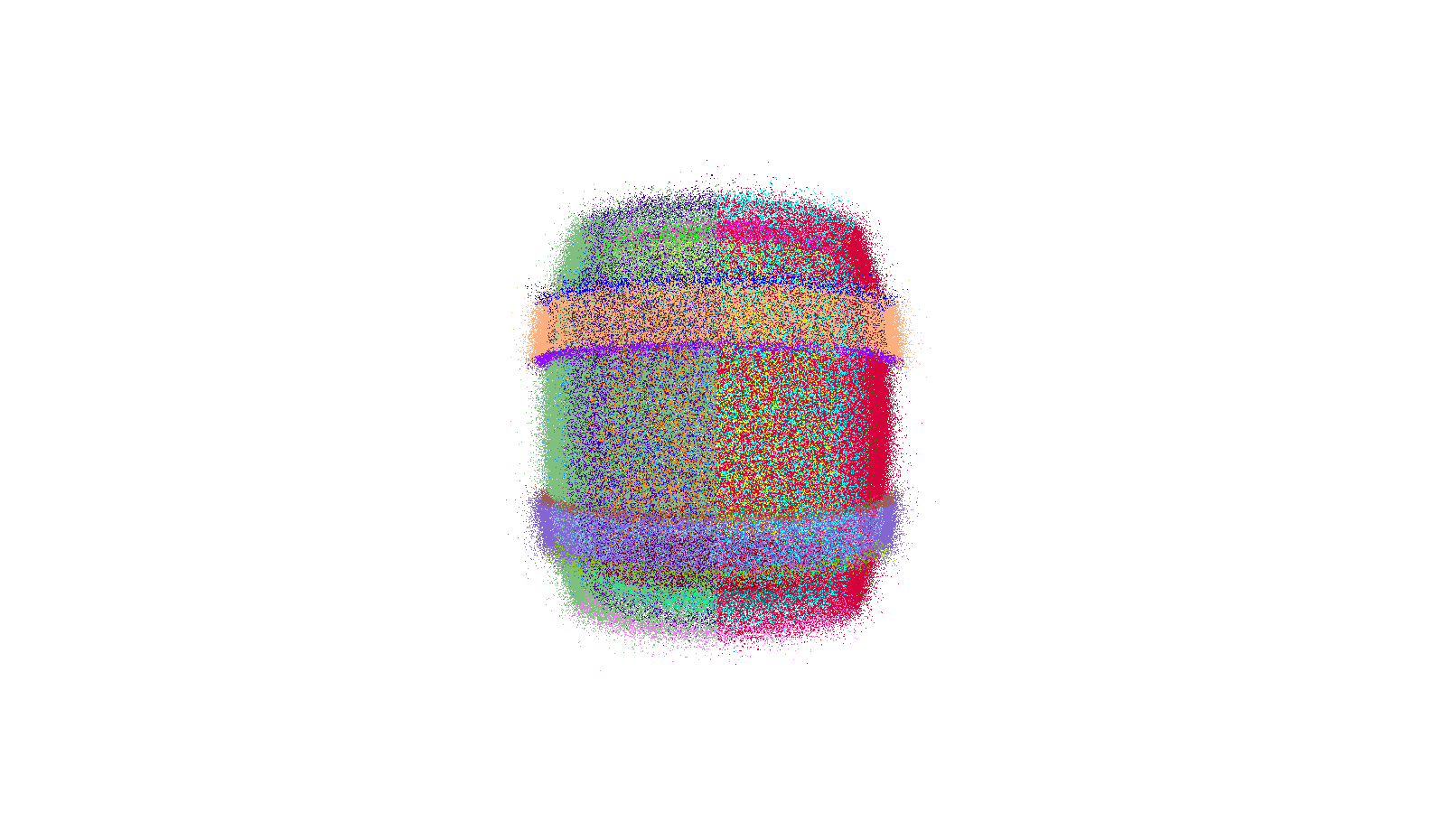}};
        \node at (0,-1.55) {\begin{tabular}{c}
        Decomposed \\
        sample points
        \end{tabular}};
        \draw[red!60,-{Triangle[width=18pt,length=5pt]},line width=10pt] (0.55,-0.8) -- (0.85,-0.8);
        \node at (1.8,0.8) {\input{images/overview/neural_implicit_surface}};
        \node at (2.55,0.2) {Implicit surface};
        \node at (1.8,-0.8) {\input{images/overview/neural_implicit_parameterization}};
        \node at (2.55,-1.4) {Implicit parameterization};
        \draw[red!60,-{Triangle[width=18pt,length=5pt]},line width=10pt] (2.5,0.8) -- (2.8,0.8);
        \draw[red!60,-{Triangle[width=18pt,length=5pt]},line width=10pt] (2.5,-0.8) -- (2.8,-0.8);
        \node at (2.55,-1.7) {$+$};
        \node at (2.1,-2.4) {\includegraphics[width=0.095\linewidth]{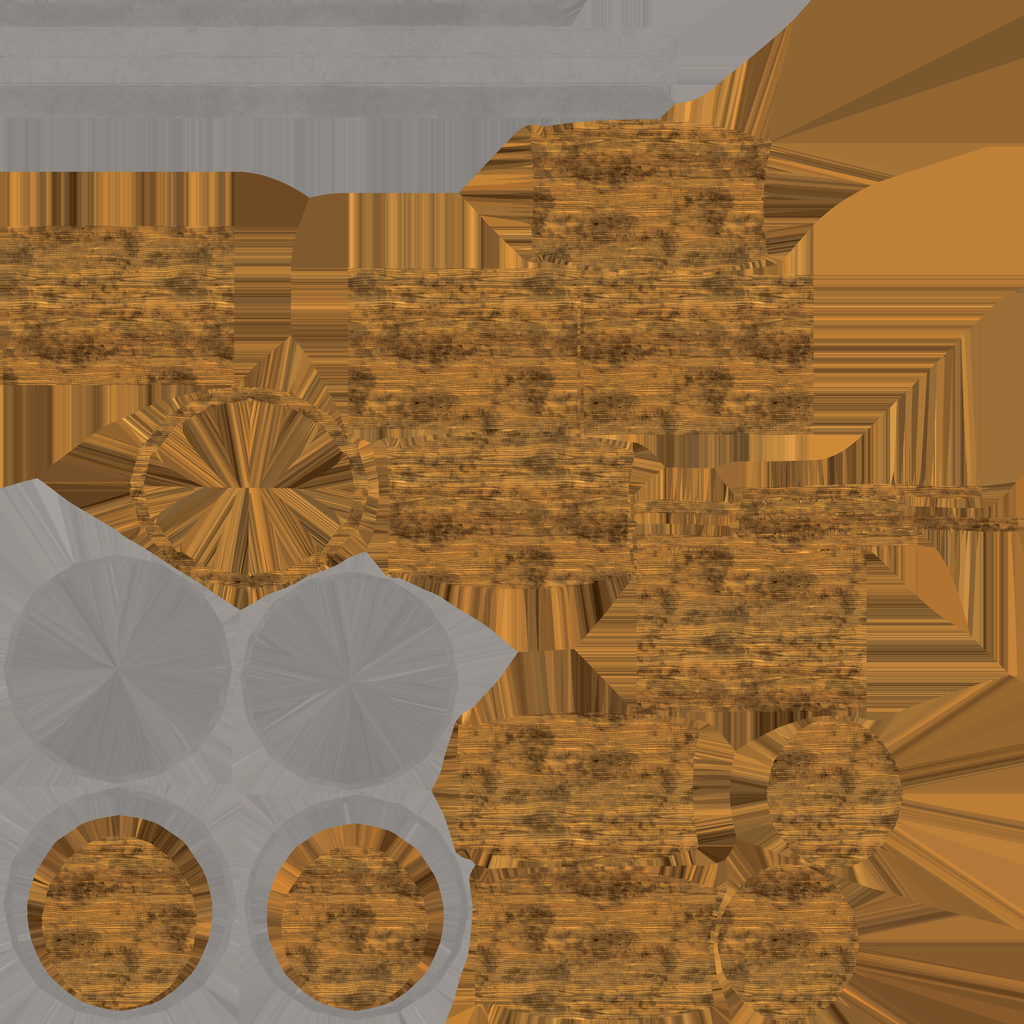}};
        \node at (3,-2.4) {\includegraphics[width=0.095\linewidth]{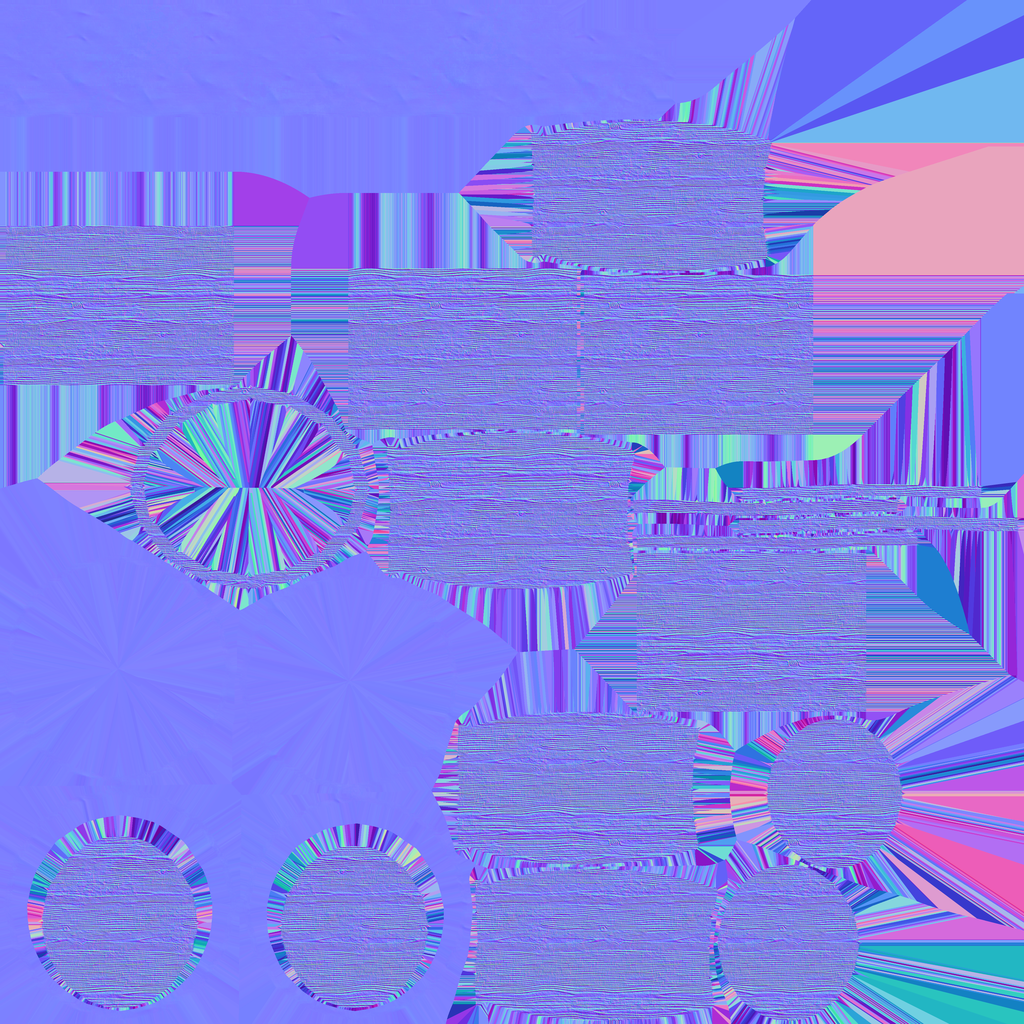}};
        \node at (2.55,-3) {Texture maps};

        \draw [densely dotted,gray,rounded corners=6,thick] (1.1,-3.2) rectangle (4,-0.3);

        \draw[decorate,decoration={brace,amplitude=5}] (4.1,0.8) -- (4.1,-1.6);
        \draw[red!60,-{Triangle[width=18pt,length=5pt]},line width=10pt] (4.4,-0.4) -- (4.7,-0.4);
    \end{tikzpicture}
    \caption{The pipeline of applying our neural implicit surface parameterization to the neural implicit surface to enable texture mapping. For each object, we train two neural representations, one representing the implicit surface and the other representing the implicit parameterization. Then, when rendering the object, we can derive the associated 2D UV coordinate for any 3D location on the zero isocontour defined by the neural implicit surface, thus allowing for texture mapping.}
    \label{fig:overview}
\end{figure}

%% file: related.tex
\section{Related work}\label{sec:related}

We briefly review the relevant literature, including neural implicit surface and appearance, surface parameterization, and overfit neural representations.

\paragraph{Neural implicit surfaces} Neural networks can approximate scalar functions $f\colon\mathbb{R}^3\to\mathbb{R}$, where surfaces are implicitly represented as zero isocontours of a level set. These functions include occupancy fields~\cite{mescheder2019occupancy,chen2019learning}, distance fields~\cite{park2019deepsdf,chibane2020neural}, or space partitioning trees~\cite{deng2020cvxnet,chen2020bsp}. Recent works show that neural networks implemented with periodic functions~\cite{sitzmann2020implicit,tancik2020fourier,mildenhall2020nerf} can improve the approximation of these scalar functions with reconstructions of high frequency details. Davies et al.~\cite{davies2020overfit} overfit neural networks to single shapes as weight-encoded representations of 3D surfaces, which can be interpreted as an efficient and lossy compression of 3D shapes. Takikawa et al.~\cite{takikawa2021neural} propose a hierarchical representation that can be adaptively scaled to different level of details, combining neural implicit surfaces with an octree feature volume, for rapid final visualization.

These works focus only on 3D geometric surface representations, precluding texture and appearance mapping.

\paragraph{Neural implicit appearance} Similar to implicit surfaces, the appearance of objects or scenes can also be implicitly modeled by a neural network as a continuous function, typically mapping from 3D locations to RGB colors~\cite{saito2019pifu,oechsle2019texture,sitzmann2019scene,niemeyer2020differentiable}. These mappings are limited to representing simple, unimodal (i.e., only diffuse color) textures. To support a greater diversity of appearance properties, implicit surface light field by Oechsle et al.~\cite{oechsle2020learning} conditions their model on lighting and viewpoint information, allowing the model to represent properties such as specular reflection and shadows. Neural radiance fields (NeRFs)~\cite{mildenhall2020nerf} take a continuous 5D input space (3D location and 2D view direction) and map them to volumetric RGB opacities that can be accumulated to form 2D images using a (differentiable and costly) volume sampling-based reconstruction.

These continuous, implicit volumetric appearance functions are inflexible and hard to manipulate. To address this limitation, NeuTex~\cite{xiang2021neutex} uses a separate network --- atop of a volumetric NeRF representation --- in an attempt to disentangle texture content into a 2D projective space parameterized by a cube map. AUV-Net~\cite{chen2022auv} learns to embed 3D surfaces into aligned 2D spaces and generates corresponding texture images, thus enabling texture transfer. While similar in spirit to our motivation, we instead directly learn for each individual object the \emph{surface parameterization} that can be used to map \emph{any} auxiliary appearance data onto our 3D surfaces during visualization.

\paragraph{Surface parameterization} Computing a surface parameterization of a 3D object is a long-standing problem in computational geometry and computer graphics, and we refer interested readers to the comprehensive survey by Sheffer et al.~\cite{sheffer2006mesh}. Briefly, a surface parameterization refers to a bijective mapping from 3D locations on an object's surface to 2D UV coordinates in a charted parameter space. Algorithms for computing planar parameterizations from explicit polygonal mesh surfaces include methods that compute surface-to-surface bijective mappings --- such as with graph embeddings~\cite{tutte1960convex} --- or those that minimize distortion metrics of the original mesh, e.g., angular~\cite{sheffer2001parameterization,sheffer2005abf++,levy2002least}, distance~\cite{sander2001texture,zigelman2002texture}, area~\cite{levy2002least,degener2003adaptable}, and boundary distortions~\cite{sawhney2017boundary}. Surfaces of greater geometric complexity introduce more distortion in these parameterizations so that, in practice, mesh cutting and chart packing~\cite{garland2001hierarchical,levy2002least} are usually necessary in order to segment the original surface into several charts suitable for independent parametric distortion minimization.

Recently, given the advances of gradient-based optimization for neural networks and the growing availability of large 3D surface datasets, parameterization has also been studied through the lens of a neural representation problem, where networks are trained to map 2D coordinates to 3D surface locations~\cite{sinha2016deep,groueix2018papier,deprelle2019learning,morreale2021neural}. These methods aim to recover the geometry of 3D surfaces in a parametric manner and cannot be directly applied to our auxiliary appearance texture mapping setting. Our model instead learns the \emph{inverse mapping}, directly reasoning about the UV maps of parameterization charts in an underlying texture atlas.

\paragraph{Overfit digital content} Recent work has shown that the weights of simple feed-forward MLPs can yield efficient compressed representations for many forms of digital content, including images (by mapping pixel locations to RGB values~\cite{dupont2021coin,strumpler2022implicit}) and 3D shapes (by mapping 3D locations to signed distances~\cite{davies2020overfit,li2022high}). The difficulty of encoding high frequency signals with MLPs can be mitigated using periodic encodings and activations~\cite{sitzmann2020implicit,tancik2020fourier,mildenhall2020nerf}. The compression speed and rate-distortion performance can be improved through meta-learned initializations~\cite{strumpler2022implicit,dupont2022coin++}.

We also rely on weight-encoded neural implicit surface parameterizations,
overfitting MLPs to objects' UV maps. Our model complements existing work on weight-encoded neural representations that focus exclusively on geometry encoding~\cite{davies2020overfit}. We further achieve a higher compression rate by pruning our model using the lottery ticket method~\cite{frankle2019lottery}.

%% file: method.tex
\section{Methodology}\label{sec:method}

The parameterization of a surface $\mathcal{S}$ is defined as a mapping from the surface points $\mathbf{p}_\mathcal{S}\in\mathcal{S}$ to UV coordinates $\mathbf{u}\in\mathbb{R}^2$, denoted as:
\begin{equation}\label{eq:uv-surface}
\operatorname{UV}_\mathcal{S}(\mathbf{p}_\mathcal{S})=\mathbf{u}.
\end{equation}
To make this function applicable jointly with an implicit surface defined by the signed distance function (SDF), we want to have both functions work on a consistent domain. Therefore, we extend the parameterization function to all query points $\mathbf{p}\in\mathbb{R}^3$ as follows:
\begin{equation}\label{eq:uv}
\operatorname{UV}(\mathbf{p})=\operatorname{UV}_\mathcal{S}\left(\argmin_{\mathbf{p}_\mathcal{S}}\left|\mathbf{p}-\mathbf{p}_\mathcal{S}\right|\right).
\end{equation}
Our objective is to train a neural network $f_\theta$ that approximates this parameterization function, such that
\begin{equation}\label{eq:neural-parameterization}
f_\theta(\mathbf{p})\approx\operatorname{UV}(\mathbf{p}).
\end{equation}


We describe our method for representing surface signal parameterization using neural networks. We first explain how we generate the training data and then the design of our neural implicit representation.

\subsection{Training set}\label{sec:method-sampling}

We sample our training set based on the importance sampling~\cite{kahn1951estimation} scheme proposed by Davies et al.~\cite{davies2020overfit}, which permits the integration of weighting functions to define the importance over the sampling domain. For each sample point, apart from the UV coordinate given by Equation~\ref{eq:uv}, we generate a decomposition component label indicating the surface segment in a texture atlas to which the point belongs.

\paragraph{Importance sampling} Given the SDF describing an object's surface that maps every point coordinate $\mathbf{p}$ in the 3D sampling space to a scalar distance $d\in\mathbb{R}$, denoted as:
\begin{equation}\label{eq:sdf}
\operatorname{SDF}(\mathbf{p})=d,
\end{equation}
and a neural network $g_{\theta^{\prime}}$ that approximates the SDF:
\begin{equation}\label{eq:neural-sdf}
g_{\theta^{\prime}}(\mathbf{p})\approx\operatorname{SDF}(\mathbf{p}),
\end{equation}
the objective of the weighting function is to assign higher weights to 3D locations that are closer to the implicitly defined surface $\operatorname{SDF}(\cdot)=0$, such that the signals closer to the surface can be better represented by neural networks during learning. Specifically, we use the following metric~\cite{davies2020overfit} to define the importance for each sample point:
\begin{equation}\label{eq:weight}
w(\mathbf{p})=e^{-\beta\left|\operatorname{SDF}(\mathbf{p})\right|},
\end{equation}
where $\beta$ is a coefficient ranging $[0, +\infty]$. In our experiments, we set $\beta = 60$. We then apply a Monte Carlo approximation to down-sample a set $S$ of $n$ points from a set of $m$ uniformly sampled points $U$, such that the mean absolute error (MAE) between the real distances and the distances predicted by the neural approximation $g_{\theta^{\prime}}$ computed over the down-sampled set is approximated to the weighted MAE over the uniformly sampled set, i.e.,
\begin{equation}\label{eq:monte-carlo-approximation}
\begin{split}
&\frac{1}{n}\sum_{\mathbf{p}\in S}{\left|\operatorname{SDF}(\mathbf{p})-g_{\theta^{\prime}}(\mathbf{p})\right|} \\
\approx&\frac{1}{m}\sum_{\mathbf{p}\in U}{\left|\operatorname{SDF}(\mathbf{p})-g_{\theta^{\prime}}(\mathbf{p})\right|}w(\mathbf{p}).
\end{split}
\end{equation}

Davies et al.~\cite{davies2020overfit} provide the full details of this sampling scheme and it applies seamlessly to our setting.

\paragraph{Spatial decomposition} Previous research on deep generative models has observed that conditioned inputs can help models learn complex structured output representations~\cite{sohn2015learning}.
We observe that the discontinuities and the piecewise nature of parameterization charts in texture atlases complicate learning; neural networks simply tend to learn smoothed boundaries, which for texture mapping specifically results in jarring visual artifacts.
To minimize this effect, we assign a unique discrete label to each region of the parameterization charts, and condition the input to our model on this decomposition signal to indicate different surface segments according to the parameterization charts.

\input{figures/decomposition}

Given a texture atlas composed of $k$ parameterization charts for $k$ different surface segments, we decompose the sampling space into $k$ components by assigning each sample point to its nearest surface segment. Figure~\ref{fig:decomposition} illustrates a set of points sampled via importance sampling around an object, with the component label --- shown as the color corresponding to the parameterization chart --- associated with each sample point. 

\subsection{Designing neural implicit representations}\label{sec:method-design}

Our neural implicit representation of the surface parameterization is a simple feed-forward MLP that takes 3D locations as input and outputs UV coordinates, composed of only fully-connected (FC) layers. We first describe the parameters of the MLP layers and then detail the overall network architecture.

\paragraph{Pre-processing and layer implementation} To improve the reconstruction quality of the MLP, we pre-process the input layer of the network with a Fourier positional encoding~\cite{mildenhall2020nerf,tancik2020fourier} and implement the hidden layers as sinusoidal representation network (SIREN) layers~\cite{sitzmann2020implicit}.

Specifically, the value $p$ on each coordinate position of the input is encoded as an array of Fourier features using the following function~\cite{mildenhall2020nerf}:
\begin{equation}\label{eq:positional-encoding}
\begin{split}
\gamma(p)=&\left(\sin(2^0\pi p),\cos(2^0\pi p),\dots,\right. \\
&\left.\sin(2^{L-1}\pi p),\cos(2^{L-1}\pi p)\right),
\end{split}
\end{equation}
where $L$ defines the number of Fourier series terms; we use $L = 5$ in our implementation.

Our FC layers additionally differ from standard FC layers as we rely on sinusoidal representations~\cite{sitzmann2020implicit}. To implement a SIREN layer, we initialize the weight matrix $\mathbf{W}$ and the bias vector $\mathbf{b}$ using a He uniform initializer~\cite{he2015delving},
before activating the layer using a sine function, as:
\begin{equation}\label{eq:sine}
\phi(\mathbf{x})=\sin\left(\omega_0\mathbf{W}\mathbf{x}+\mathbf{b}\right),
\end{equation}
where $\mathbf{x}$ refers to the layer's input tensor and $\omega_0$ scales the angular frequencies. We use $\omega_0 = 1$.

\input{figures/architecture}

\paragraph{Network architecture} We design a two-stage network architecture that handles both component predictions and UV coordinate predictions at the same time; see Figure~\ref{fig:architecture} for the designed architecture of our model. We use one MLP network, termed \texttt{point2component}, to predict the component label to which the input point belongs, and another MLP network, termed \texttt{point2UV}, to predict the UV coordinate associated with the input point. Both \texttt{point2component} and \texttt{point2UV} are composed of $N$ feed-forward FC layers with each having $H$ hidden units. We set $N=8$ and $H=64$ in our implementation, which we found to provide a balanced trade-off between model complexity and prediction error, as per our ablation study.

The first stage, \texttt{point2component}, takes a 3D point coordinate as input and outputs a $k$-sized vector where each entry corresponds to the probability the current point belongs to a particular component. We then convert this to an integer $c$ that represents the predicted component label corresponding to the highest probability component.
Then, we concatenate the component prediction $c$ with the point coordinate and use the concatenated vector as the input to \texttt{point2UV}, which outputs UV coordinate predictions.
During training, \texttt{point2component} and \texttt{point2UV} are being updated in parallel separately. Once the two neural networks are trained, we use the entire concatenated architecture for predicting the surface parameterization.

\input{figures/color_comparison}

Since \texttt{point2component} performs classification, its last layer uses a softmax activation and our training objective minimizes cross-entropy loss. The last layer of \texttt{point2UV} uses a sigmoid activation unit and the objective of \texttt{point2UV} is to minimize the MAE between real and predicted UV coordinates.

%% file: figures/decomposition.tex
\begin{figure}
    \centering
    \begin{tikzpicture}
        \node at (0,0) {\includegraphics[width=0.5\linewidth,trim={270 100 270 100},clip]{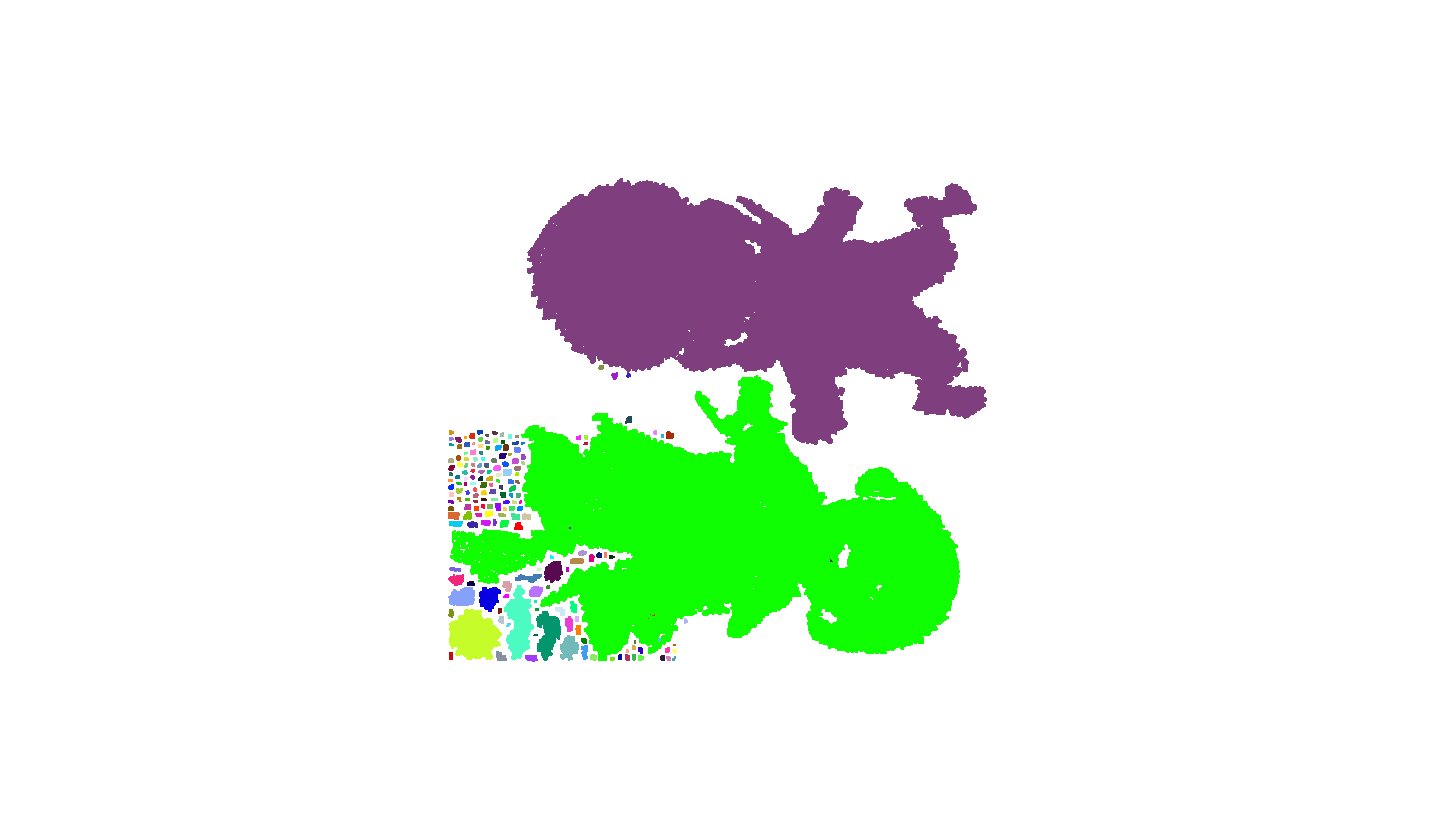}};
        \node at (3.2,0) {\includegraphics[width=0.5\linewidth,trim={180 90 350 90},clip]{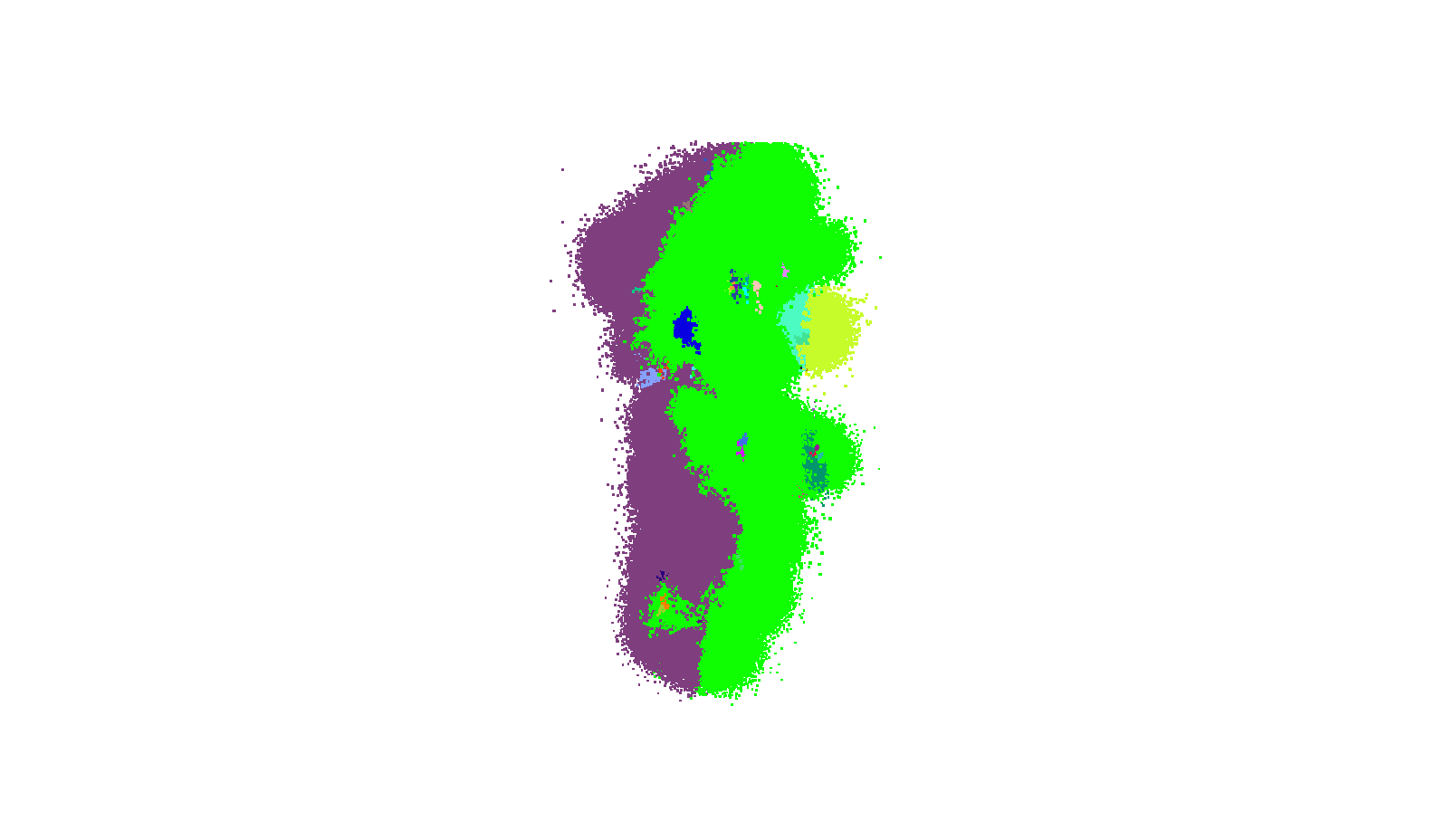}};
        \draw[thick,stealth-stealth] (0.6,0.6) .. controls (1.2,1.6) and (2.6,1.6) .. (3.25,0.8);
        \draw[thick,stealth-stealth] (0.2,-0.6) .. controls (1.2,-1.6) and (2.8,-1.6) .. (4,-0.4);
    \end{tikzpicture}
    \caption{Decomposing the sample points according to the parameterization charts in the texture atlas, where the points in each decomposition component (right) are plotted in the color of their corresponding parameterization chart (left).}
    \label{fig:decomposition}
\end{figure}

%% file: figures/architecture.tex
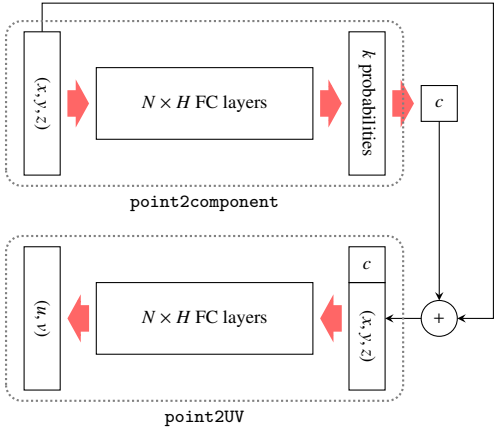
\begin{figure}
    \centering
    \begin{tikzpicture}[scale=0.95]
        \draw (0,0) rectangle (0.5,2) node[pos=.5,rotate=270] {$(x,y,z)$};
        \draw (1,0.5) rectangle (4,1.5) node[pos=.5] {$N\times H$ FC layers};
        \draw (4.5,0) rectangle (5,2) node[pos=.5,rotate=270] {$k$ probabilities};
        \draw (5.5,0.75) rectangle (6,1.25) node[pos=.5] {$c$};

        \draw[red!60,-{Triangle[width=18pt,length=5pt]},line width=10pt] (0.6,1) -- (0.9,1);
        \draw[red!60,-{Triangle[width=18pt,length=5pt]},line width=10pt] (4.1,1) -- (4.4,1);
        \draw[red!60,-{Triangle[width=18pt,length=5pt]},line width=10pt] (5.1,1) -- (5.4,1);

        \draw (5.75,-2) circle [radius=0.25] node {$+$};

        \draw (4.5,-3) rectangle (5,-1.5) node[pos=.5,rotate=270] {$(x,y,z)$};
        \draw (4.5,-1.5) rectangle (5,-1) node[pos=.5] {$c$};
        \draw (1,-2.5) rectangle (4,-1.5) node[pos=.5] {$N\times H$ FC layers};
        \draw (0,-3) rectangle (0.5,-1) node[pos=.5,rotate=270] {$(u,v)$};

        \draw[red!60,-{Triangle[width=18pt,length=5pt]},line width=10pt] (4.4,-2) -- (4.1,-2);
        \draw[red!60,-{Triangle[width=18pt,length=5pt]},line width=10pt] (0.9,-2) -- (0.6,-2);

        \draw[-stealth] (0.25,2) |- (6.5,2.4) |- (6,-2);
        \draw[-stealth] (5.75,0.75) -- (5.75,-1.75);
        \draw[-stealth] (5.5,-2) -- (5,-2);

        \draw [densely dotted,gray,rounded corners=6,thick] (-0.25,-0.15) rectangle (5.25,2.15);
        \node at (2.5,-0.4) {\texttt{point2component}};

        \draw [densely dotted,gray,rounded corners=6,thick] (-0.25,-3.15) rectangle (5.25,-0.85);
        \node at (2.5,-3.4) {\texttt{point2UV}};
    \end{tikzpicture}
    \caption{The network architecture of the proposed model.}
    \label{fig:architecture}
\end{figure}

%% file: figures/color_comparison.tex
\begin{figure*}
    \centering
    \begin{subfigure}{\linewidth}
        \centering
        \begin{tikzpicture}[spy using outlines={circle,orange,magnification=5,size=1.5cm,connect spies}]
            \node at (0,0) {\includegraphics[height=55pt,trim={200 70 200 70},clip]{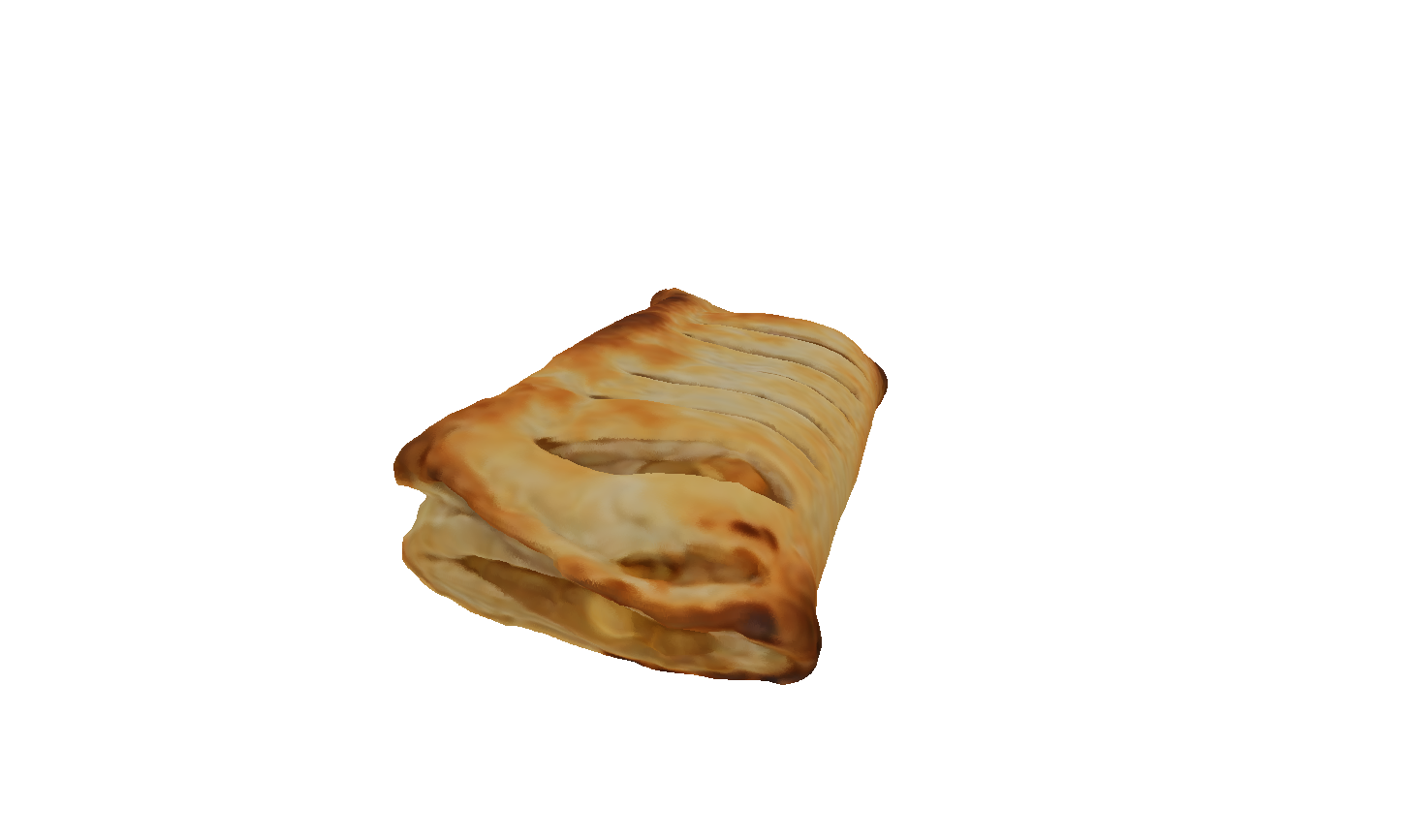}};
            \spy on (0,-0.5) in node [left] at (1.9,0);
            \node at (3.2,0) {\includegraphics[height=55pt,trim={210 70 190 45},clip]{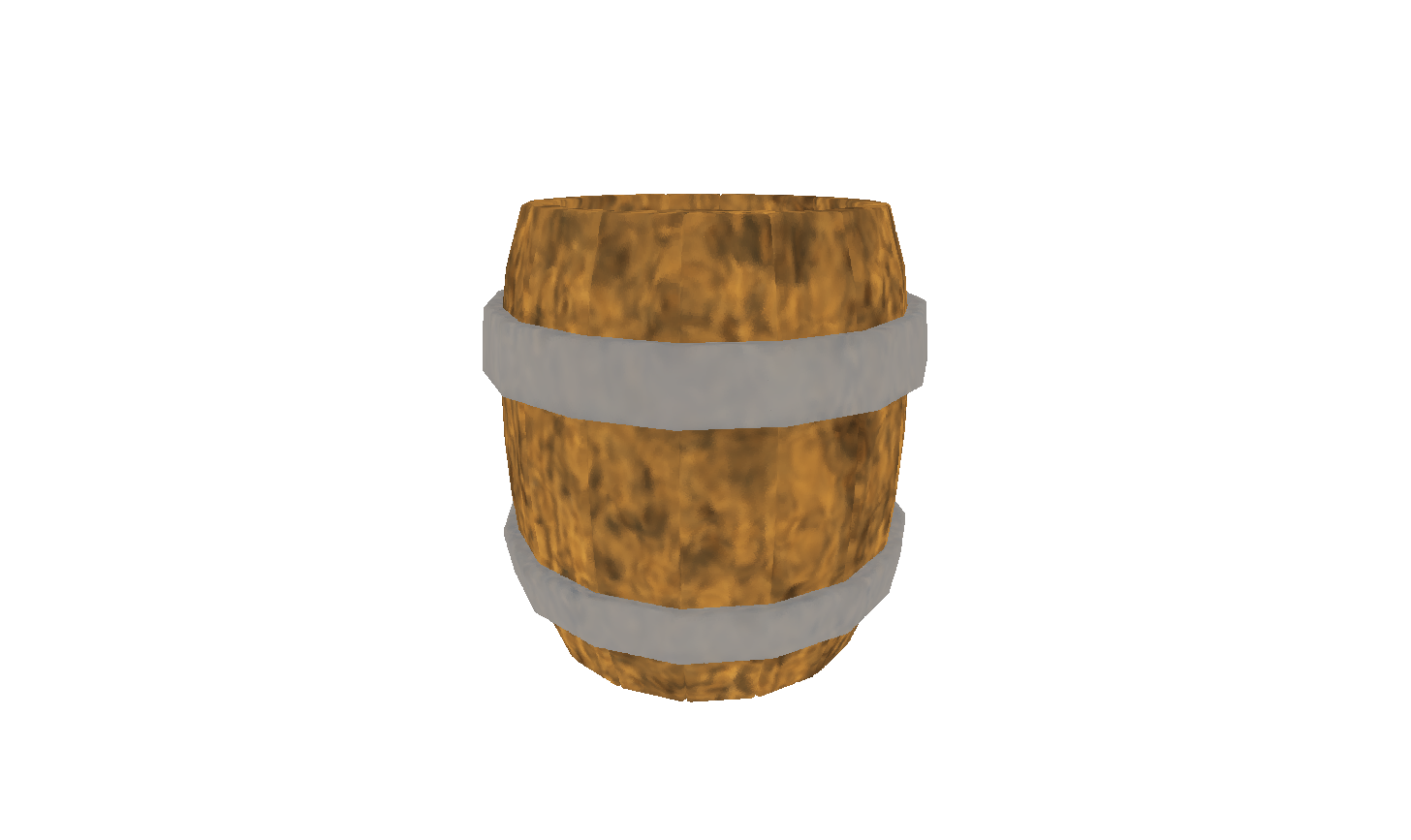}};
            \spy on (3.2,-0.3) in node [left] at (5.1,0);
            \node at (6.4,0) {\includegraphics[height=55pt,trim={200 70 200 110},clip]{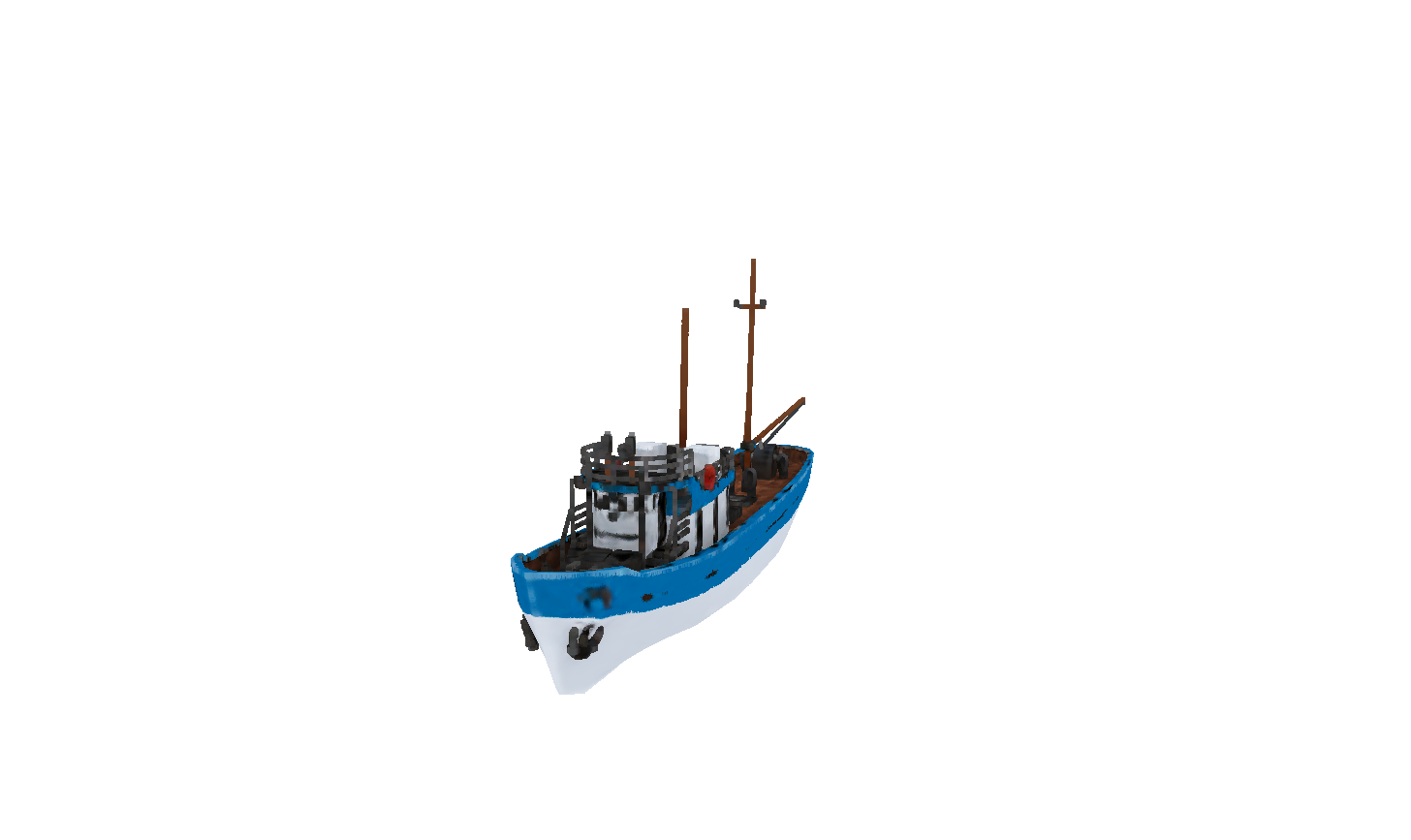}};
            \spy on (6.05,-0.35) in node [left] at (8.3,0);
            \node at (9.6,0) {\includegraphics[height=55pt,trim={200 90 200 70},clip]{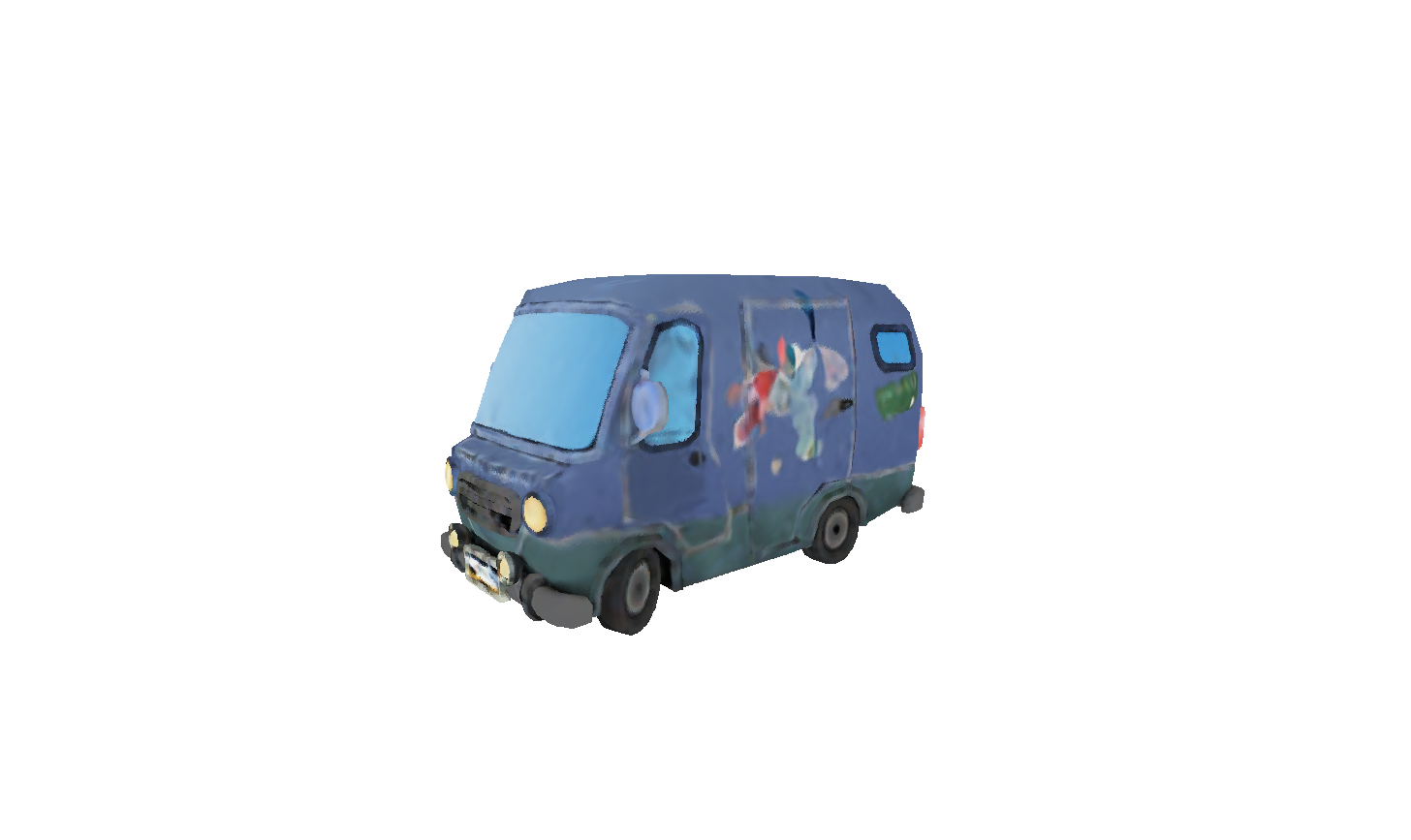}};
            \spy on (9.9,0.06) in node [left] at (11.5,0);
            \node at (12.8,0) {\includegraphics[height=55pt,trim={210 45 190 60},clip]{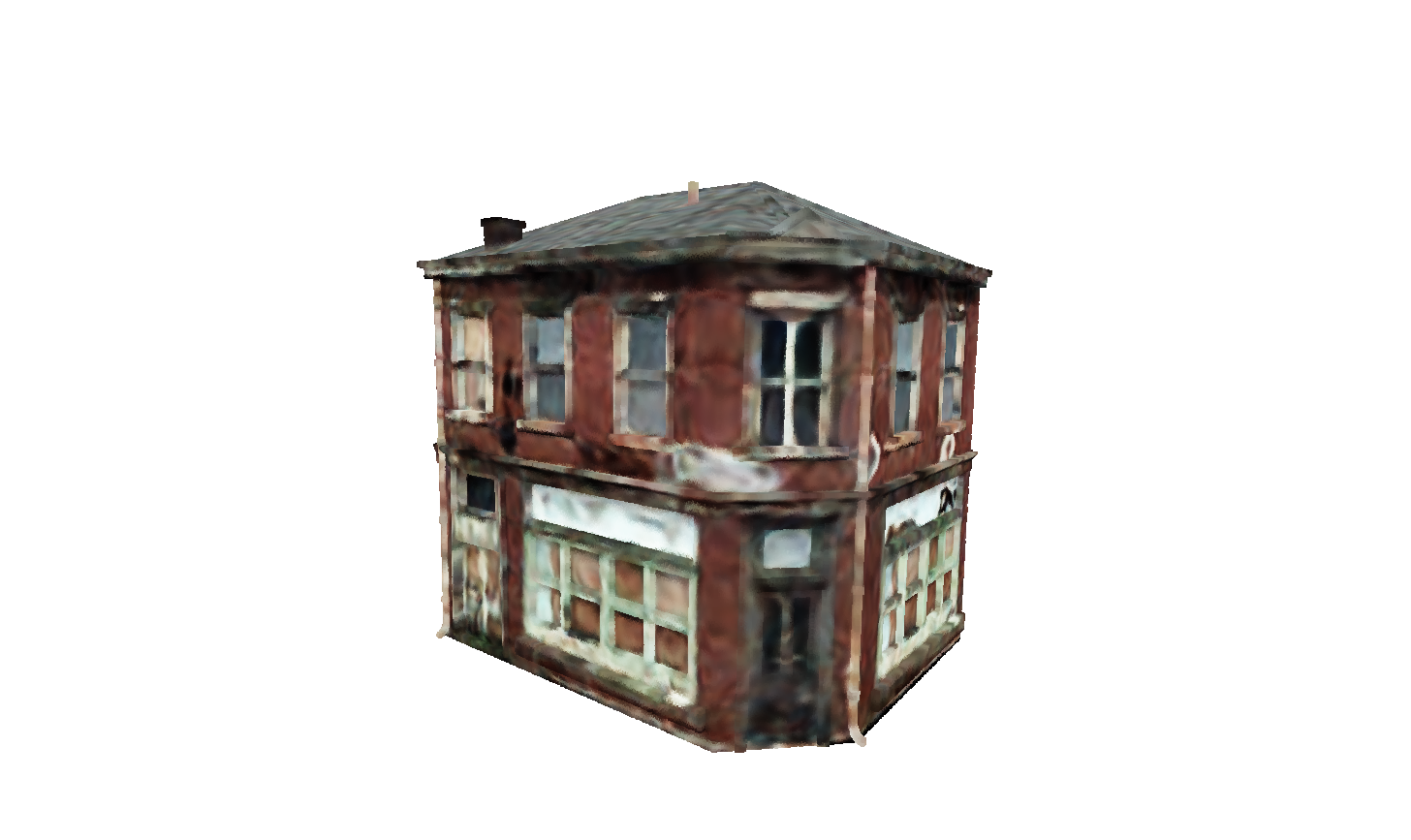}};
            \spy on (12.6,-0.25) in node [left] at (14.7,0);
        \end{tikzpicture}
        \caption{OverfitColor}
        \label{fig:color-comparision-a}
    \end{subfigure}
    \\
    \begin{subfigure}{\linewidth}
        \centering
        \begin{tikzpicture}[spy using outlines={circle,orange,magnification=5,size=1.5cm,connect spies}]
            \node at (0,0) {\includegraphics[height=55pt,trim={200 70 200 70},clip]{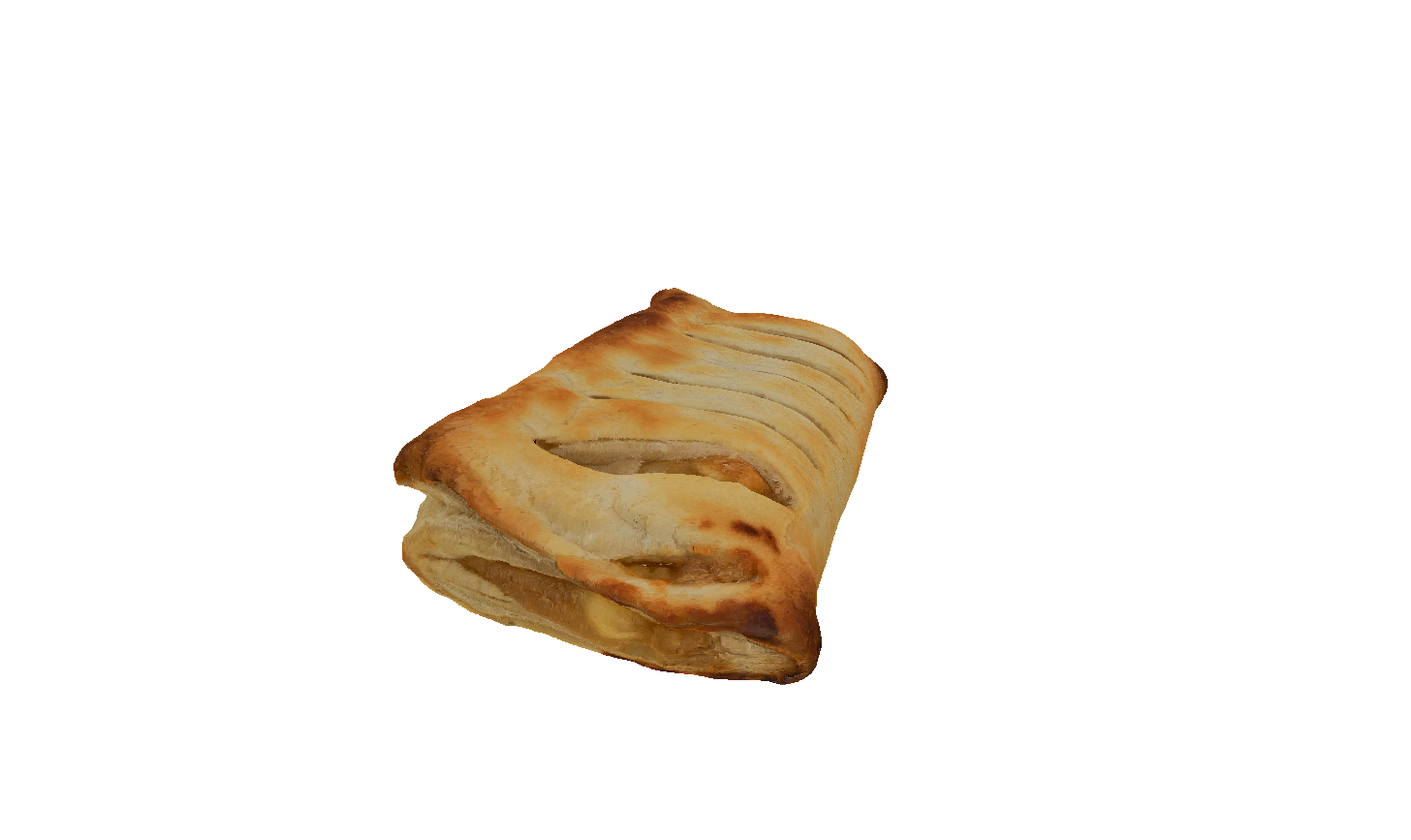}};
            \spy on (0,-0.5) in node [left] at (1.9,0);
            \node at (3.2,0) {\includegraphics[height=55pt,trim={210 70 190 45},clip]{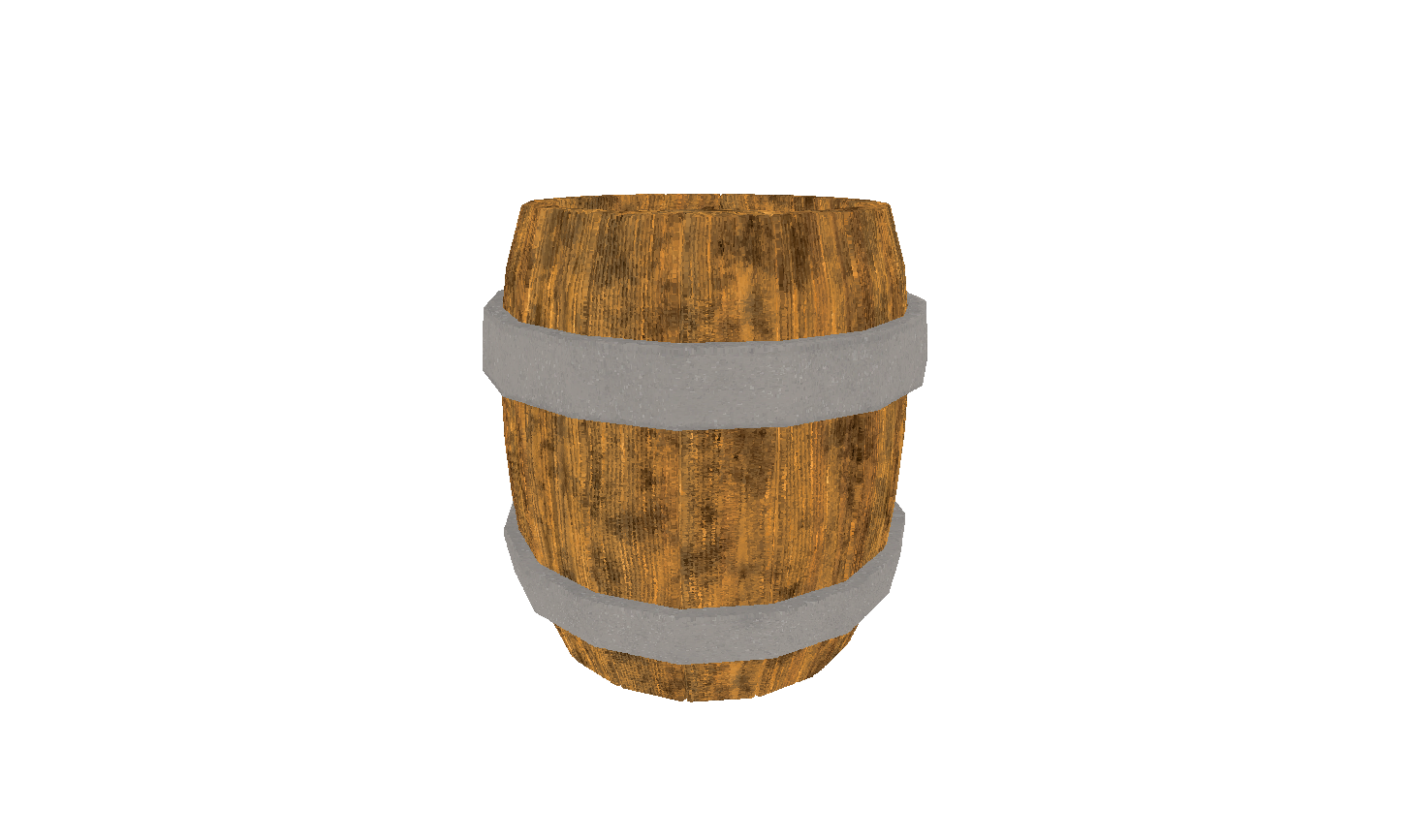}};
            \spy on (3.2,-0.3) in node [left] at (5.1,0);
            \node at (6.4,0) {\includegraphics[height=55pt,trim={200 70 200 110},clip]{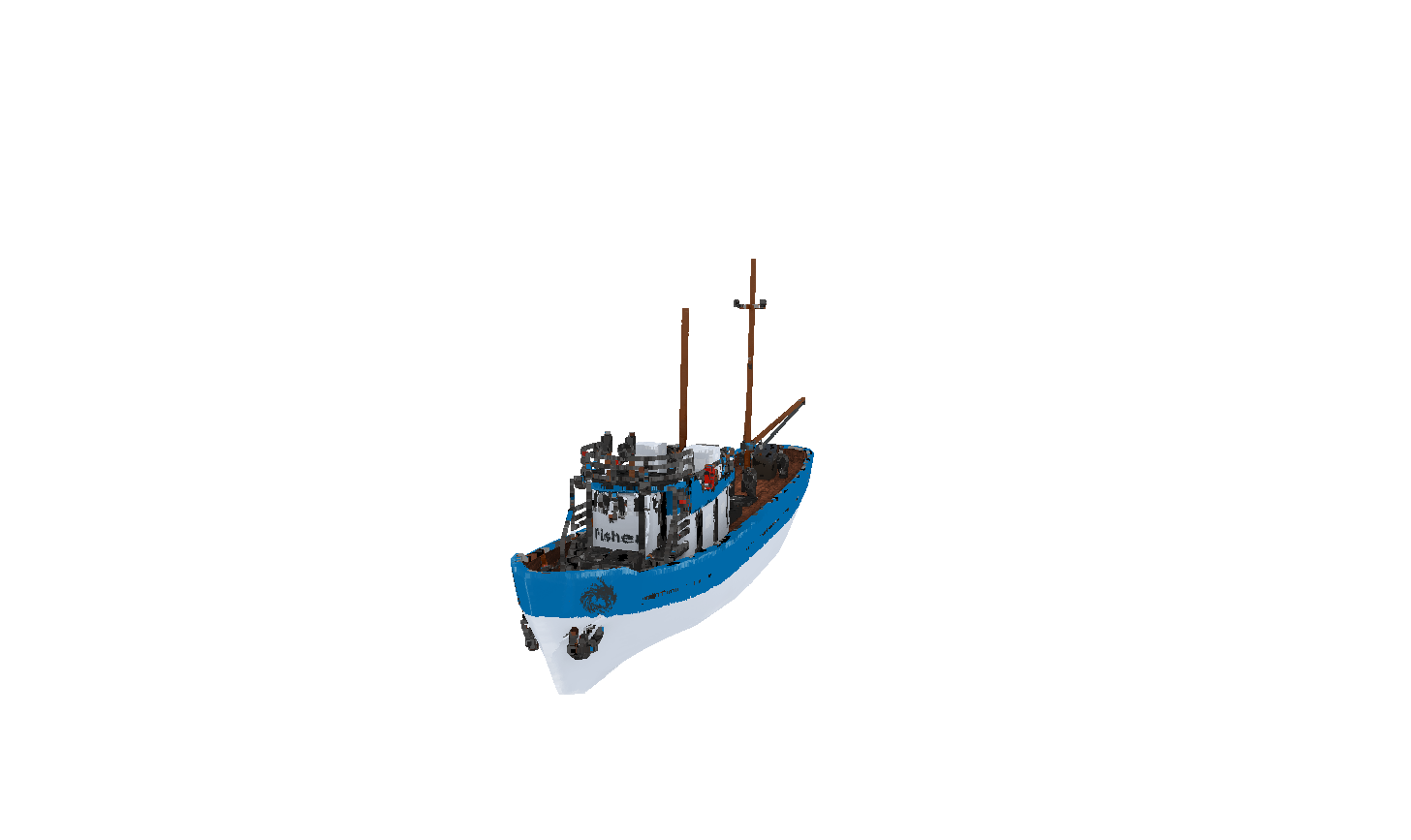}};
            \spy on (6.05,-0.35) in node [left] at (8.3,0);
            \node at (9.6,0) {\includegraphics[height=55pt,trim={200 90 200 70},clip]{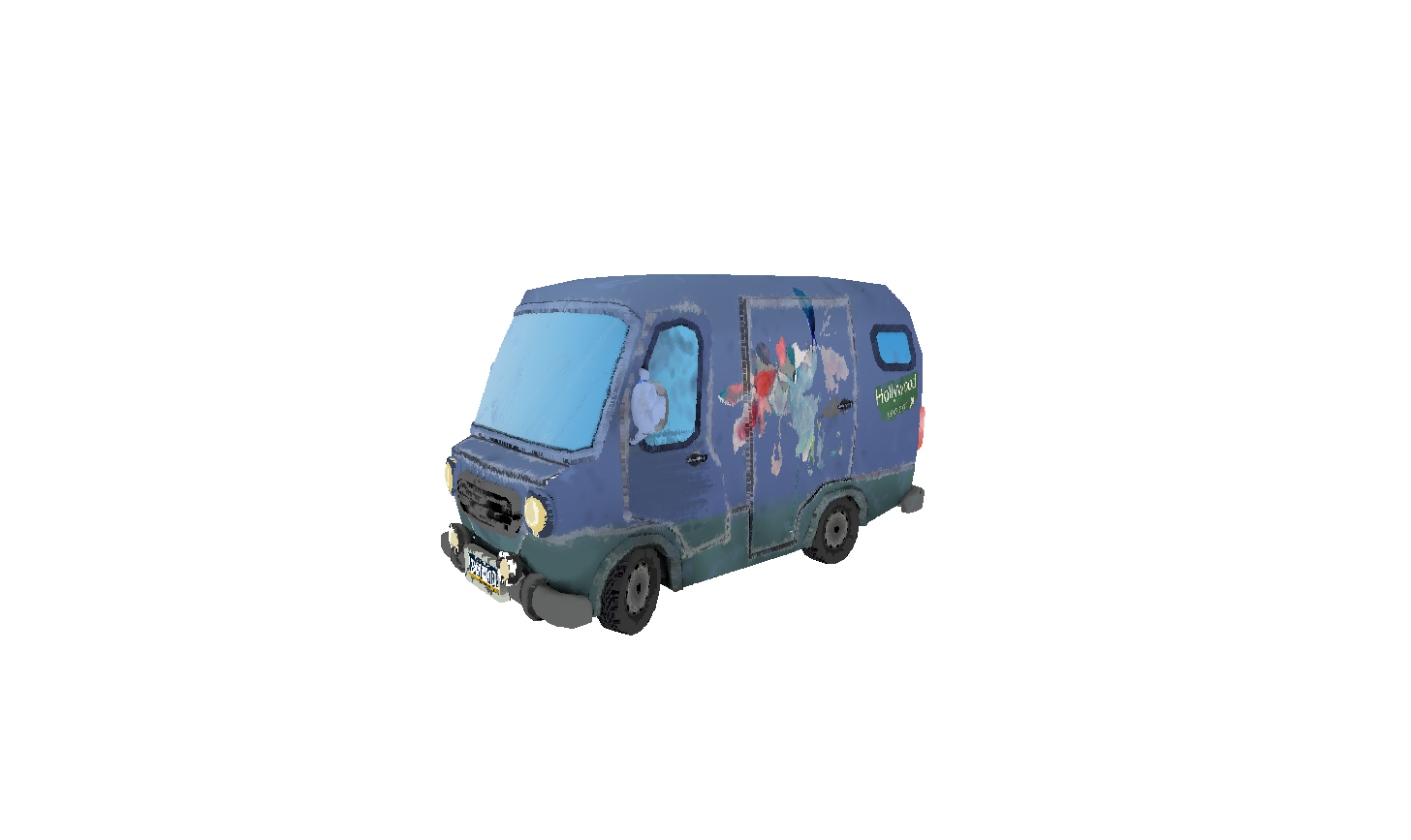}};
            \spy on (9.9,0.06) in node [left] at (11.5,0);
            \node at (12.8,0) {\includegraphics[height=55pt,trim={210 45 190 60},clip]{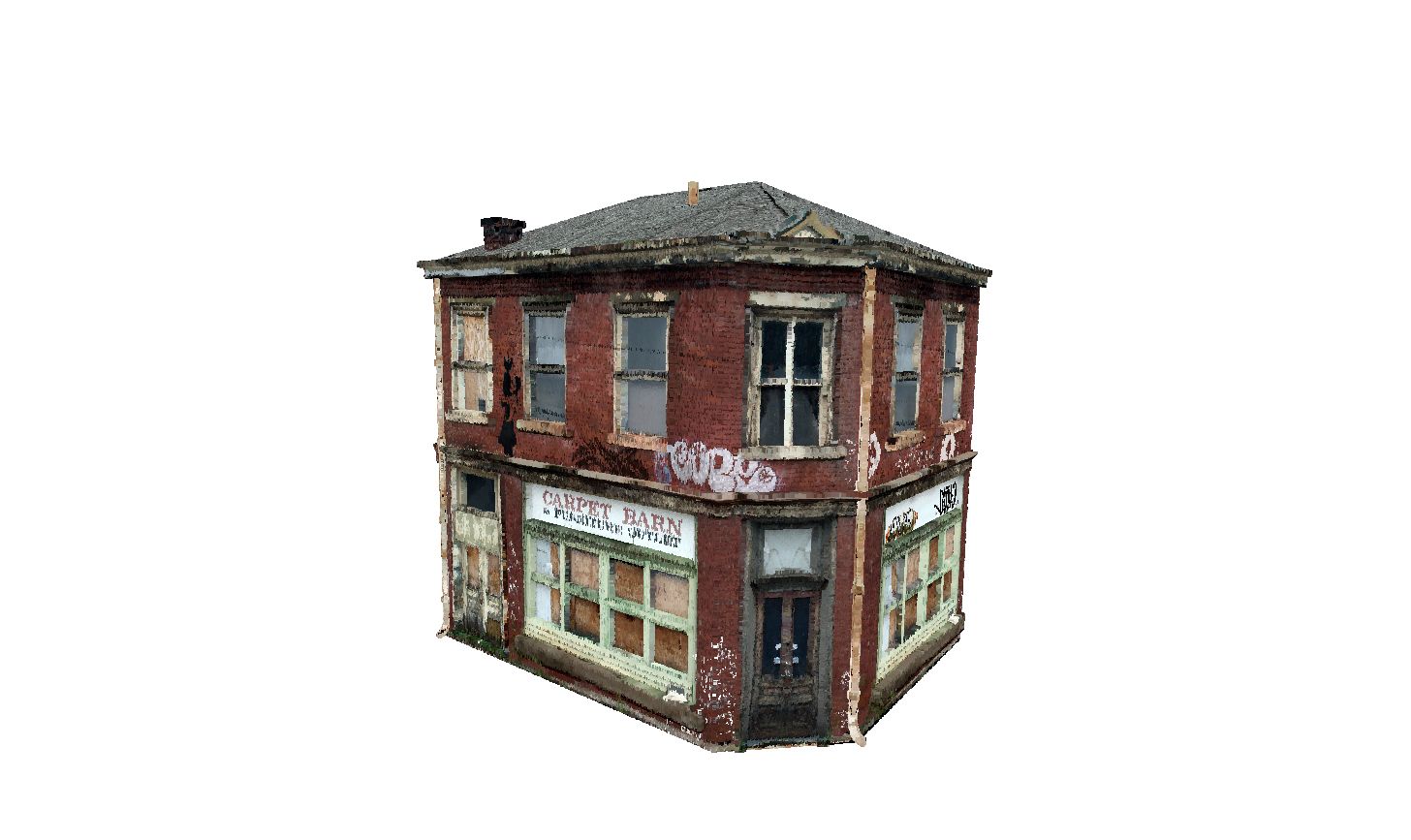}};
            \spy on (12.6,-0.25) in node [left] at (14.7,0);
        \end{tikzpicture}
        \caption{Ours}
        \label{fig:color-comparision-b}
    \end{subfigure}
    \\
    \begin{subfigure}{\linewidth}
        \centering
        \begin{tikzpicture}[spy using outlines={circle,orange,magnification=5,size=1.5cm,connect spies}]
            \node at (0,0) {\includegraphics[height=55pt,trim={200 70 200 70},clip]{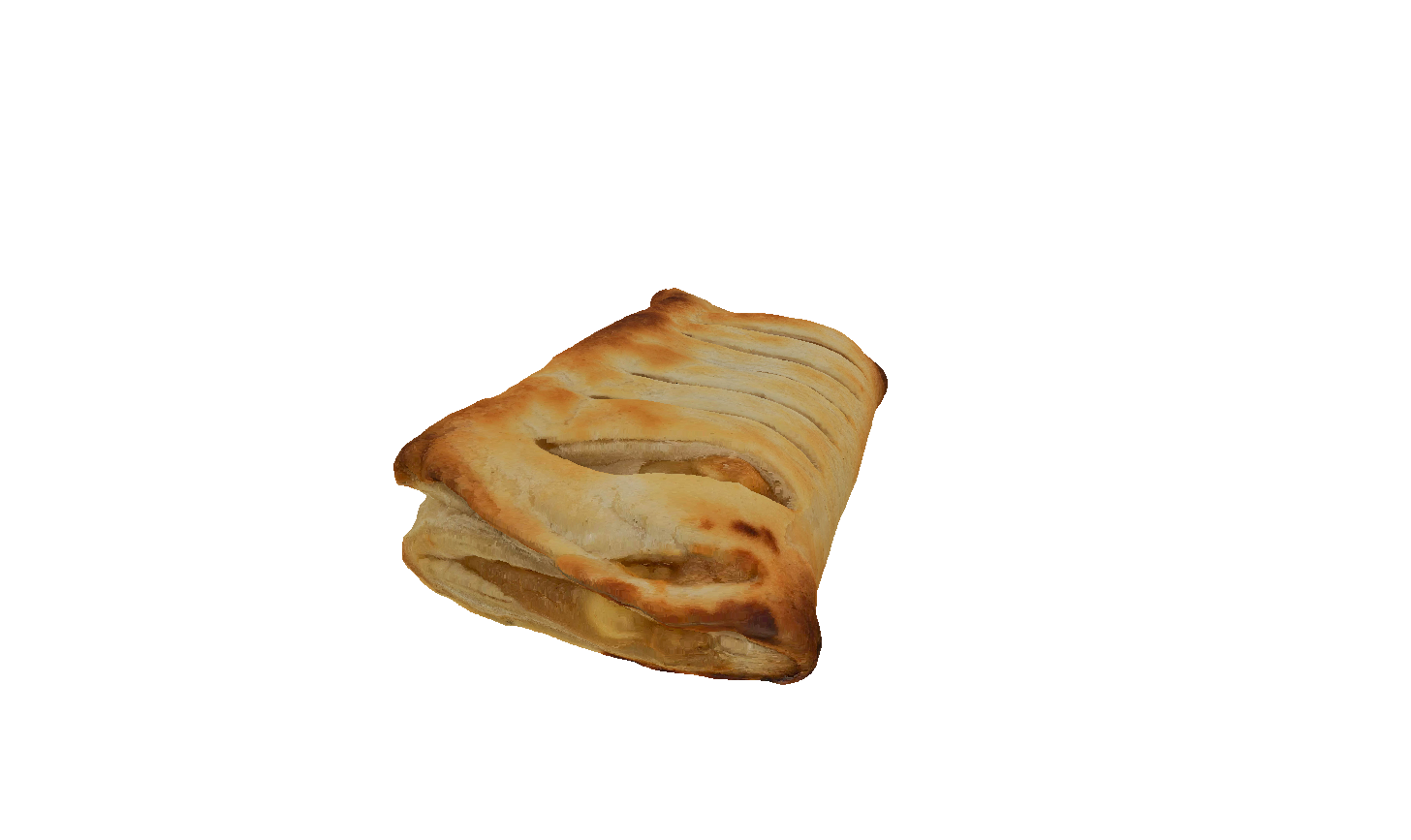}};
            \spy on (0,-0.5) in node [left] at (1.9,0);
            \node at (3.2,0) {\includegraphics[height=55pt,trim={210 70 190 45},clip]{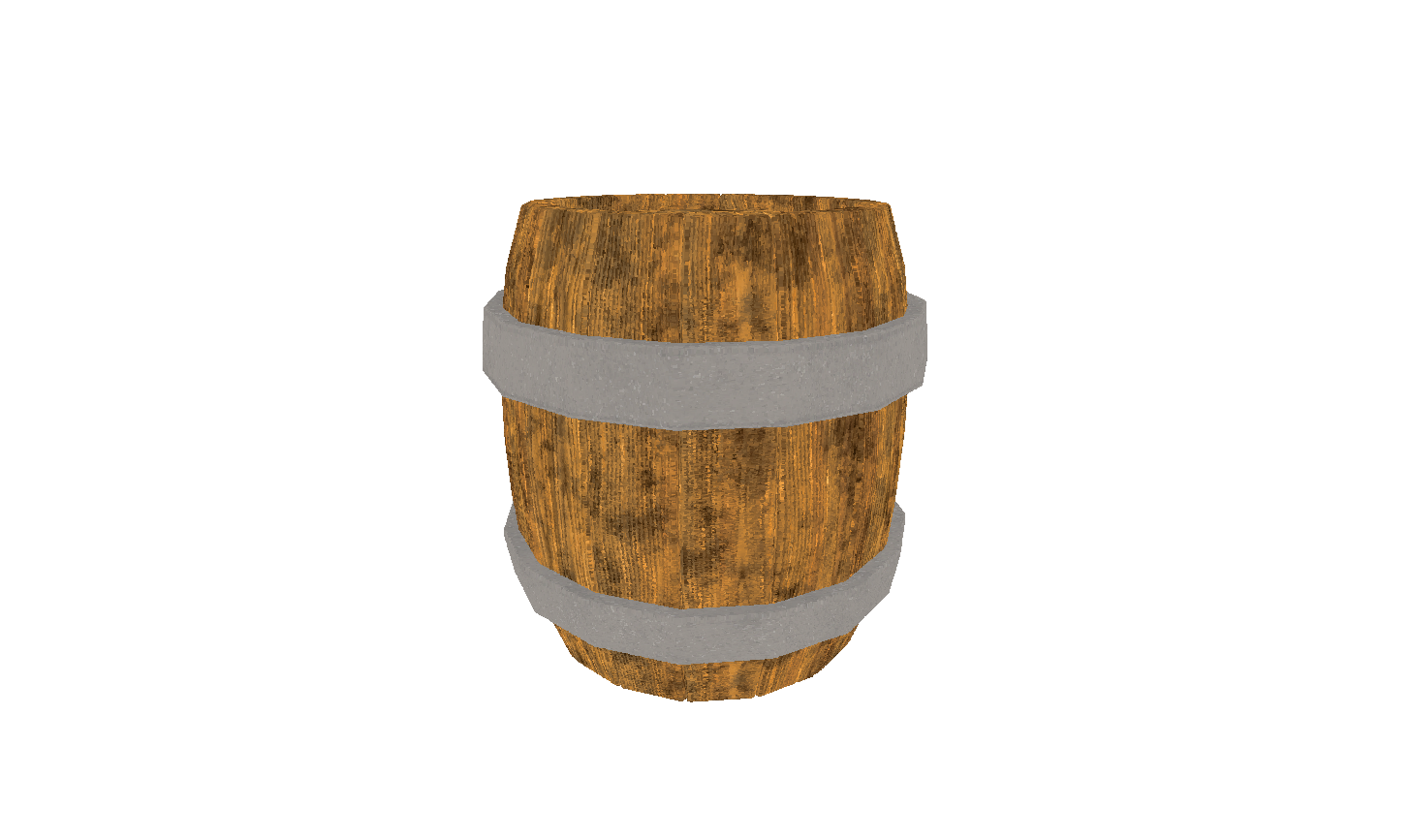}};
            \spy on (3.2,-0.3) in node [left] at (5.1,0);
            \node at (6.4,0) {\includegraphics[height=55pt,trim={200 70 200 110},clip]{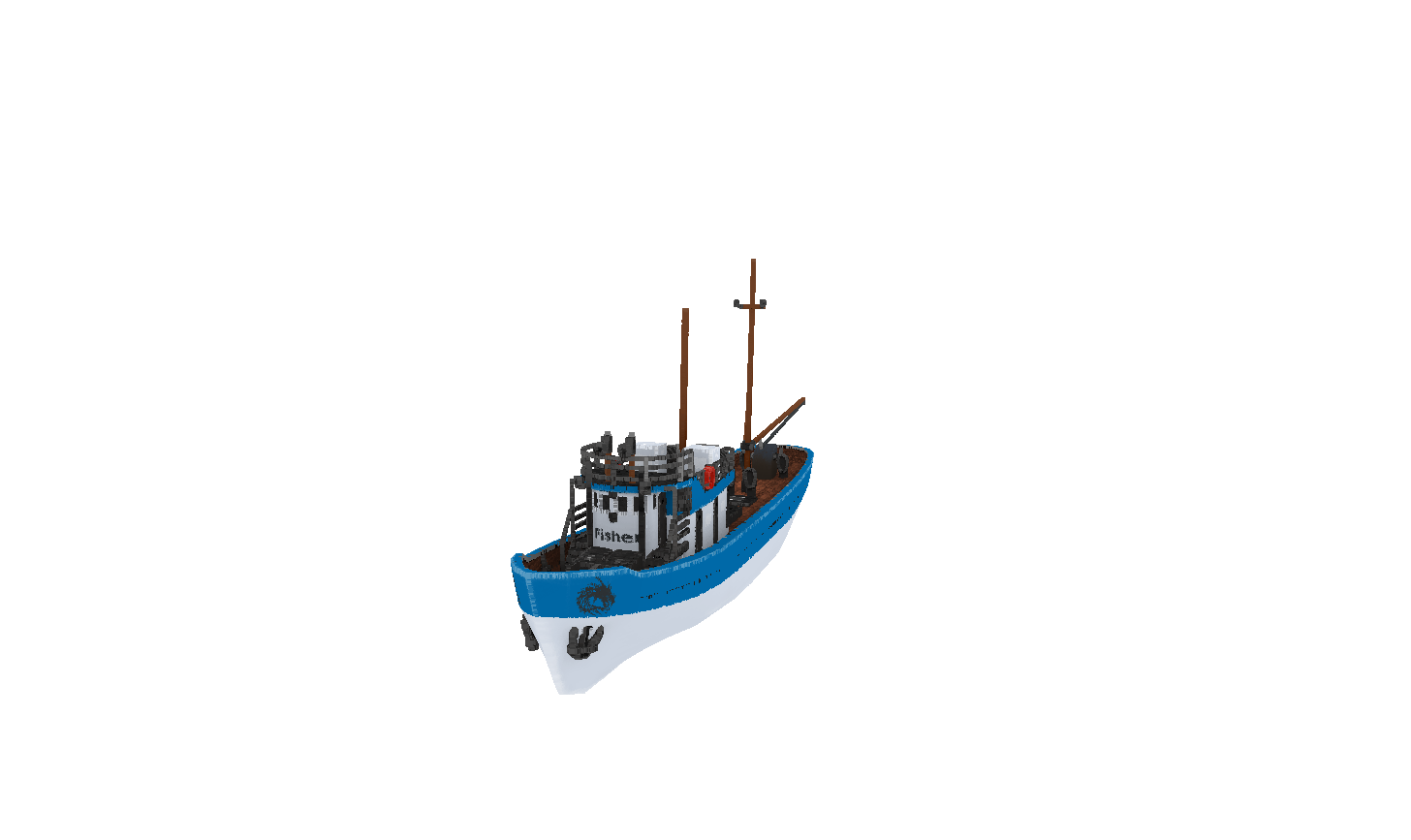}};
            \spy on (6.05,-0.35) in node [left] at (8.3,0);
            \node at (9.6,0) {\includegraphics[height=55pt,trim={200 90 200 70},clip]{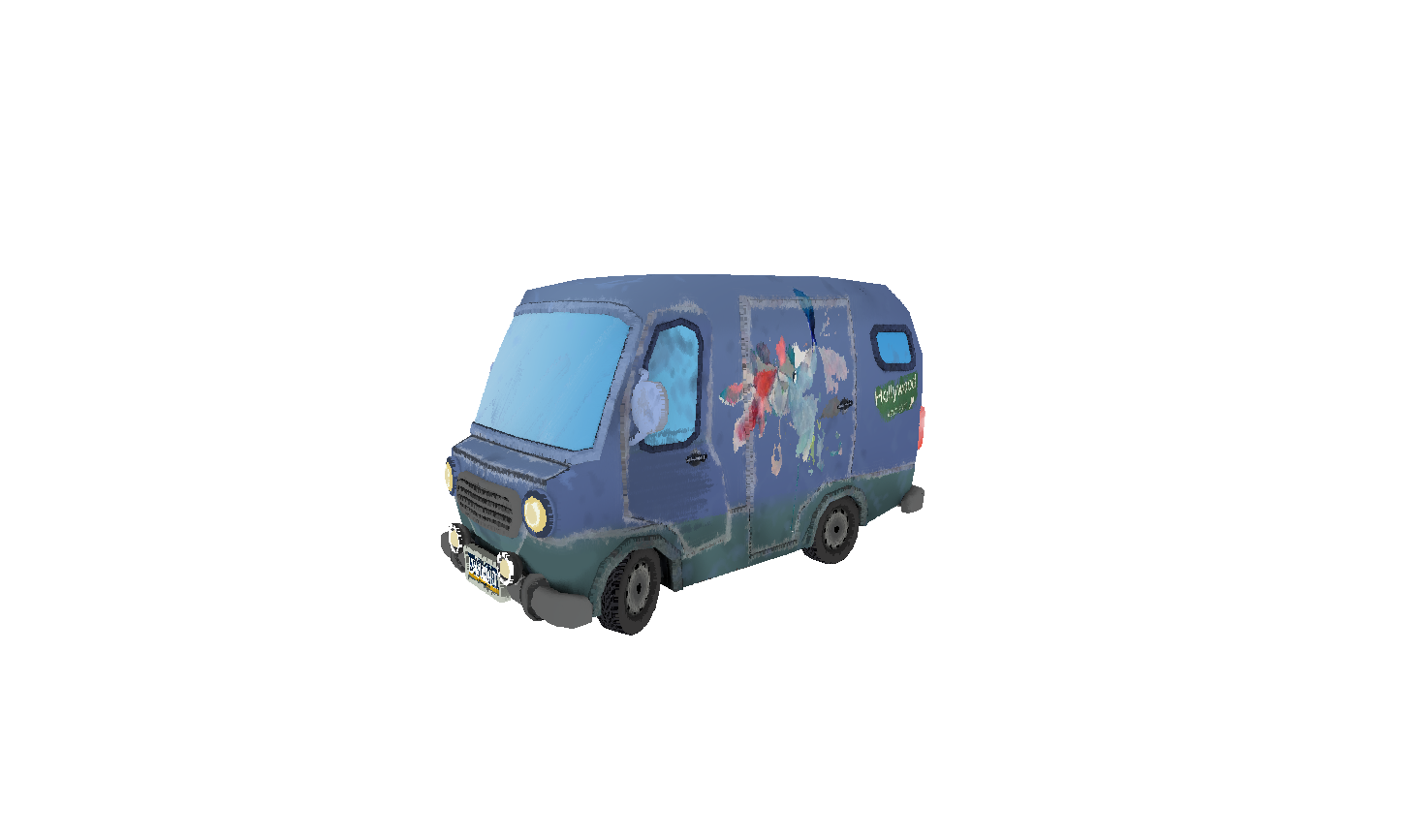}};
            \spy on (9.9,0.06) in node [left] at (11.5,0);
            \node at (12.8,0) {\includegraphics[height=55pt,trim={210 45 190 60},clip]{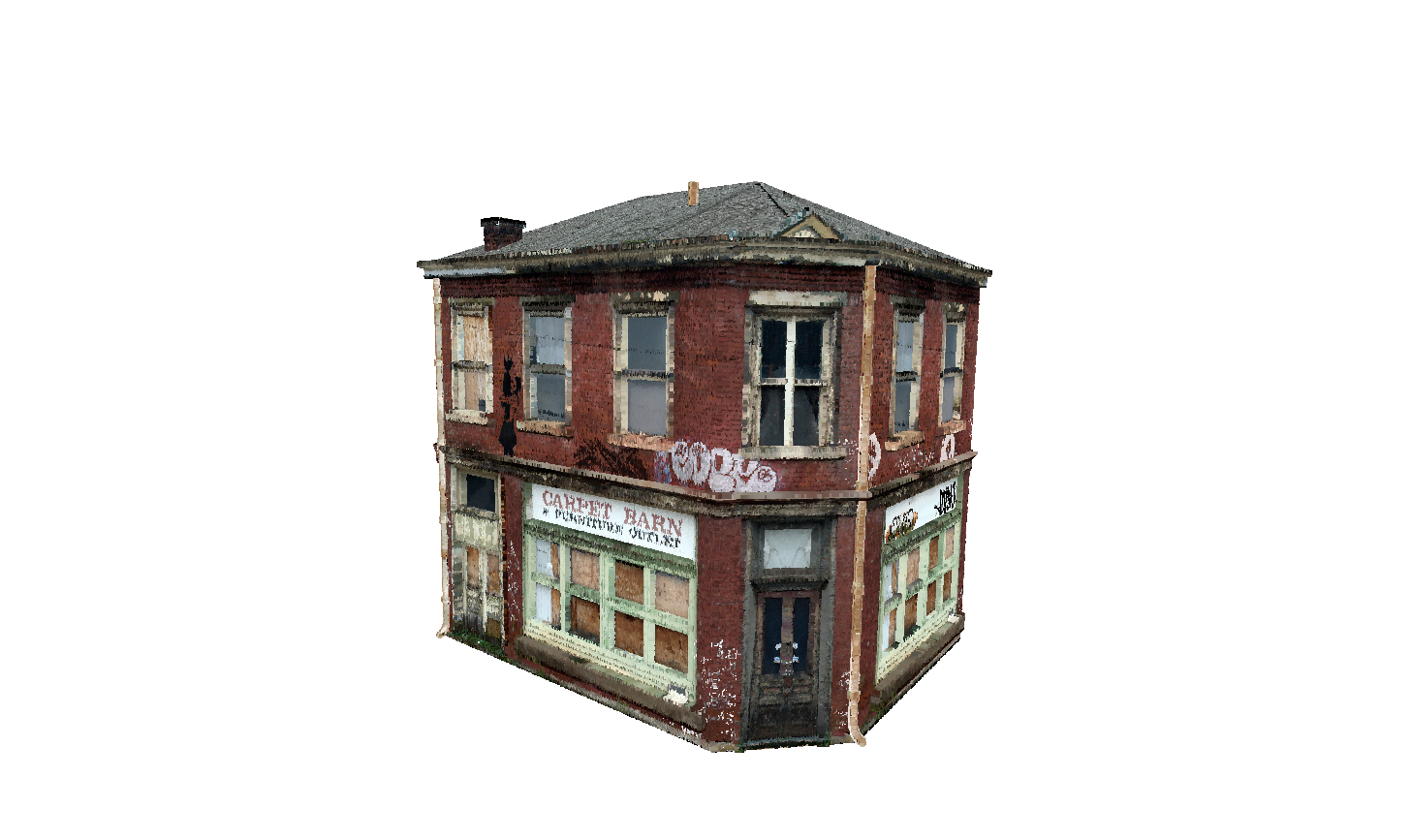}};
            \spy on (12.6,-0.25) in node [left] at (14.7,0);
        \end{tikzpicture}
        \caption{GT}
        \label{fig:color-comparision-c}
    \end{subfigure}
    \caption{The diffuse color of the surface sample points, with zoom-in views shown on the right of each example. Our method can preserve more high frequency details in the diffuse mapping results, compared to learning the surface color directly as a continuous function.}
    \label{fig:color-comparision}
\end{figure*}

%% file: results.tex
\section{Results}\label{sec:results}

We detail our experimental methodology before presenting various texture mapped results and a simple editing scenario. We conclude by discussing how the model is compressed. Our implementation details can be found in our code\footnote{\url{https://github.com/IsaacGuan/NISP}}.

\subsection{Experimental setup}\label{sec:results-setup}

We train our model to represent the surface parameterization of objects. To represent objects' geometry, we train for each object a model by Davies et al.~\cite{davies2020overfit}, which we call \emph{OverfitSDF} for brevity. OverfitSDF also uses $8$ hidden layers with $64$ units per layer. Here we describe the setup for training the neural implicit representations.

\paragraph{Experimental data} We use $16$ 3D objects from TurboSquid\footnote{\url{https://www.turbosquid.com/}} to train our model. Every object is represented as a triangle mesh in the Wavefront \texttt{.obj} format and each has auxiliary appearance image texture file(s). We normalize object sizes to a unit bounding sphere prior to training. For each object, we generate our training data by sampling $10^6$ points in the unit sphere using the importance sampling scheme, and we split these samples into training batches of $2048$ points each. For test and validation, we generate another set of $10^6$ uniformly sampled points on the object's surface.

\paragraph{Optimizer and hyper-parameters} We train all the neural networks using the Adam optimizer~\cite{kingma2015adam}. For \texttt{point2component} and \texttt{point2UV}, we set the learning rate to $5\times 10^{-4}$. For OverfitSDF, the learning rate is $10^{-4}$. We train the models until convergence for $2000$ epochs.

\paragraph{Visualization} We use the sphere tracing algorithm~\cite{hart1996sphere} to visualize our objects, where we march along camera rays by the distance to the object surface at each step location, until the current distance to the surface is smaller than a user-definable threshold $\epsilon$; our results use $\epsilon=10^{-4}$.

\paragraph{Machine configuration and timing} We trained our neural networks on an NVIDIA GeForce RTX 2070 SUPER GPU with $\SI{8}{\giga\byte}$ of memory and CUDA version 10.1. Training our model requires an average of $\SI{1.1}{\hour}$ per object, roughly twice the time needed to train the geometry-only OverfitSDF model. Given $10^4$ query positions, OverfitSDF takes $\SI{0.3}{\second}$ to predict the signed distances and our model takes $\SI{0.6}{\second}$ to predict the UV coordinates.

\subsection{Texture mapping and editing}\label{sec:results-texture}

We demonstrate three proof-of-concept applications facilitated by the surface parameterization learned using our method: diffuse texture mapping, normal texture mapping, and (post-training) texture editing. We also quantitatively evaluate the distortion of our learned parameterizations and compare the texture mapping results by our method to the synthesized views by the state-of-the-art neural appearance models.

\input{figures/uv_comparison}

\paragraph{Diffuse mapping} To benchmark the utility and fidelity of diffuse texture mapping using our neural surface parameterization, we train an OverfitSDF-style baseline model to represent the surface color of objects as a continuous function in the 3D space --- i.e., directly mapping 3D locations to RGB colors using an MLP network. 
We call this baseline model \emph{OverfitColor}, and configure and train this model similarly as described before, including pre-processing the input with Fourier positional encoding. 
As shown in Figure~\ref{fig:color-comparision}, we compare the color on the surface sample points generated by diffuse mapping using our model (Figure~\ref{fig:color-comparision-b}) to those learned by OverfitColor (Figure~\ref{fig:color-comparision-a}) and to the ground truth (GT) color (Figure~\ref{fig:color-comparision-c}). We clearly observe that our results more faithfully preserve the all-frequency texture detail, such as the lettering on the building's shop sign.

To illustrate the advantages of conditioning the input of neural networks with spatial decomposition when learning surface parameterization signals, we train a \texttt{point2UV} network without the conditioned inputs provided by \texttt{point2component}, i.e., mapping 3D locations directly to UV coordinates. Figure~\ref{fig:uv-comparision} compares the UV coordinates predicted by \texttt{point2UV} with and without conditioned inputs, and we observe --- from the scatter charts of the UV coordinates --- that, without input conditions, \texttt{point2UV} fails to provide accurate predictions in regions close to the boundary of each parameterization chart. This results in visual artifacts during, e.g., diffuse texture mapping, such as distortions on the boat and the anomalies on the characters' hands (Figure~\ref{fig:uv-comparision-a}). With conditioned inputs, however, these distortions are largely mitigated (Figure~\ref{fig:uv-comparision-b}), resulting in almost identical diffuse mapping results compared to the GT parameterization.

\input{figures/normal_mapping}

\paragraph{Normal mapping} The surface parameterization our model learns also allows users to easily apply normal maps to object surfaces implicitly defined by the SDF, adding more visual details to the surface that a neural implicit surface cannot capture. Figure~\ref{fig:normal-mapping} illustrates our rendered results using normal mapping on neural implicit surfaces: here, we can easily add geometric detail, such as the barrel's wood grain and the house's brick wall patterns (Figure~\ref{fig:normal-mapping-b}), using content creation paradigms familiar to digital artists.

\input{figures/texture_editing}

\paragraph{Texture editing} As our model learns directly from the original texture space, users can easily edit original texture images to change the appearance of the neural implicit surface \emph{after} it has been trained. Figure~\ref{fig:texture-editing} demonstrates examples of editing the diffuse maps applied to our neural implicit surfaces, where we apply a photo filter to the texture image of the watermelon and paint letters on the roof of the van.

\paragraph{Distortion metric} We quantitatively evaluate the distortion of the learned parameterizations according to the metric described by Sorkine et al.~\cite{sorkine2002bounded}. Given the affine mapping between the original triangle mesh and its corresponding parameterization, the distortion $\delta$ caused to each triangle is measured using the largest and smallest singular values $\sigma_\text{max}$ and $\sigma_\text{min}$ of the Jacobian matrix of this transformation, written as:
\begin{equation}\label{eq:distortion-metric}
\delta=\max\left(\sigma_\text{max},\frac{1}{\sigma_\text{min}}\right).
\end{equation}

\input{tables/distortion_metric}

We feed the vertices of objects' triangle meshes into the neural networks to generate the parameterized meshes. For each parameterization, we compute the mean value $\delta_\text{mean}$, the maximum value $\delta_\text{max}$, and the standard deviation $\delta_\text{std}$ of the distortions caused to all triangles in the mesh. Table~\ref{tab:distortion-metric} demonstrates this quantitative evaluation for the learned and GT parameterizations in Figure~\ref{fig:uv-comparision}. We see how much the distortion is improved using conditioned input of our model.

\input{figures/model_comparison}

\paragraph{Model comparison} We compare our texture mapping results to the synthesized views by two different neural appearance models, namely, the scene representation network (SRN)~\cite{sitzmann2019scene} and NeRF~\cite{mildenhall2020nerf}. Figure~\ref{fig:model-comparison} illustrates this comparison. We further compute the mean squared error (MSE), peak signal-to-noise ratio (PSNR), and structural similarity index (SSIM) between GT and the synthesized views as well as our rendered results, as summarized in Table~\ref{tab:model-comparison-quantitative}. We can see that our method outperforms these neural appearance models in preserving high frequency geometry and appearance details. Both SRN and NeRF require a higher GPU memory usage, as compared to our method. Using our experimental device (a GPU with $\SI{8}{\giga\byte}$ of memory) and to train the models at a reasonable speed, we have to reduce the training views to a lower resolution (e.g., $128 \times 128$). We train both models until convergence for $200\,000$ epochs, taking around $\SI{8}{\hour}$ for SRN and $\SI{4}{\hour}$ for NeRF. 
However, our method incurs substantially smaller GPU memory and computational cost, with a quicker training time.
With the auxiliary texture maps, we can render our texture mapping results using traditional rendering pipelines at arbitrary high resolutions (e.g., $1024 \times 1024$ or higher).

\input{tables/model_comparison_quantitative}

\subsection{Model compression}\label{sec:results-compression}

We can also interpret our model along with OverfitSDF as a compression strategy for large 3D objects represented as meshes.
The average size of the \texttt{.obj} files of the $16$ experimental objects is $\SI{13.4}{\mega\byte}$, while the network weights of our model require $\SI{332}{\kilo\byte}$ per object and the weights of OverfitSDF require $\SI{155}{\kilo\byte}$ per object, which total to $\SI{487}{\kilo\byte}$ for the entire geometry and surface parameterization representation. This already yields a compression rate of $1:27$, but can be further compressed by applying the lottery ticket method~\cite{frankle2019lottery}, i.e., by seeking out sparse trainable subnetworks --- called winning tickets --- from the originally trained neural model. These (much more compact) subnetworks reach the same test accuracy as the original network.

Following the lottery ticket training methodology~\cite{frankle2019lottery}, we identify winning tickets iteratively, i.e., in each training epoch, pruning and freezing $q^{\frac{1}{t}}\,\%$ of the weights that survive the previous epoch from the original model (trained for $t$ epochs), 
to obtain a final pruning rate of $q\,\%$. We use three strategies to decide the weights to be pruned or kept:
\begin{enumerate*}[label=(\roman*)]
\item prune the smallest magnitude weights at each prunable layer;
\item prune the smallest weights across all prunable layers;
\item keep the weights that have the largest magnitude from a pre-trained model~\cite{zhou2019deconstructing}.
\end{enumerate*}
We refer to these three strategies as \texttt{smallest}, \texttt{smallest\char`_global}, and \texttt{large\char`_final}.

\input{figures/model_pruning}

\input{figures/model_pruning_visual}

We demonstrate the results of these pruning strategies on the Mario object, where \texttt{point2component} and \texttt{point2UV} have $43\,454$ and $31\,874$ trainable weights, respectively. We use $10$ pruning rates ranging from $\SI{20}{\percent}$ to $\SI{99}{\percent}$ and use the test accuracy and MAE as evaluation metrics for \texttt{point2component} and \texttt{point2UV}. Figure~\ref{fig:model-pruning} summarizes our findings: for \texttt{point2component}, the test accuracy does not drop significantly until a pruning rate of $\SI{84.4}{\percent}$, and for \texttt{point2UV}, the test MAE stays around $5\times 10^{-4}$ until a pruning rate of $\SI{64.4}{\percent}$, using the pruning strategy \texttt{smallest}. Figure~\ref{fig:model-pruning-visual} visualizes the UV predictions made by the pruned networks as scatter charts, where we see the predictions will not distort remarkably until pruning \texttt{point2component} at $\SI{84.4}{\percent}$ or pruning \texttt{point2component} at $\SI{64.4}{\percent}$.

\input{tables/model_compression}

We can save the weights of the pruned networks as sparse matrices, as pruned weights are set to zero. Table~\ref{tab:model-compression} compares the sizes of the weights of the unpruned and pruned models, where \texttt{point2component} is pruned at $\SI{84.4}{\percent}$ and \texttt{point2UV} is pruned at $\SI{64.4}{\percent}$ using the \texttt{smallest} strategy. The pruned model is compressed using different coding formats of sparse matrices, namely, the coordinate format (COO), the compressed sparse column format (CSC), the compressed sparse row format (CSR), and the diagonal storage format (DIA). We observe that our model can be further reduced from $\SI{303.5}{\kilo\byte}$ to roughly $\SI{126}{\kilo\byte}$.

%% file: figures/uv_comparison.tex
\begin{figure*}[t!]
    \centering
    \begin{subfigure}{\linewidth}
        \centering
        \begin{tikzpicture}[spy using outlines={circle,orange,magnification=5,size=1.5cm, connect spies}]
            \node at (0.5,0) {\includegraphics[height=60pt,trim={250 90 250 90},clip]{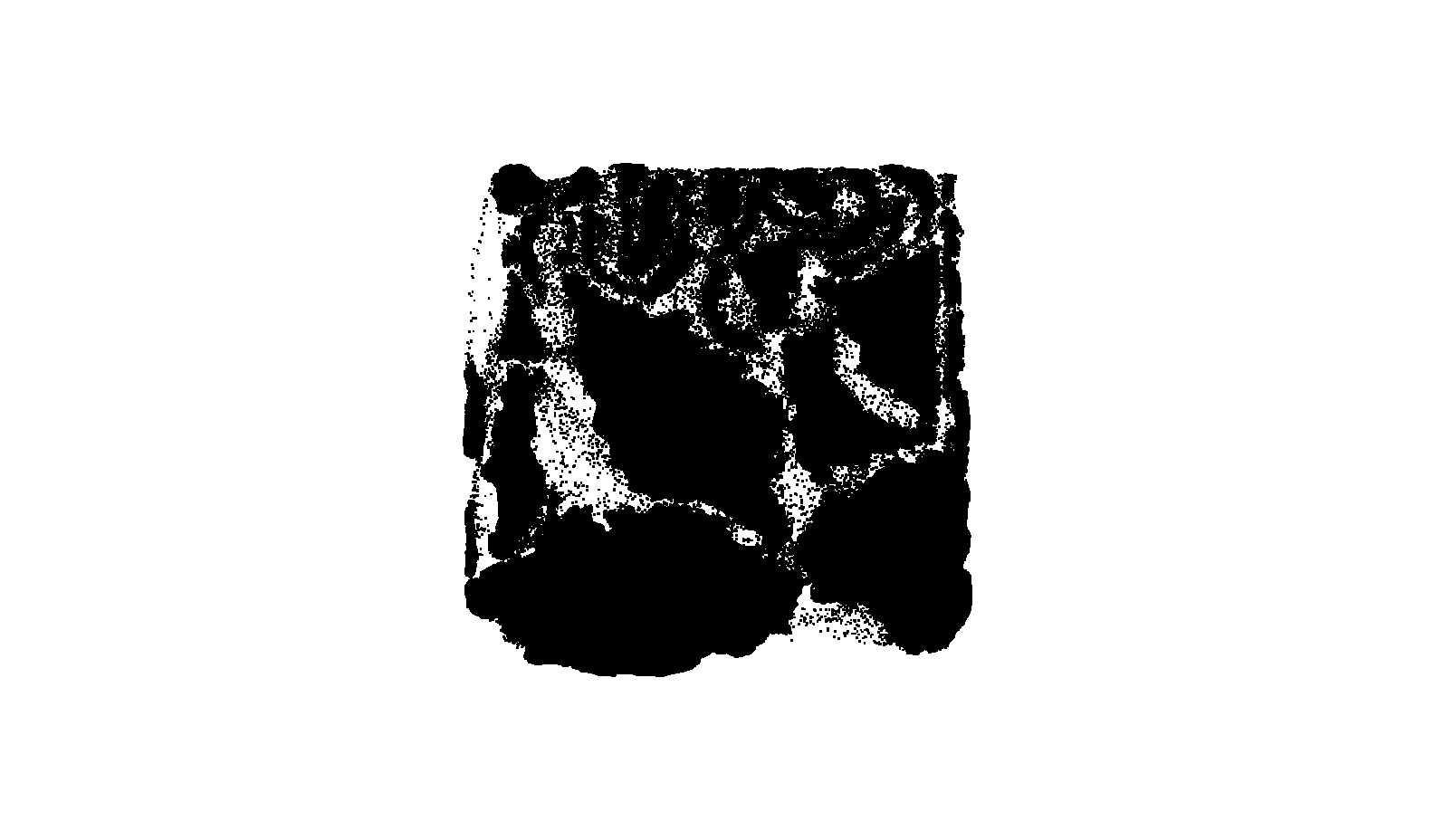}};
            \node at (0,-2) {\includegraphics[height=55pt,trim={200 70 200 70},clip]{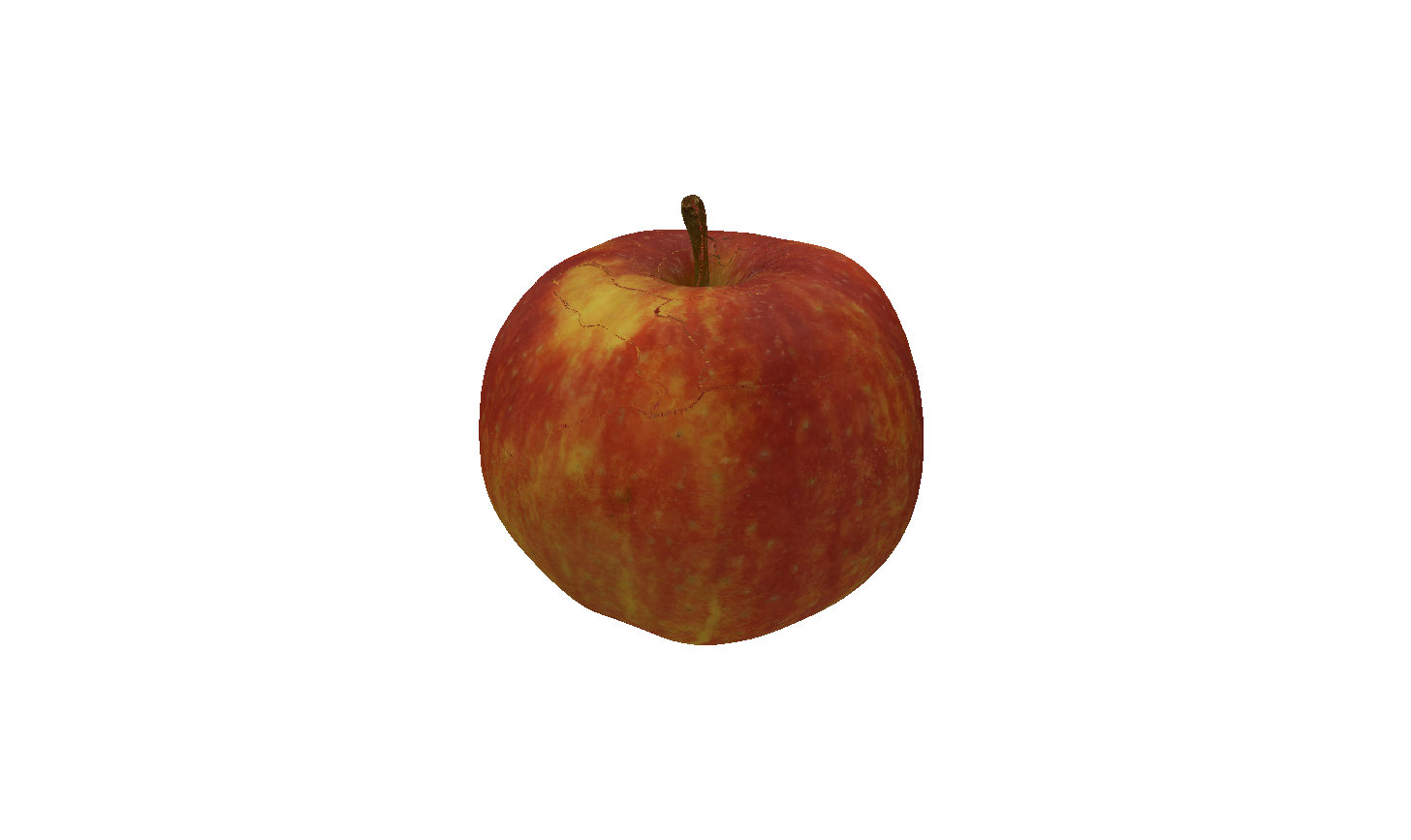}};
            \spy on (-0.2,-1.6) in node [left] at (1.9,-2);
            \node at (3.7,0) {\includegraphics[height=60pt,trim={250 90 250 90},clip]{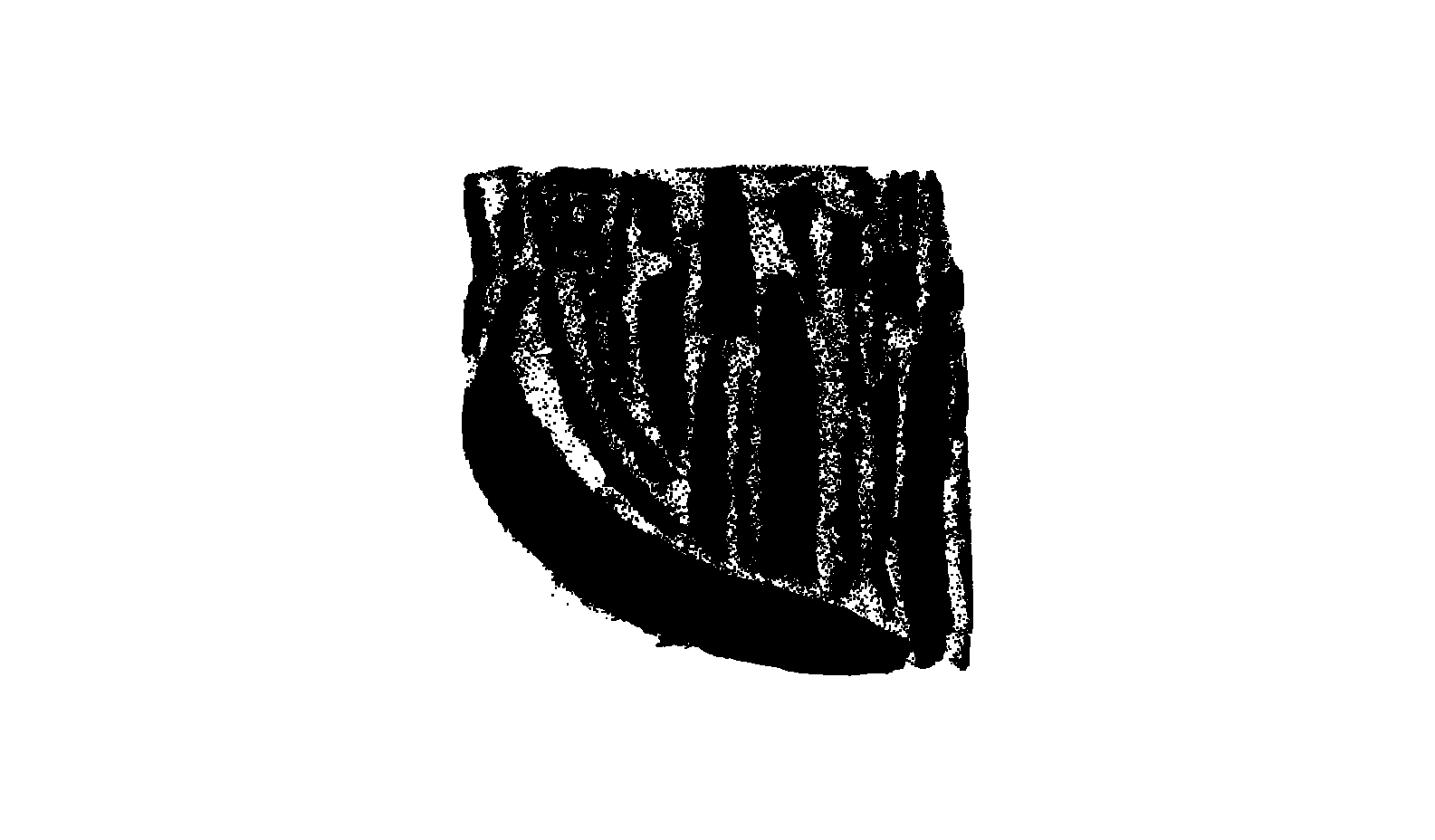}};
            \node at (3.2,-2) {\includegraphics[height=55pt,trim={200 70 200 70},clip]{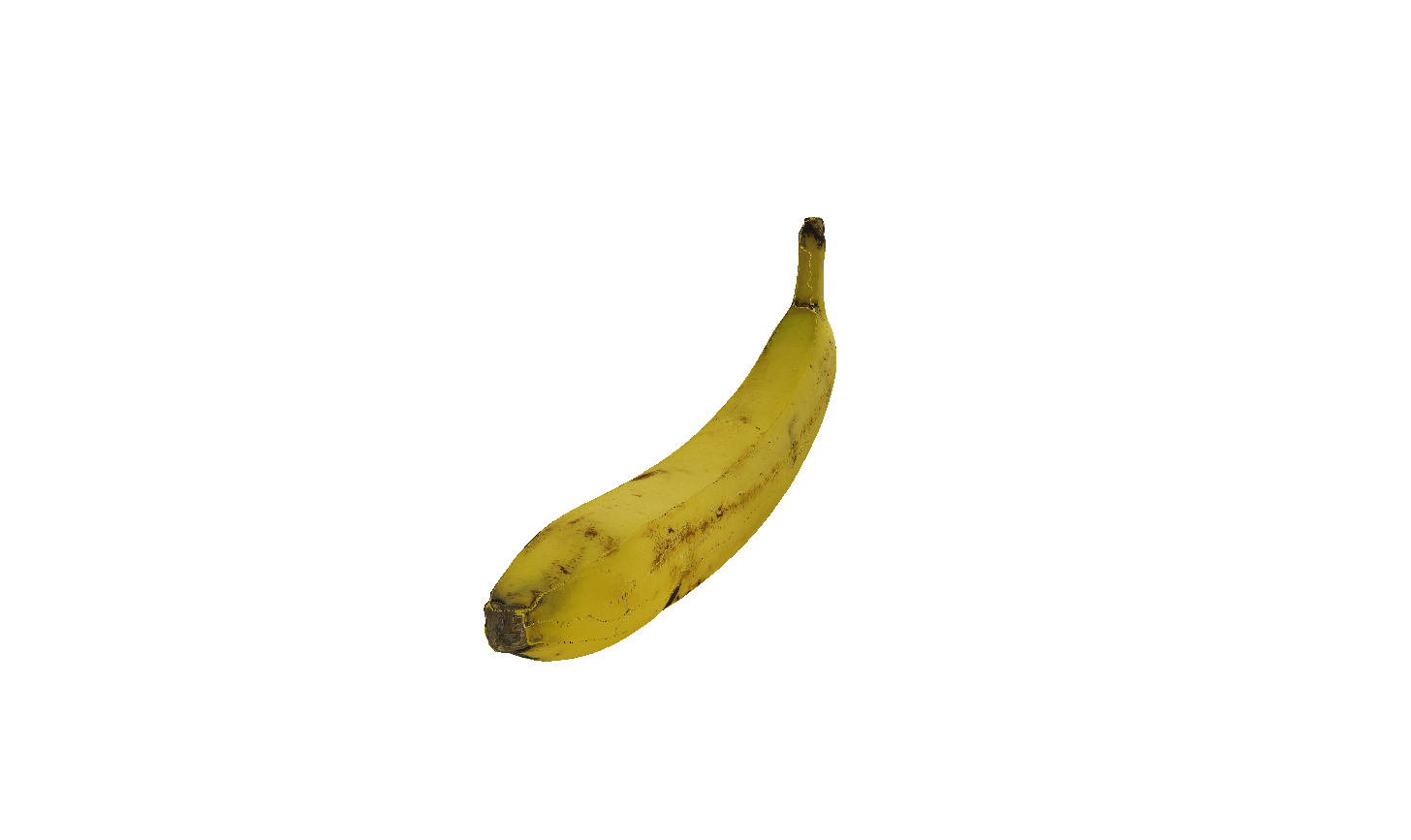}};
            \spy on (2.65,-2.65) in node [left] at (5.1,-2);
            \node at (6.9,0) {\includegraphics[height=60pt,trim={250 90 250 90},clip]{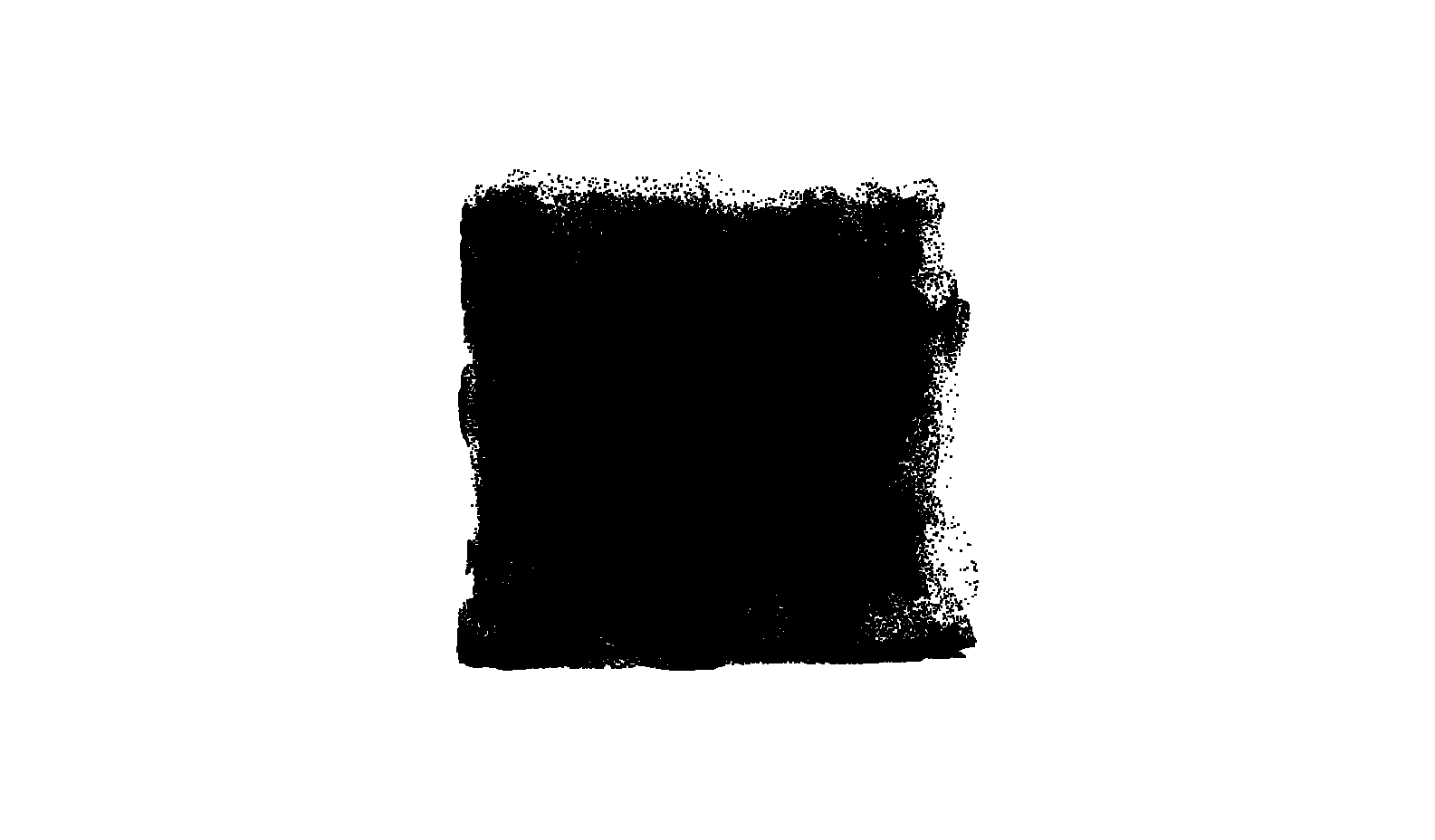}};
            \node at (6.4,-2) {\includegraphics[height=55pt,trim={200 70 200 110},clip]{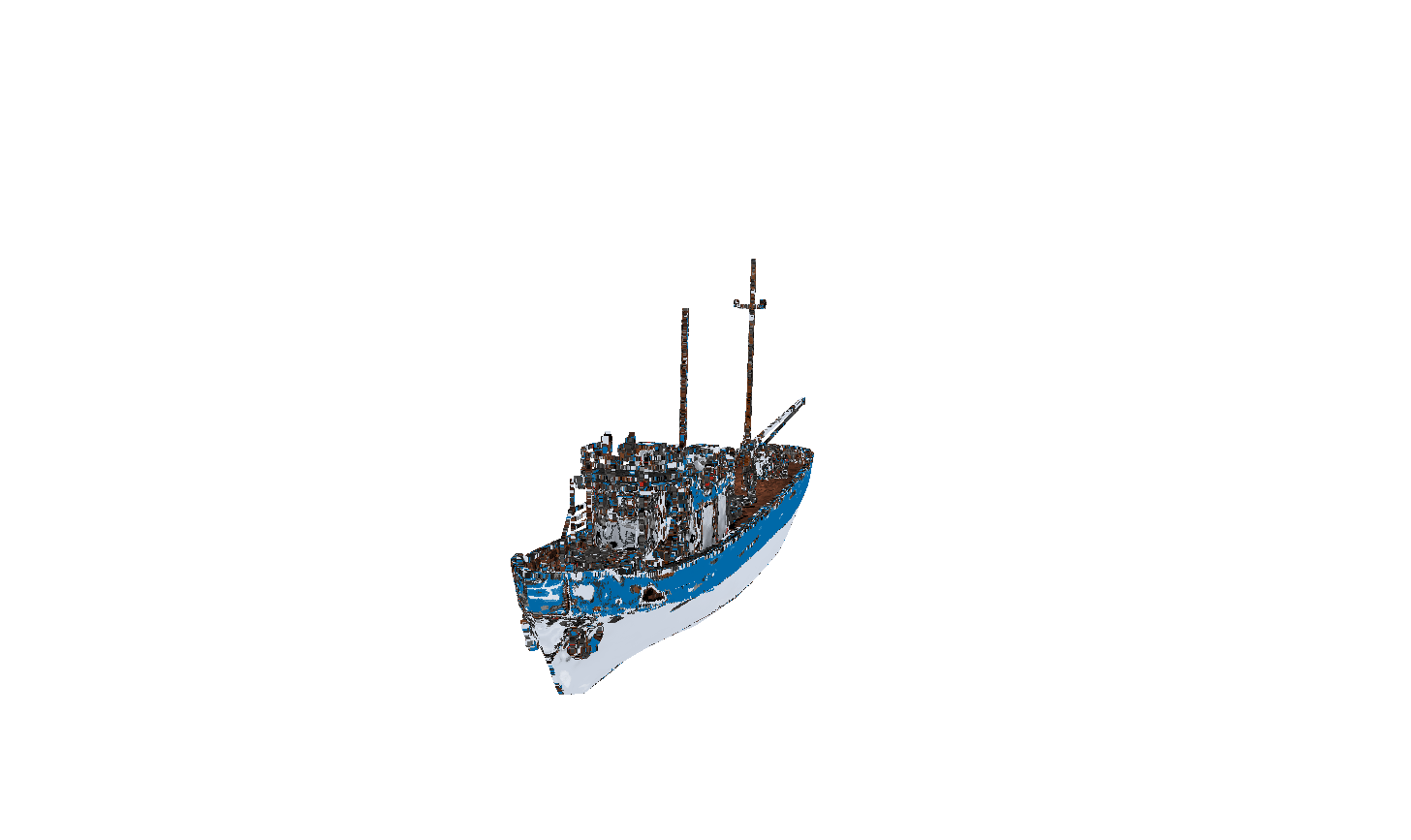}};
            \spy on (6.05,-2.35) in node [left] at (8.3,-2);
            \node at (10.1,0) {\includegraphics[height=60pt,trim={250 90 250 90},clip]{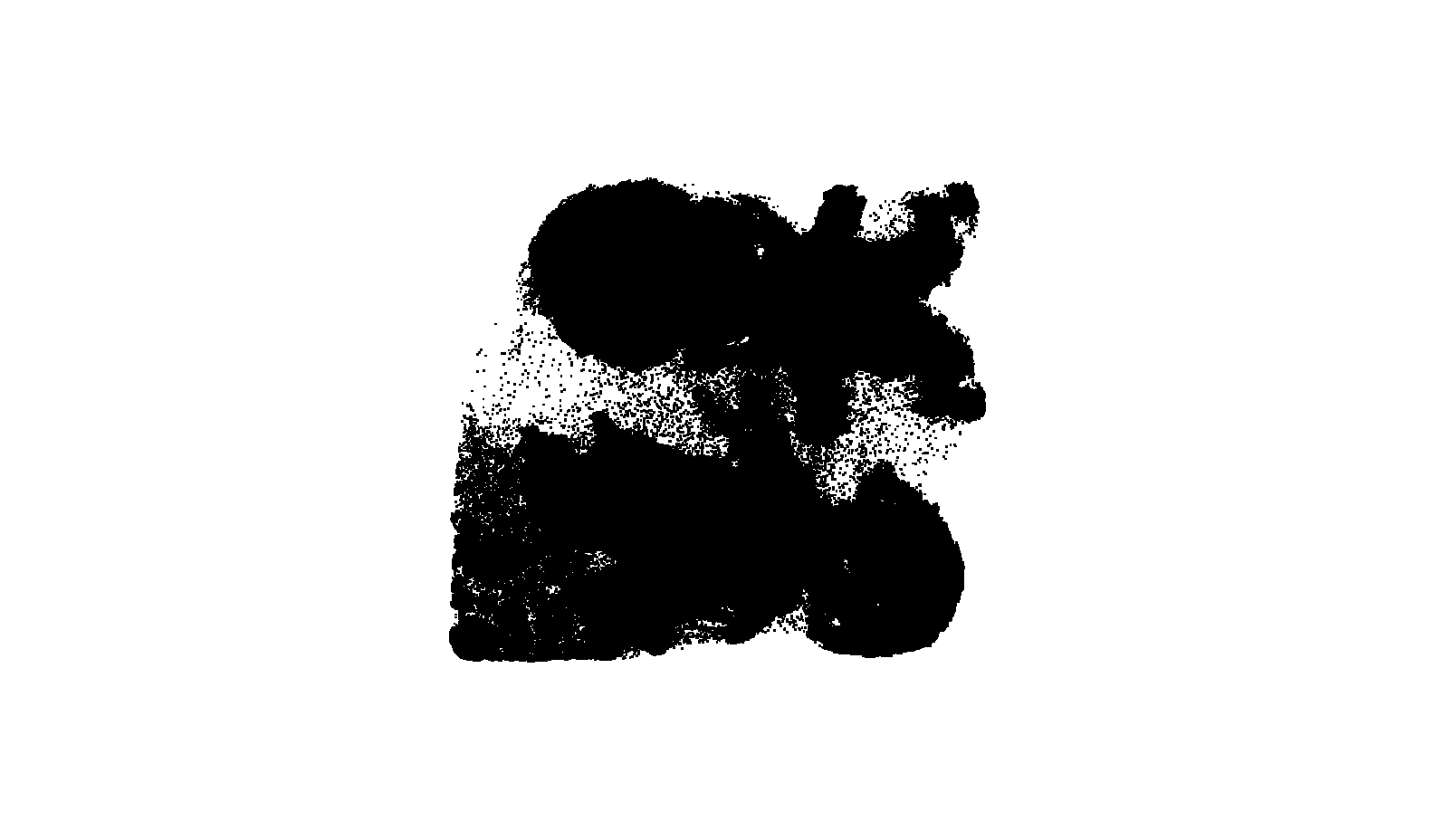}};
            \node at (9.6,-2) {\includegraphics[height=55pt,trim={200 55 200 55},clip]{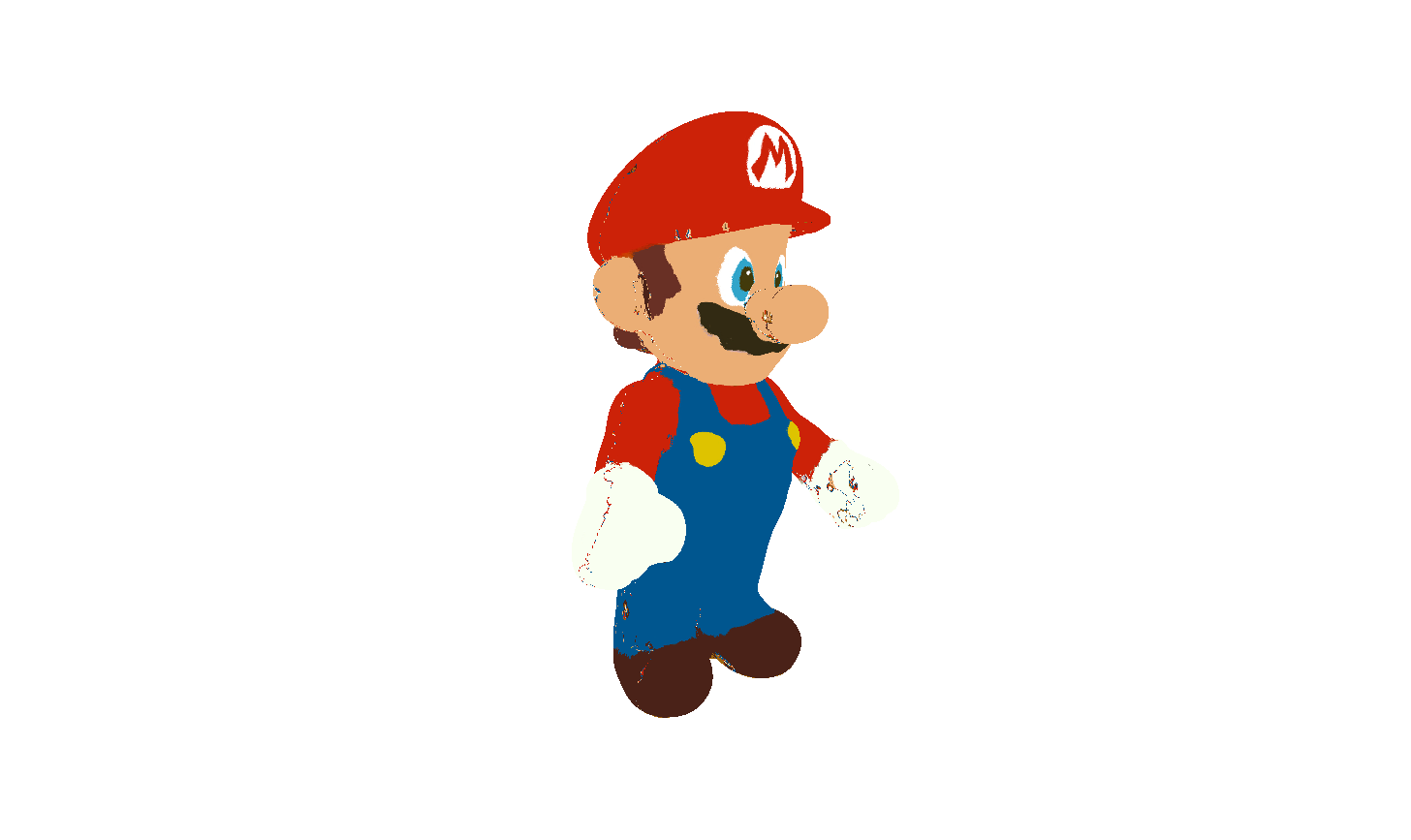}};
            \spy on (10,-2.2) in node [left] at (11.5,-2);
            \node at (13.3,0) {\includegraphics[height=60pt,trim={250 90 250 90},clip]{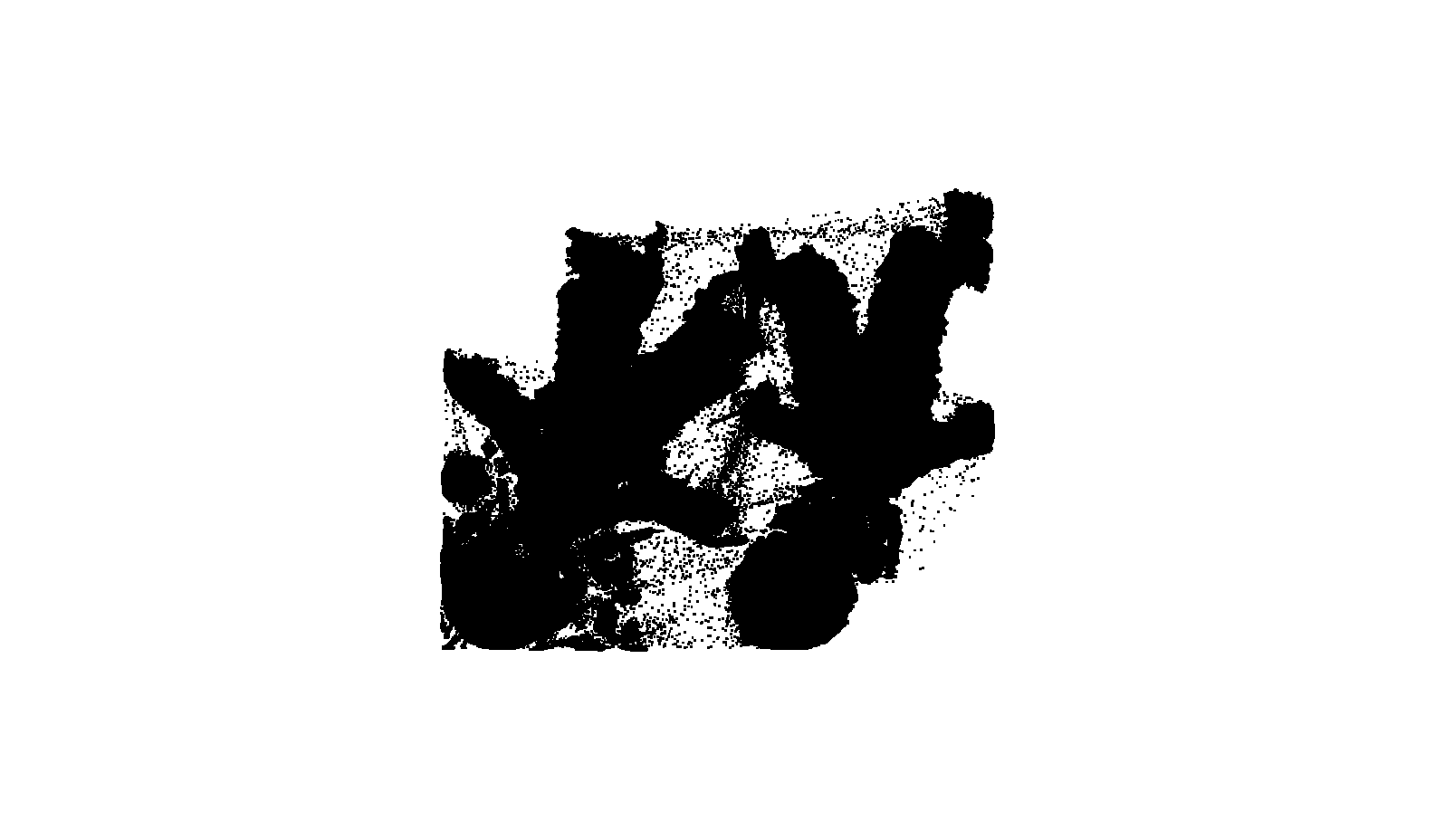}};
            \node at (12.8,-2) {\includegraphics[height=55pt,trim={200 55 200 55},clip]{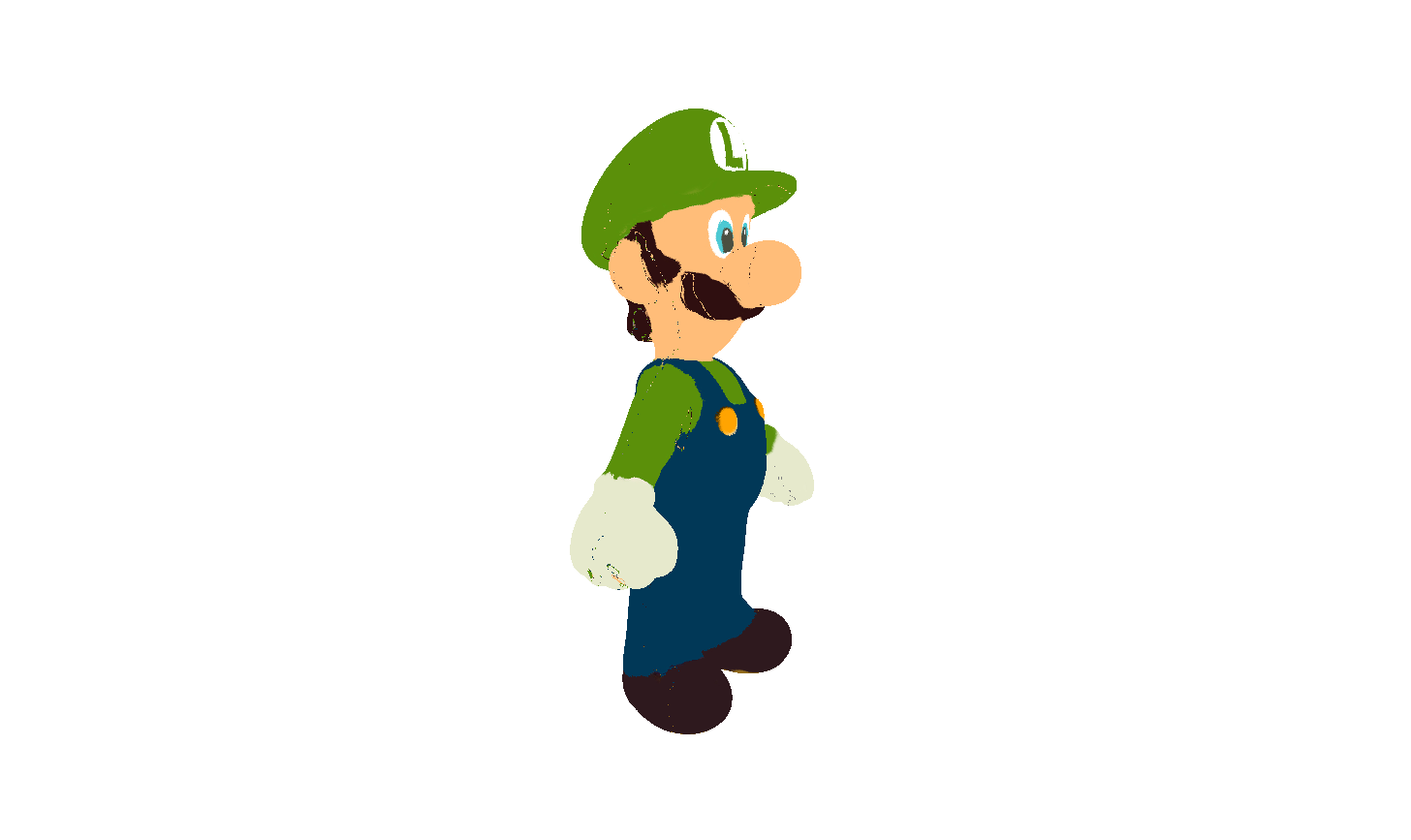}};
            \spy on (12.5,-2.4) in node [left] at (14.7,-2);
        \end{tikzpicture}
        \caption{\texttt{point2UV}}
        \label{fig:uv-comparision-a}
    \end{subfigure}
    \\
    \begin{subfigure}{\linewidth}
        \centering
        \begin{tikzpicture}[spy using outlines={circle,orange,magnification=5,size=1.5cm, connect spies}]
            \node at (0.5,0) {\includegraphics[height=60pt,trim={250 90 250 90},clip]{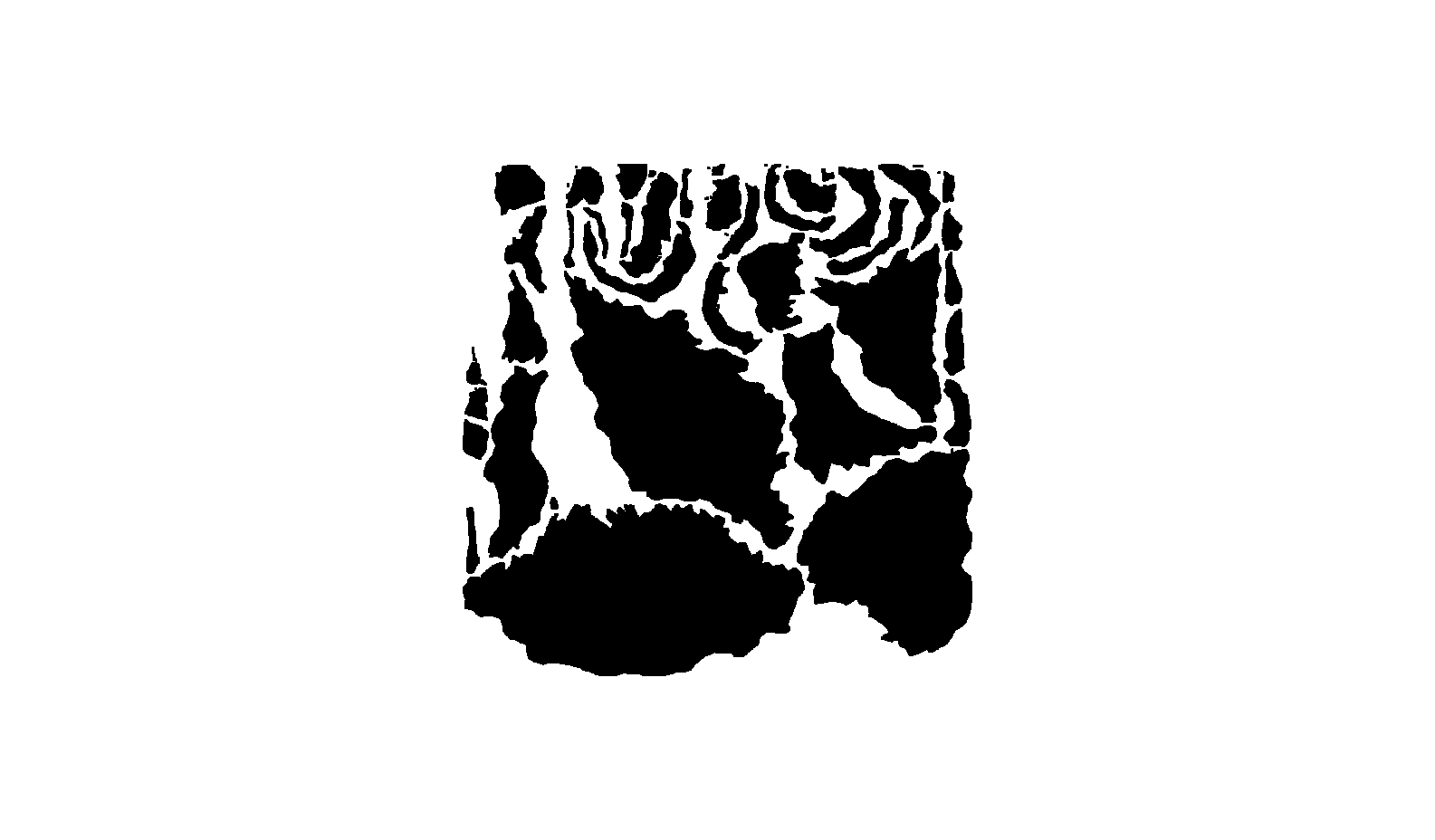}};
            \node at (0,-2) {\includegraphics[height=55pt,trim={200 70 200 70},clip]{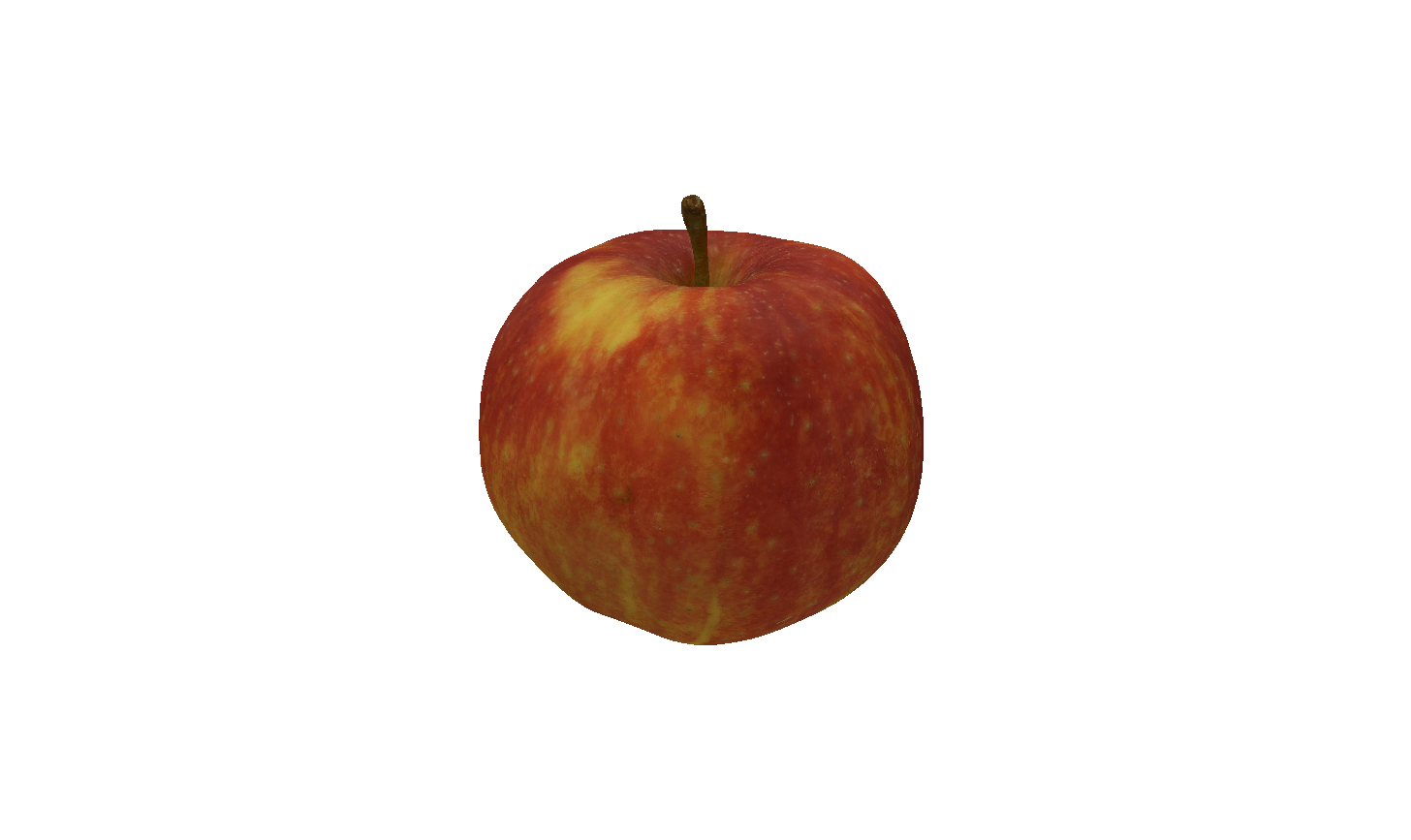}};
            \spy on (-0.2,-1.6) in node [left] at (1.9,-2);
            \node at (3.7,0) {\includegraphics[height=60pt,trim={250 90 250 90},clip]{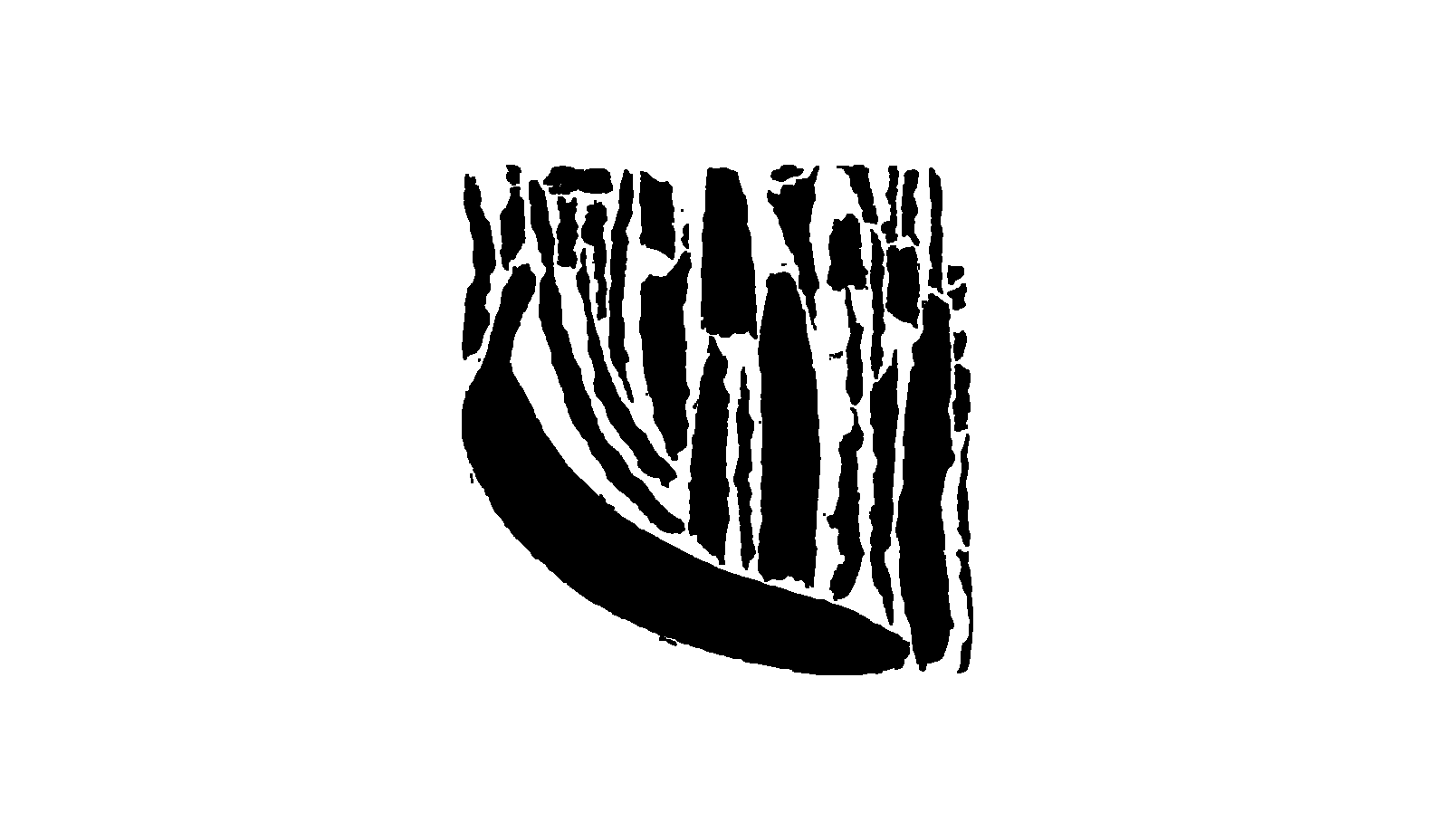}};
            \node at (3.2,-2) {\includegraphics[height=55pt,trim={200 70 200 70},clip]{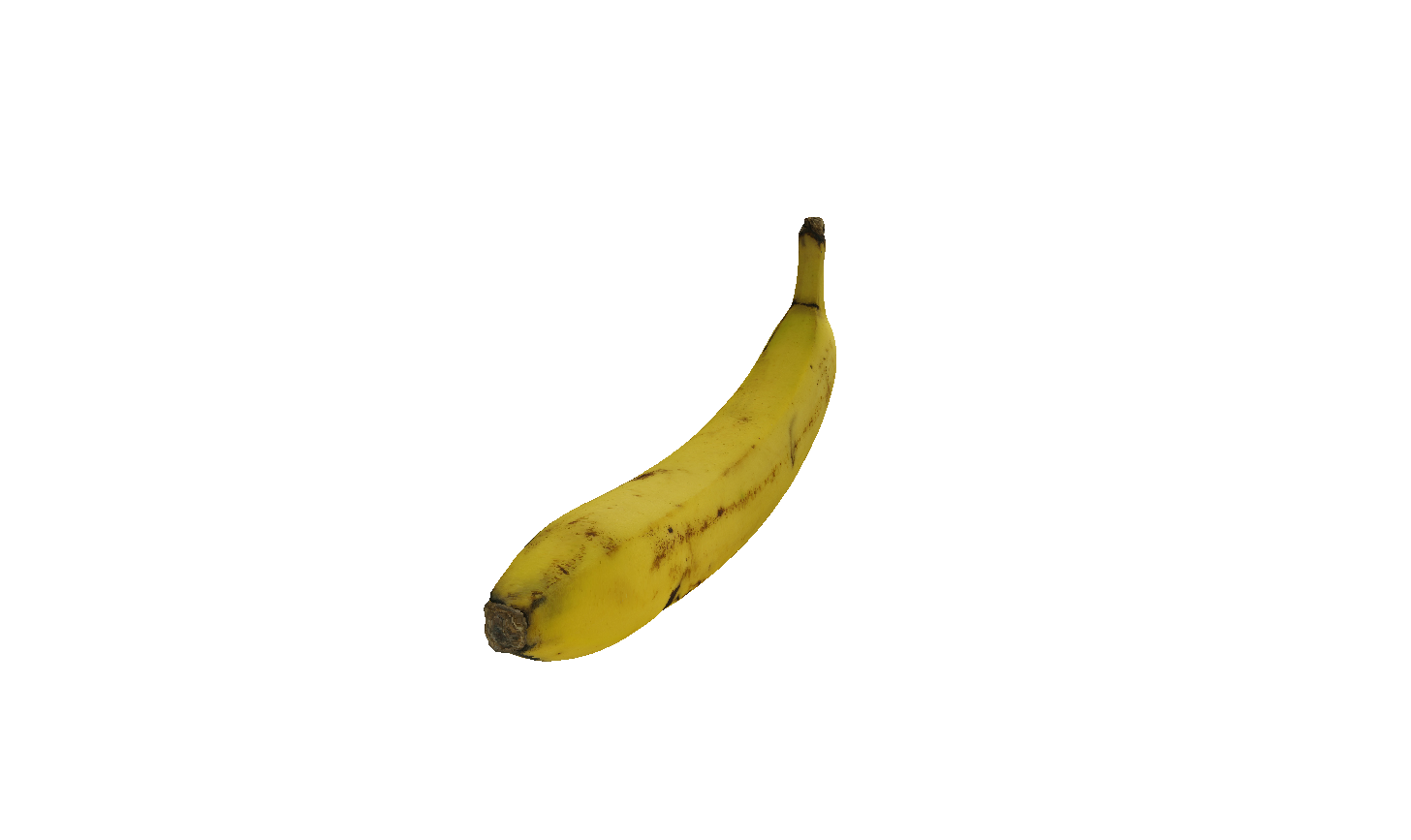}};
            \spy on (2.65,-2.65) in node [left] at (5.1,-2);
            \node at (6.9,0) {\includegraphics[height=60pt,trim={250 90 250 90},clip]{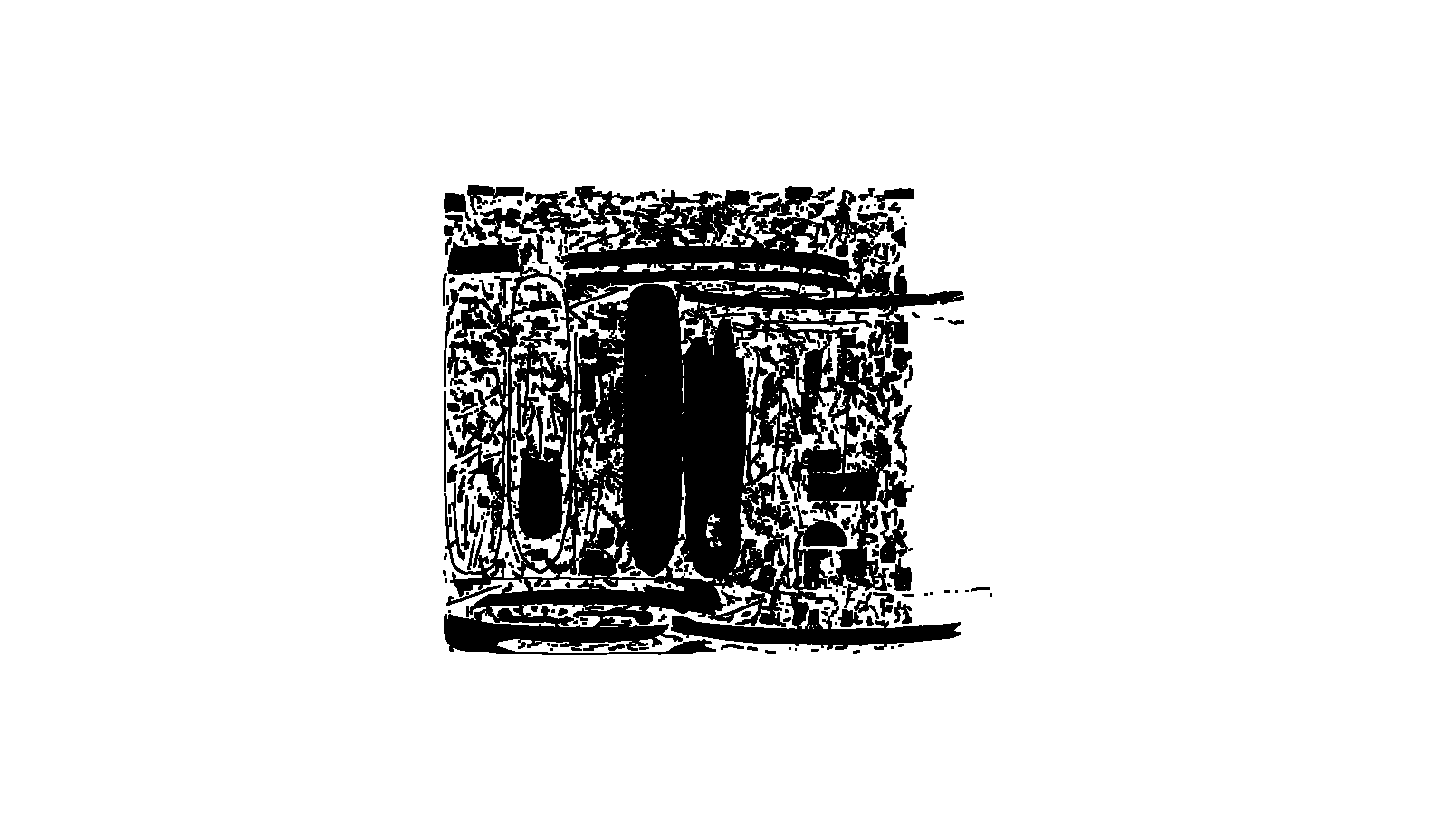}};
            \node at (6.4,-2) {\includegraphics[height=55pt,trim={200 70 200 110},clip]{images/decomposed-uv/boat.png}};
            \spy on (6.05,-2.35) in node [left] at (8.3,-2);
            \node at (10.1,0) {\includegraphics[height=60pt,trim={250 90 250 90},clip]{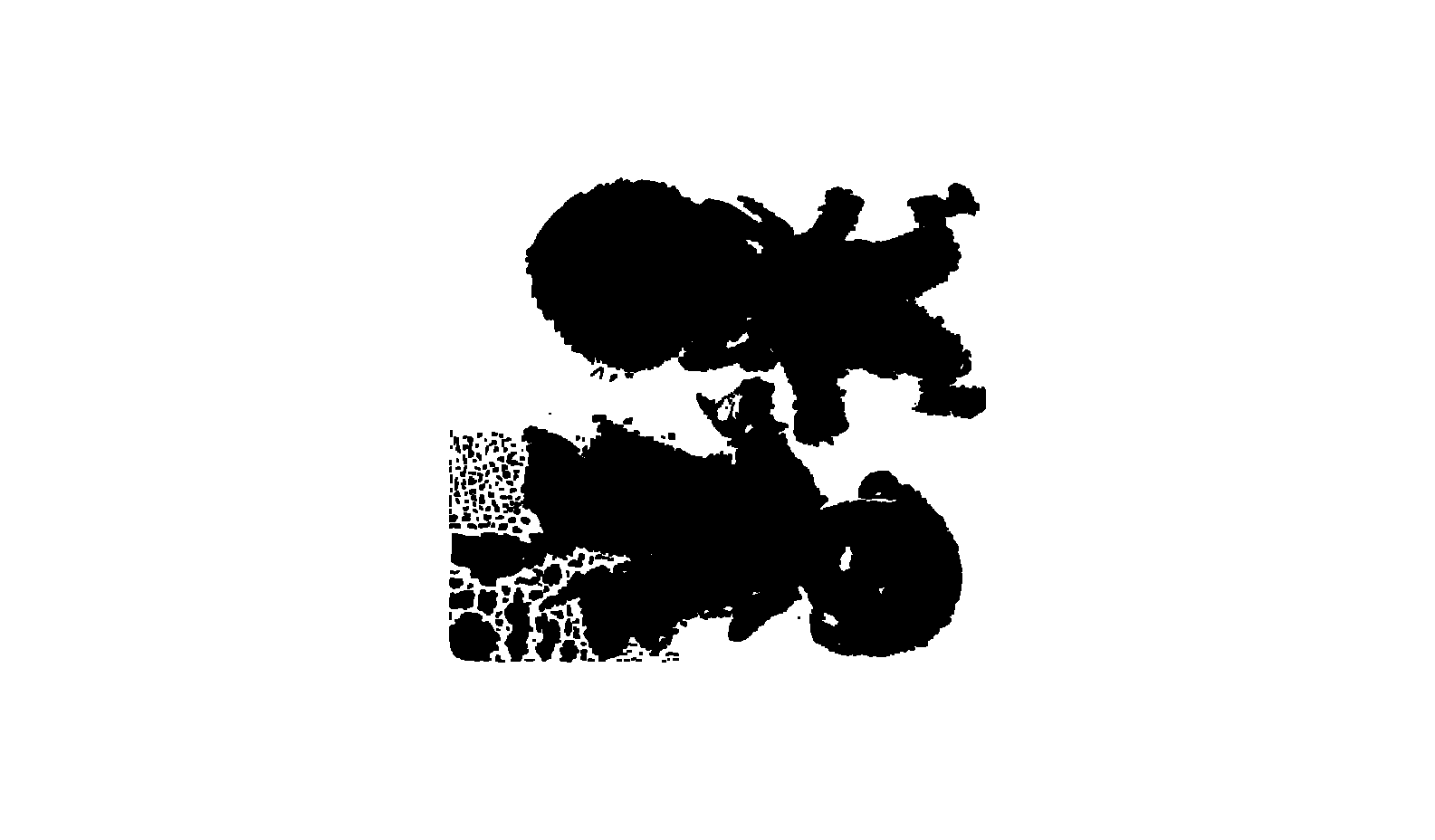}};
            \node at (9.6,-2) {\includegraphics[height=55pt,trim={200 55 200 55},clip]{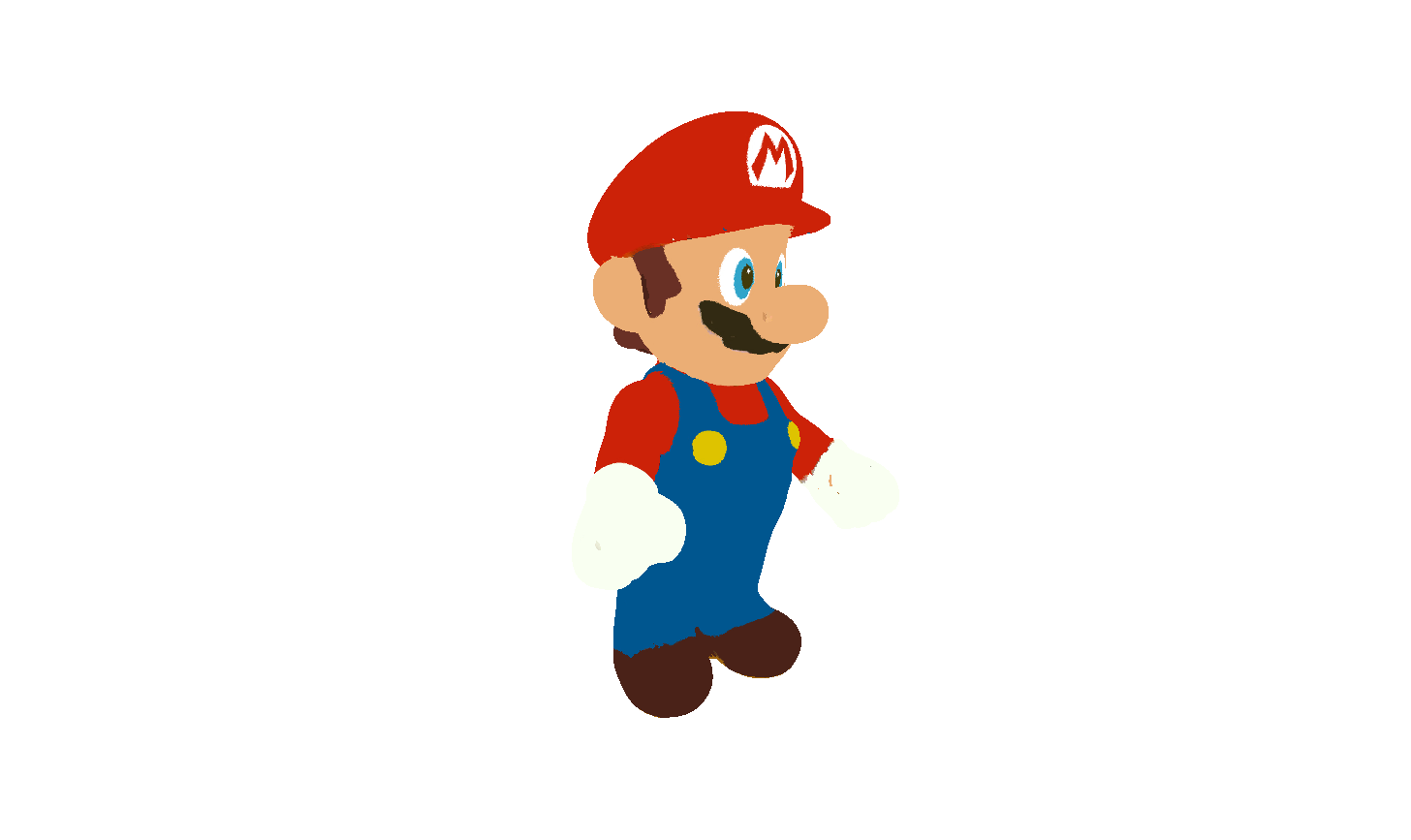}};
            \spy on (10,-2.2) in node [left] at (11.5,-2);
            \node at (13.3,0) {\includegraphics[height=60pt,trim={250 90 250 90},clip]{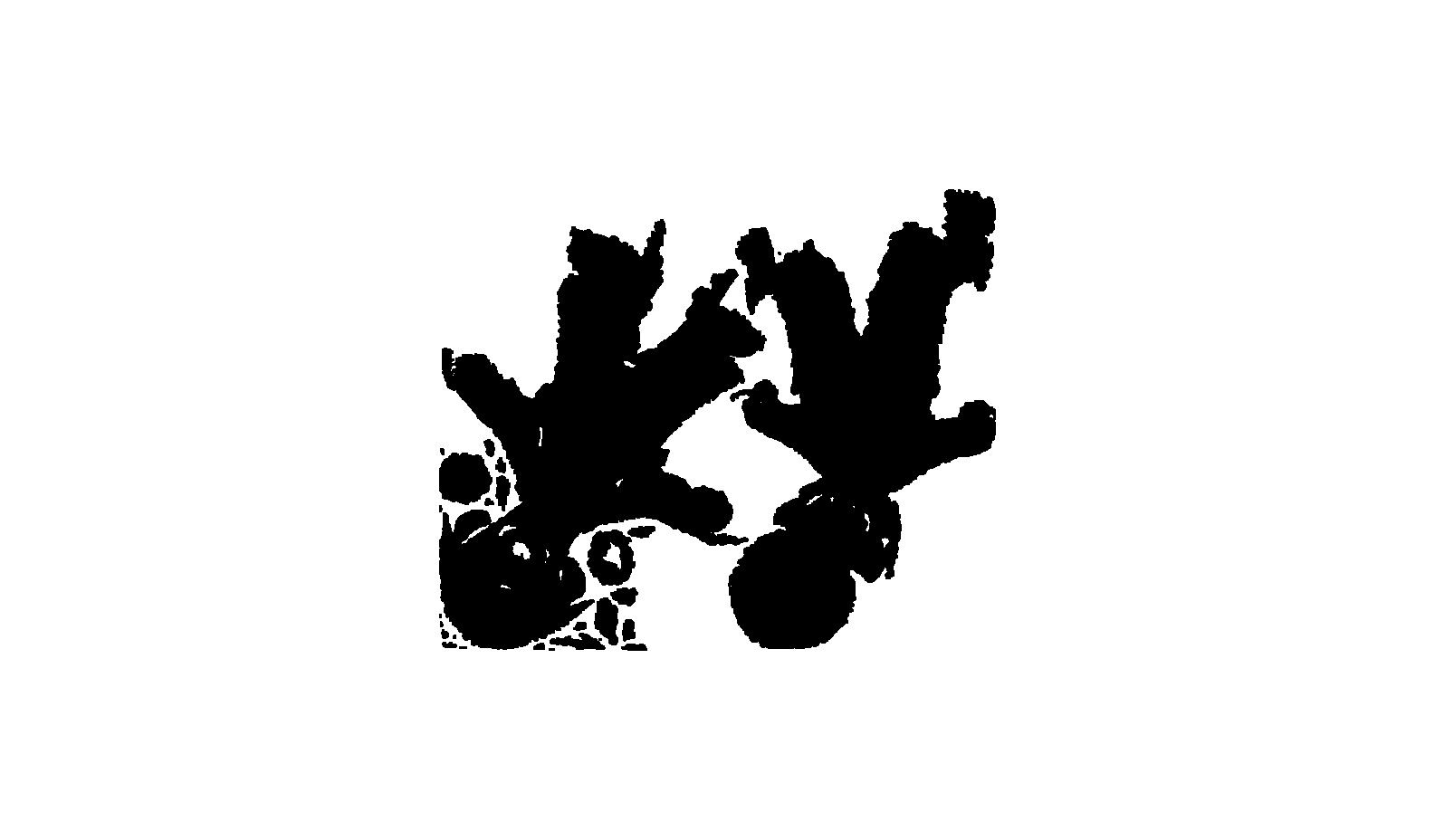}};
            \node at (12.8,-2) {\includegraphics[height=55pt,trim={200 55 200 55},clip]{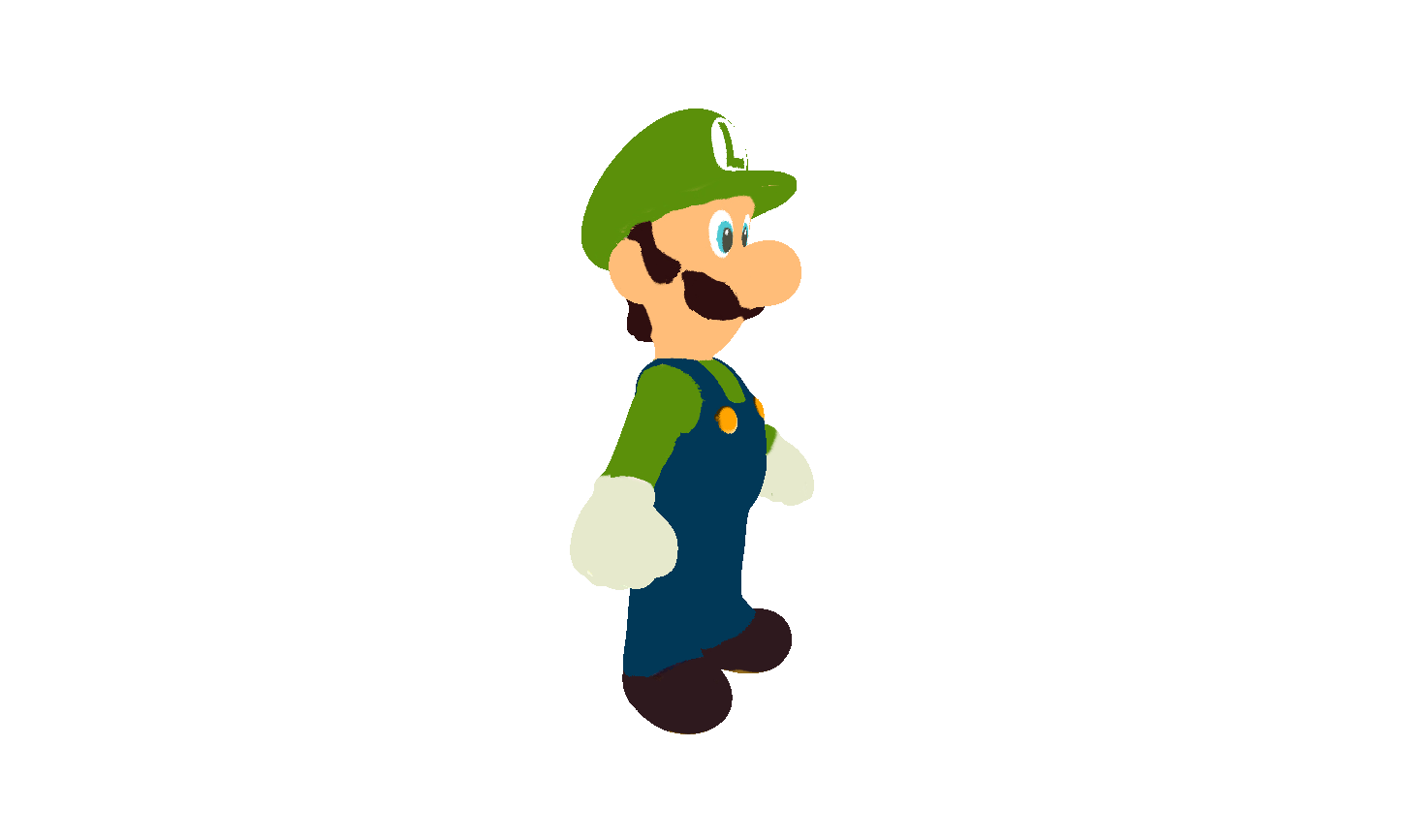}};
            \spy on (12.5,-2.4) in node [left] at (14.7,-2);
        \end{tikzpicture}
        \caption{Ours (\texttt{point2component} + \texttt{point2UV})}
        \label{fig:uv-comparision-b}
    \end{subfigure}
    \\
    \begin{subfigure}{\linewidth}
        \centering
        \begin{tikzpicture}[spy using outlines={circle,orange,magnification=5,size=1.5cm, connect spies}]
            \node at (0.5,0) {\includegraphics[height=60pt,trim={250 90 250 90},clip]{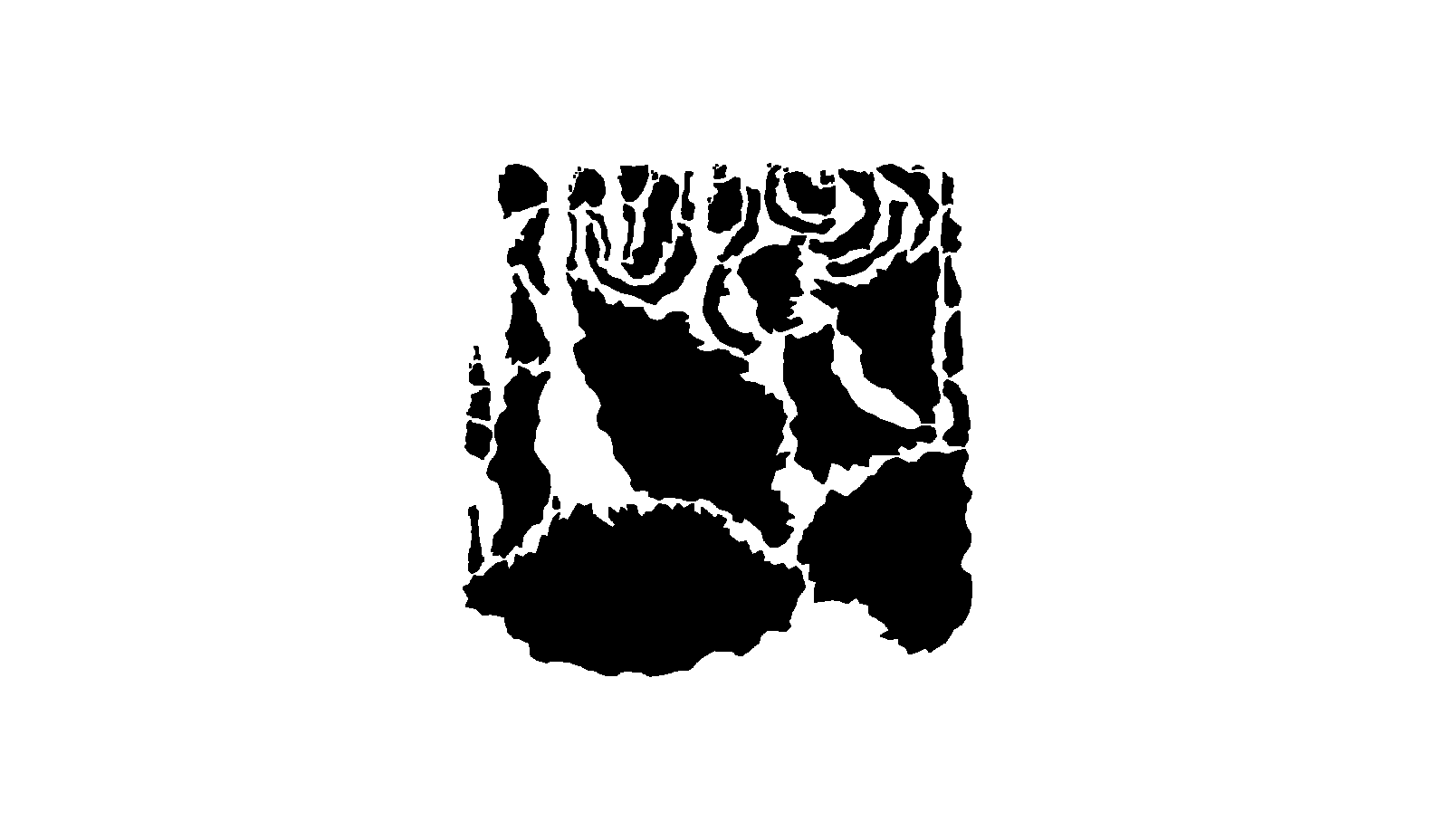}};
            \node at (0,-2) {\includegraphics[height=55pt,trim={200 70 200 70},clip]{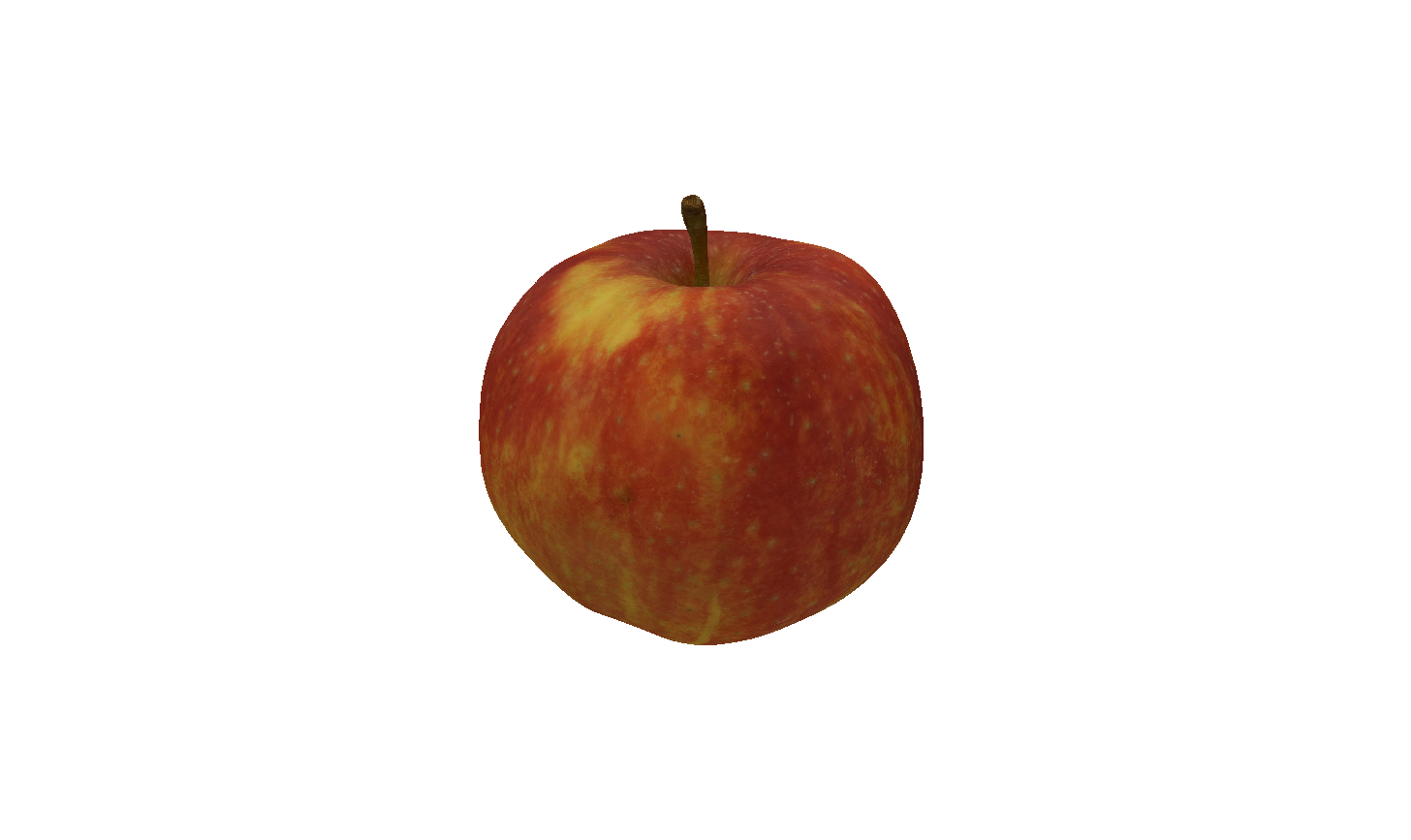}};
            \spy on (-0.2,-1.6) in node [left] at (1.9,-2);
            \node at (3.7,0) {\includegraphics[height=60pt,trim={250 90 250 90},clip]{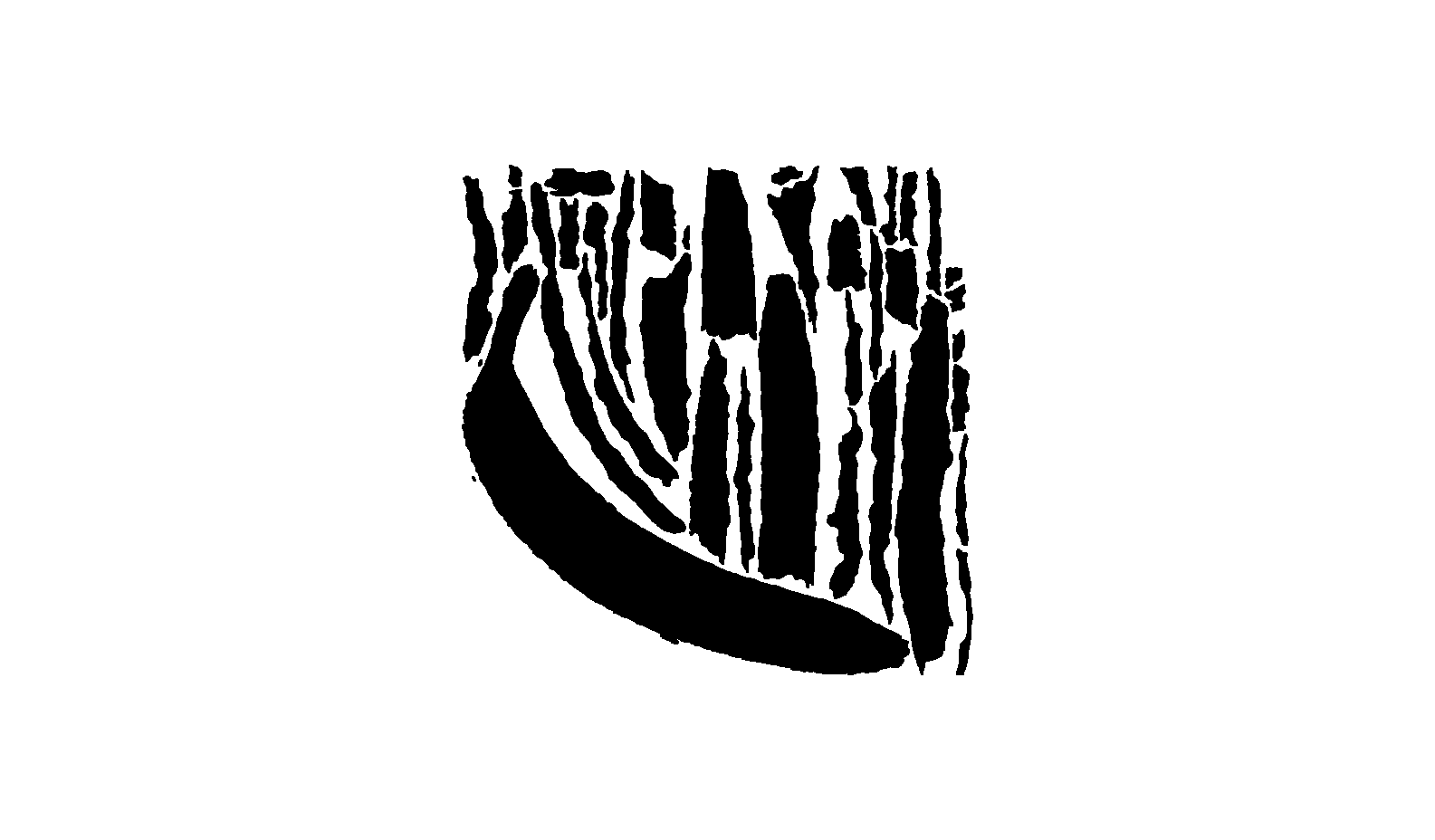}};
            \node at (3.2,-2) {\includegraphics[height=55pt,trim={200 70 200 70},clip]{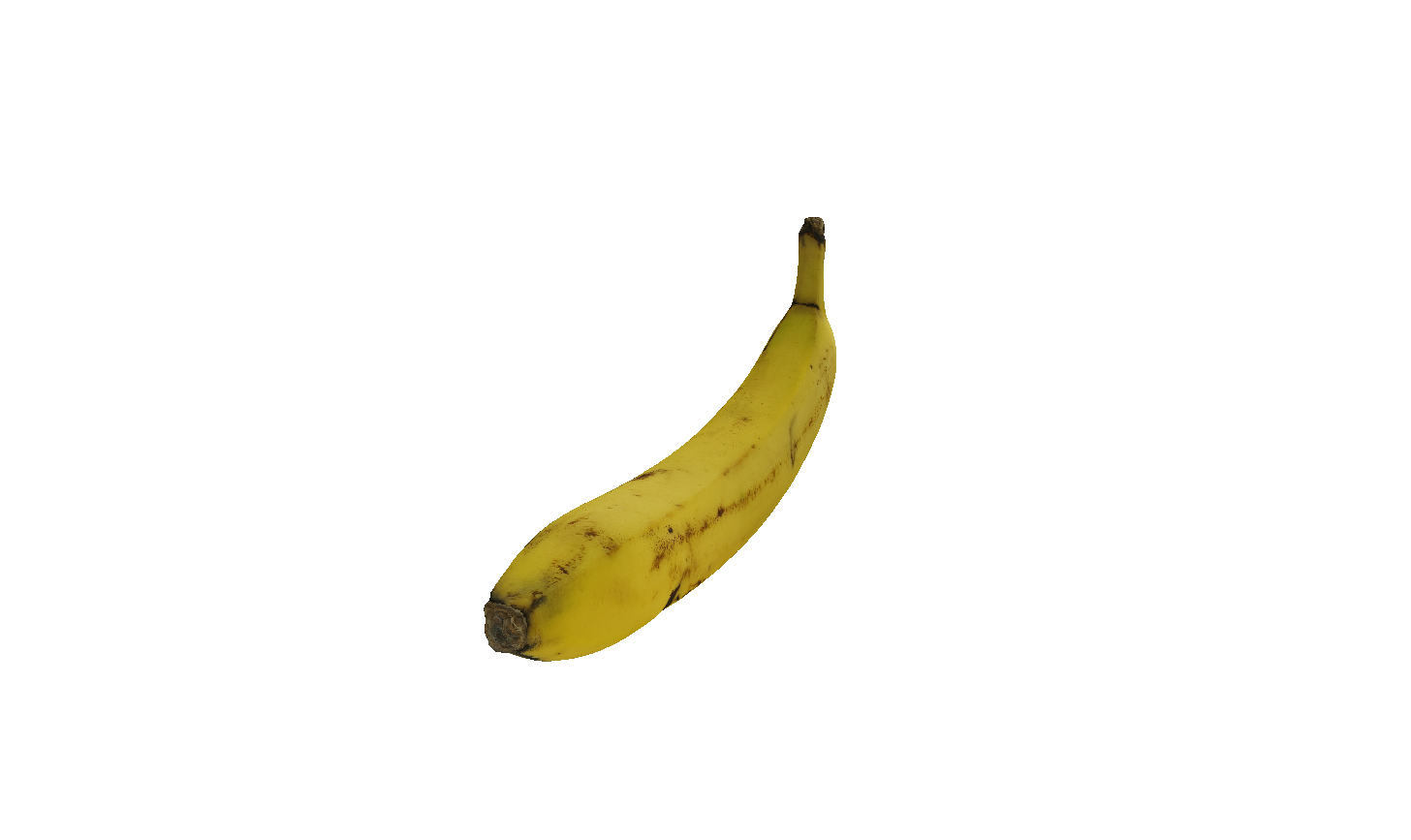}};
            \spy on (2.65,-2.65) in node [left] at (5.1,-2);
            \node at (6.9,0) {\includegraphics[height=60pt,trim={250 90 250 90},clip]{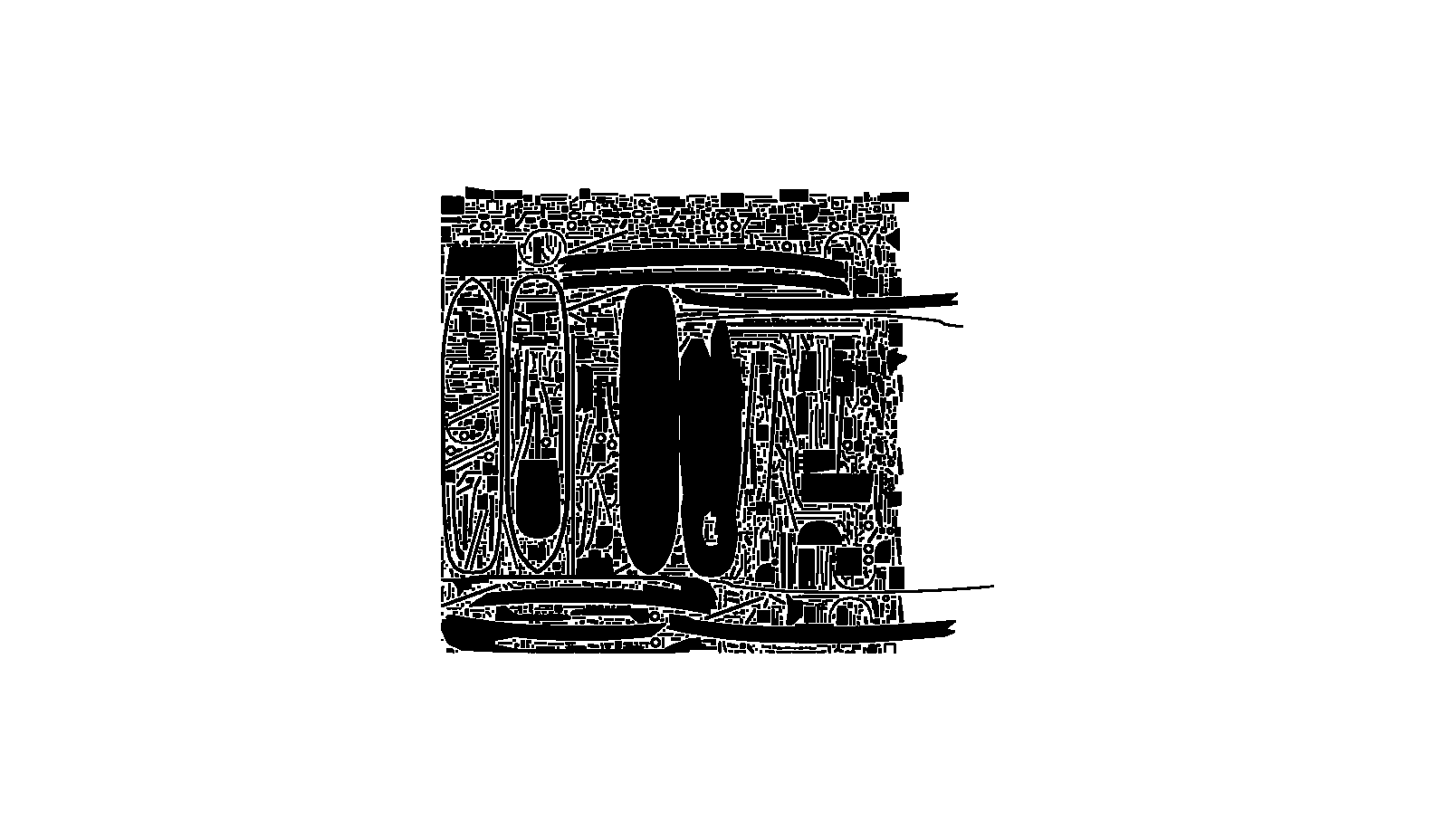}};
            \node at (6.4,-2) {\includegraphics[height=55pt,trim={200 70 200 110},clip]{images/gt/boat.png}};
            \spy on (6.05,-2.35) in node [left] at (8.3,-2);
            \node at (10.1,0) {\includegraphics[height=60pt,trim={250 90 250 90},clip]{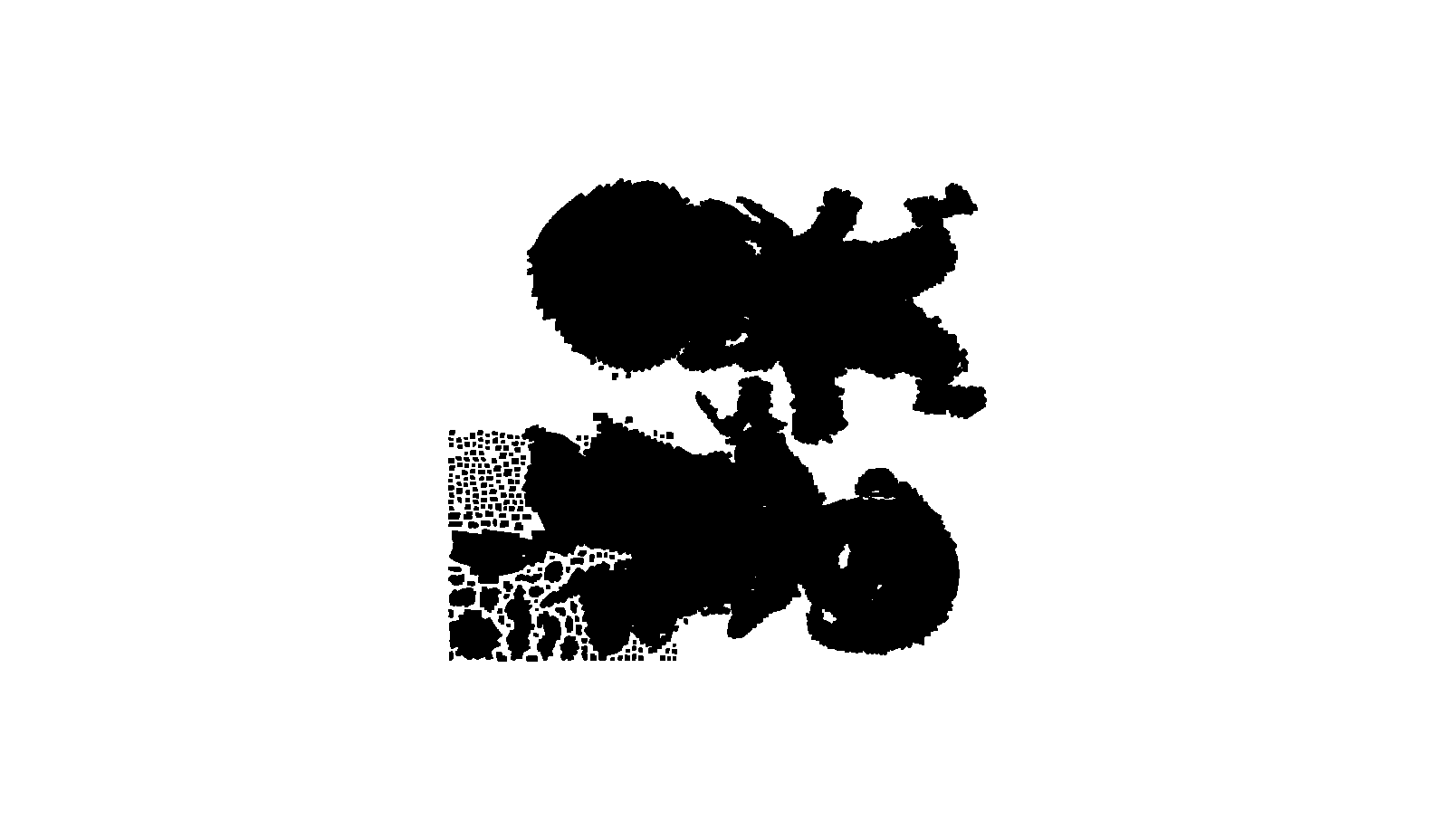}};
            \node at (9.6,-2) {\includegraphics[height=55pt,trim={200 55 200 55},clip]{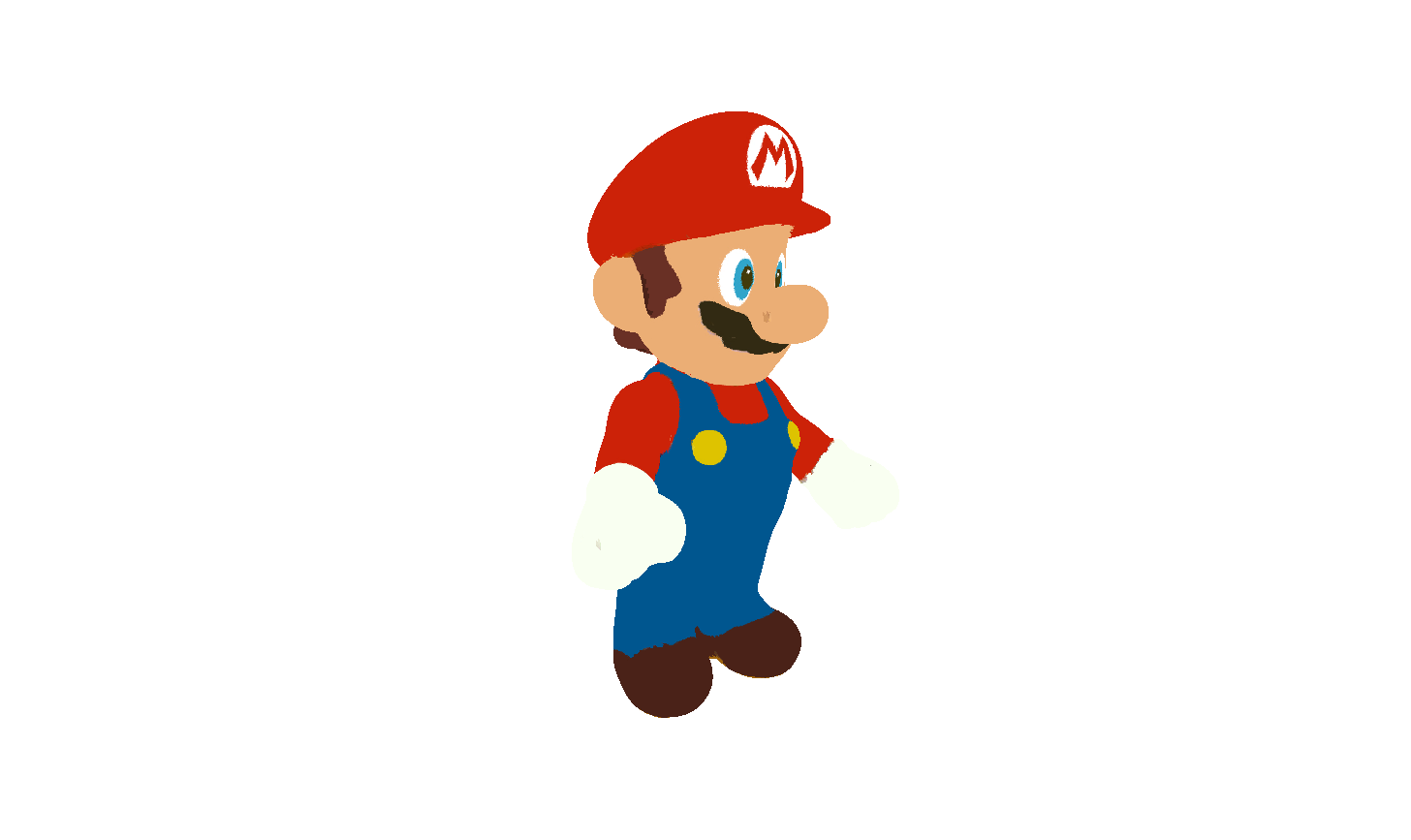}};
            \spy on (10,-2.2) in node [left] at (11.5,-2);
            \node at (13.3,0) {\includegraphics[height=60pt,trim={250 90 250 90},clip]{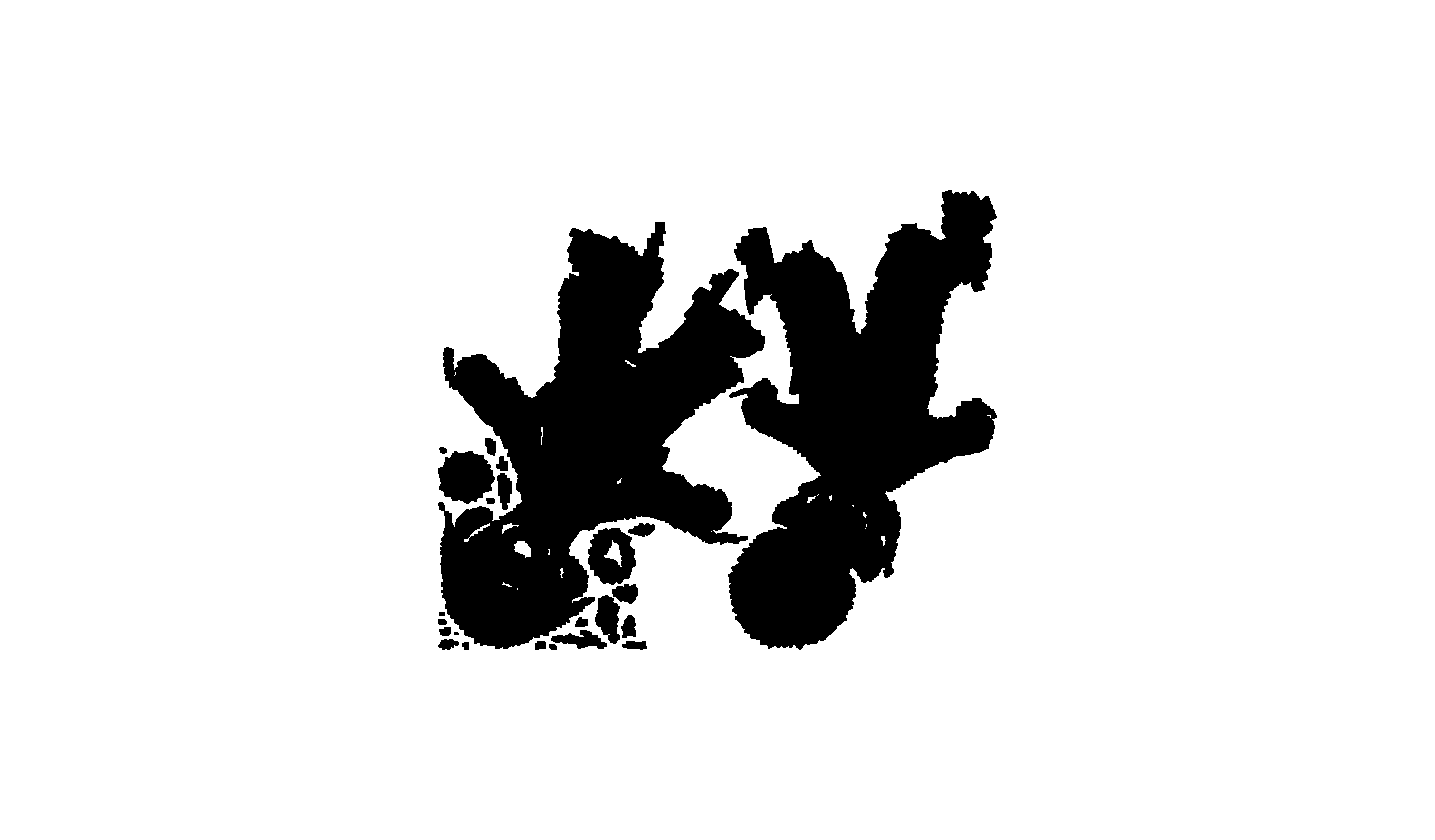}};
            \node at (12.8,-2) {\includegraphics[height=55pt,trim={200 55 200 55},clip]{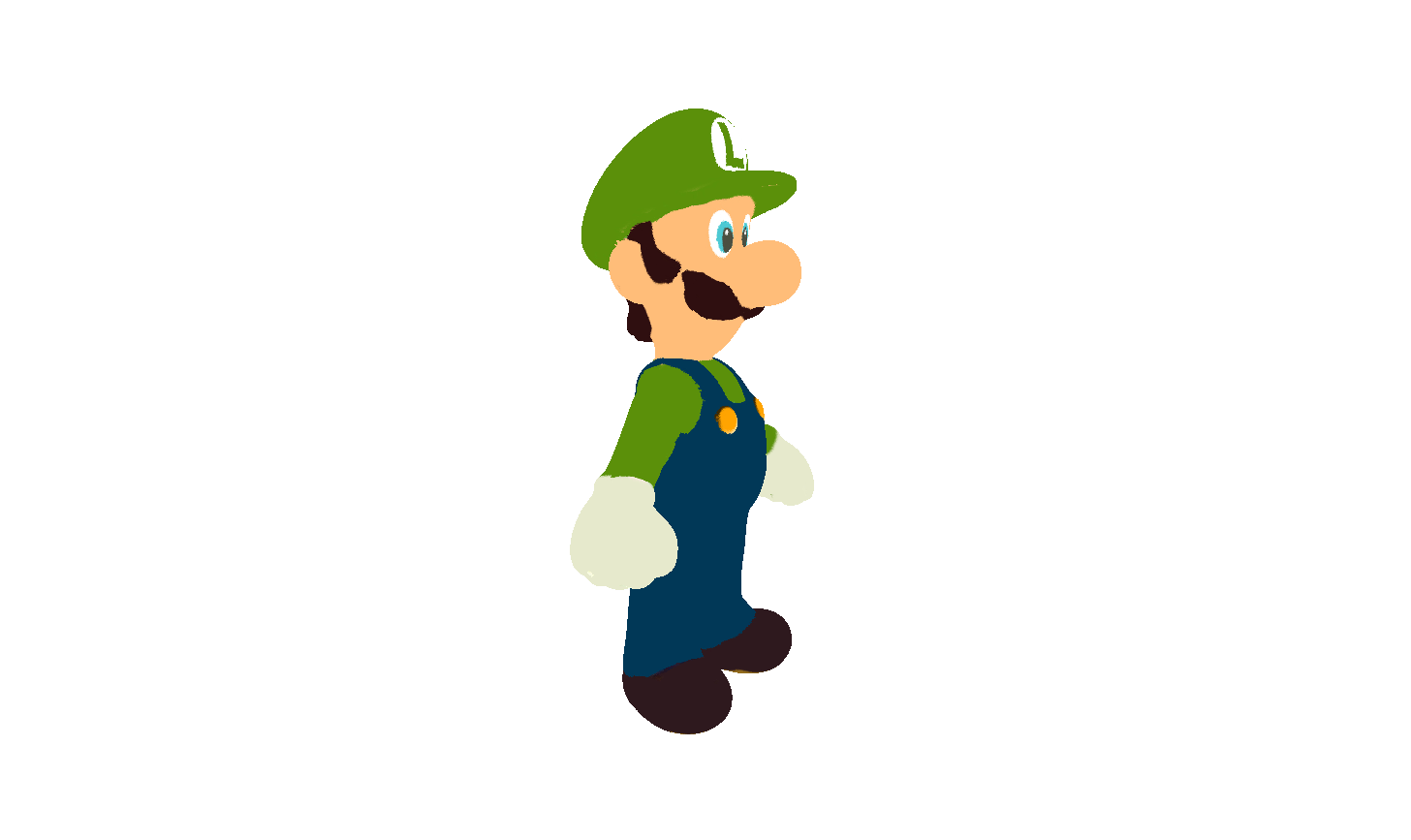}};
            \spy on (12.5,-2.4) in node [left] at (14.7,-2);
        \end{tikzpicture}
        \caption{GT}
        \label{fig:uv-comparision-c}
    \end{subfigure}
    \caption{The UV coordinates of the surface sample points, plotted as scatter charts, with the corresponding diffuse mapping results along with the zoom-in views shown on the bottom of each example. With conditioned input to \texttt{point2UV}, the noisy results in the UV predictions can be largely mitigated, resulting in almost identical diffuse mapping results to the GT.}
    \label{fig:uv-comparision}
\end{figure*}

%% file: figures/normal_mapping.tex
\begin{figure}[t!]
    \centering
    \begin{subfigure}{\linewidth}
        \centering
        \begin{tikzpicture}
            \node at (0,0) {\includegraphics[height=45pt,trim={50 50 50 50},clip]{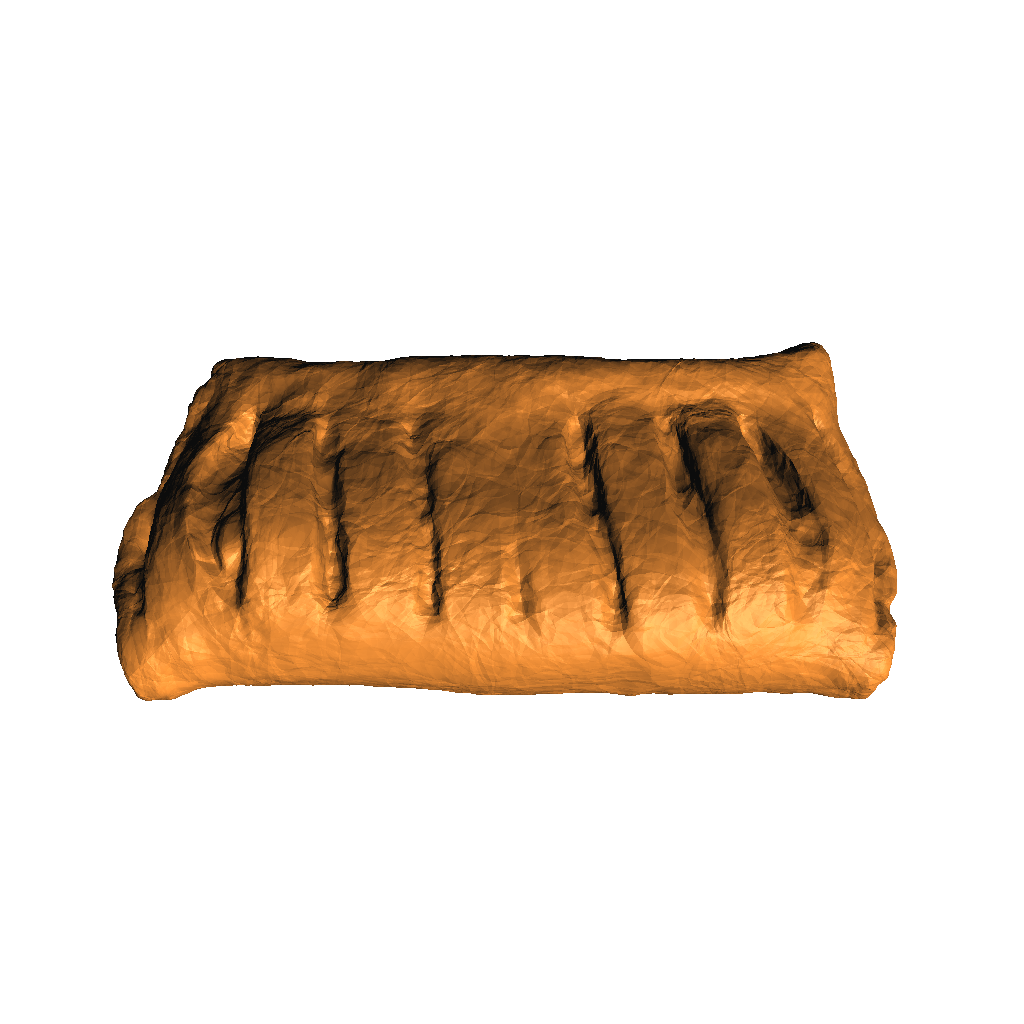}};
            \node at (1.7,0) {\includegraphics[height=45pt,trim={50 50 50 50},clip]{images/normal-mapping/barrel_sdf_normal.png}};
            \node at (3.4,0) {\includegraphics[height=45pt,trim={50 50 50 50},clip]{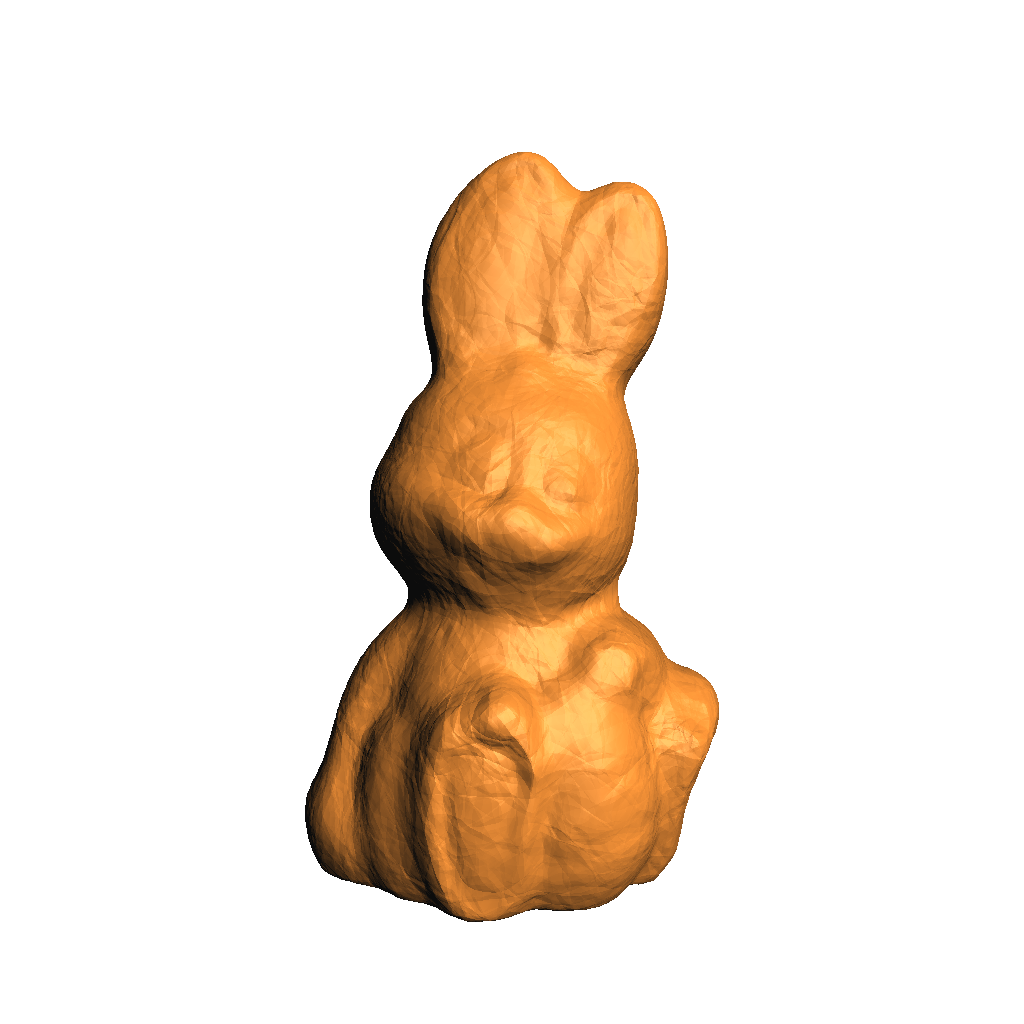}};
            \node at (5.1,0) {\includegraphics[height=45pt,trim={50 50 50 50},clip]{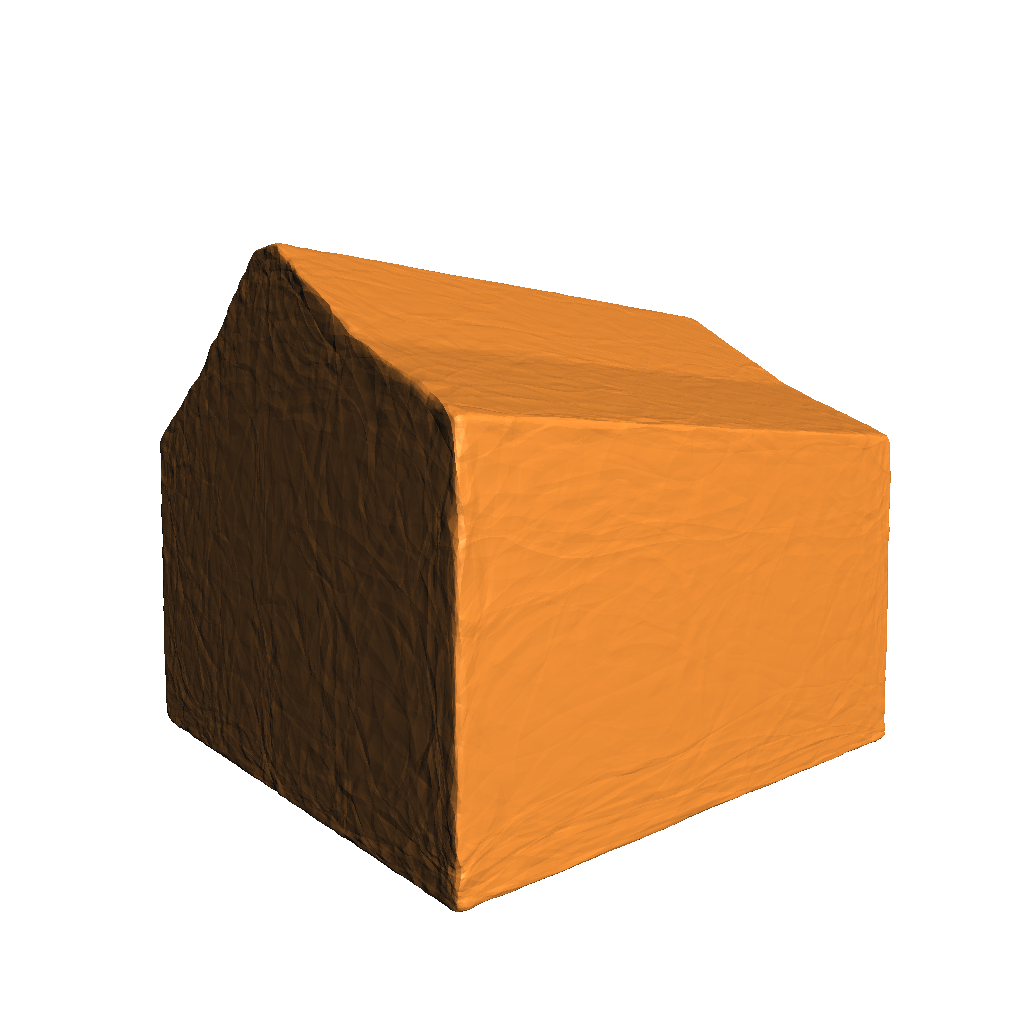}};
            \node at (0,-1.7) {\includegraphics[height=45pt,trim={50 50 50 50},clip]{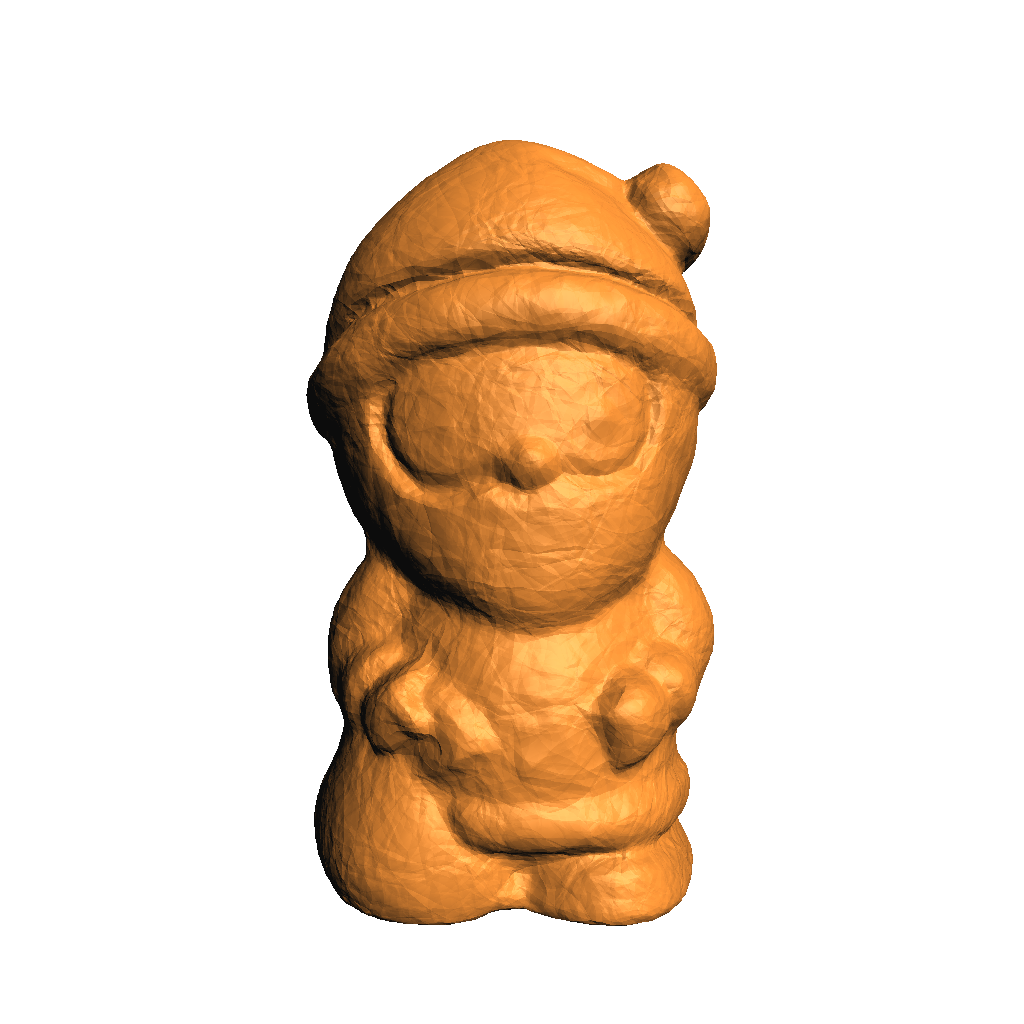}};
            \node at (1.7,-1.7) {\includegraphics[height=45pt,trim={50 50 50 50},clip]{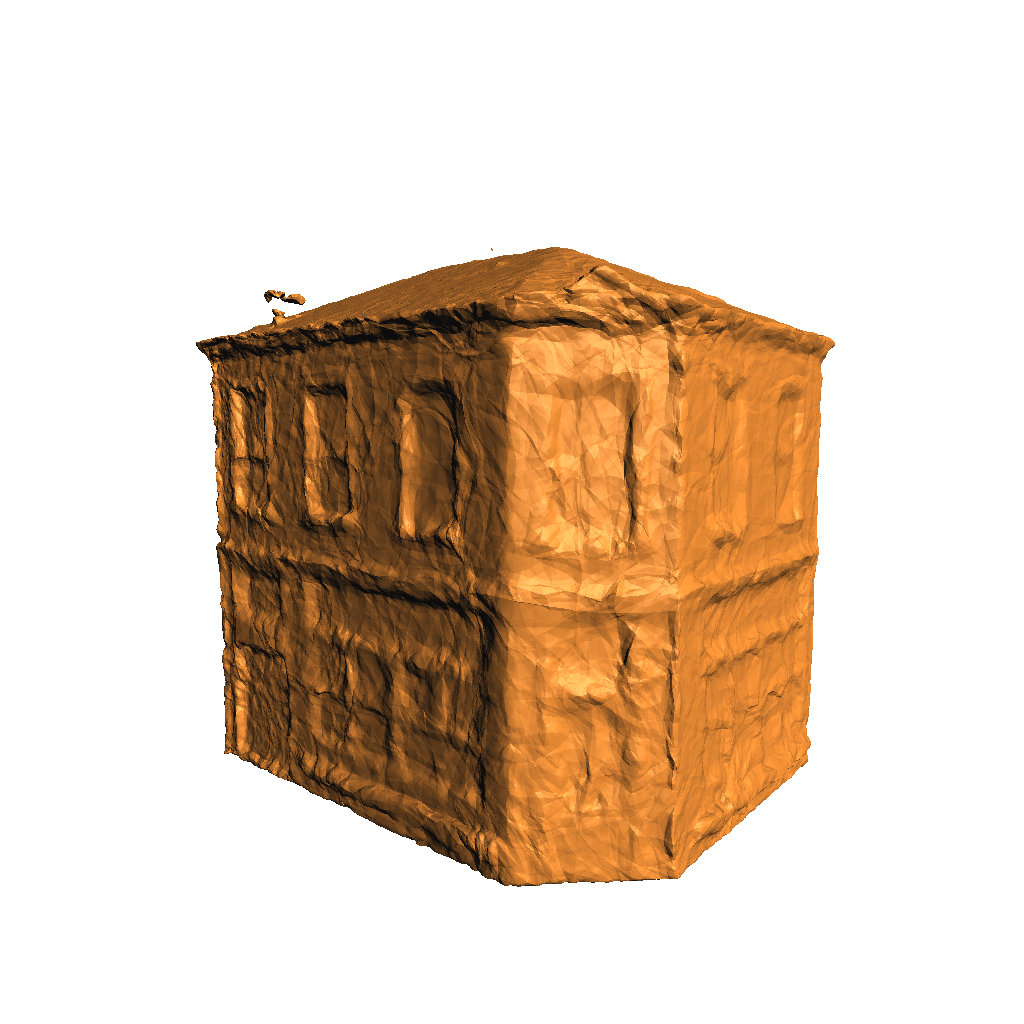}};
            \node at (3.4,-1.7) {\includegraphics[height=45pt,trim={50 50 50 50},clip]{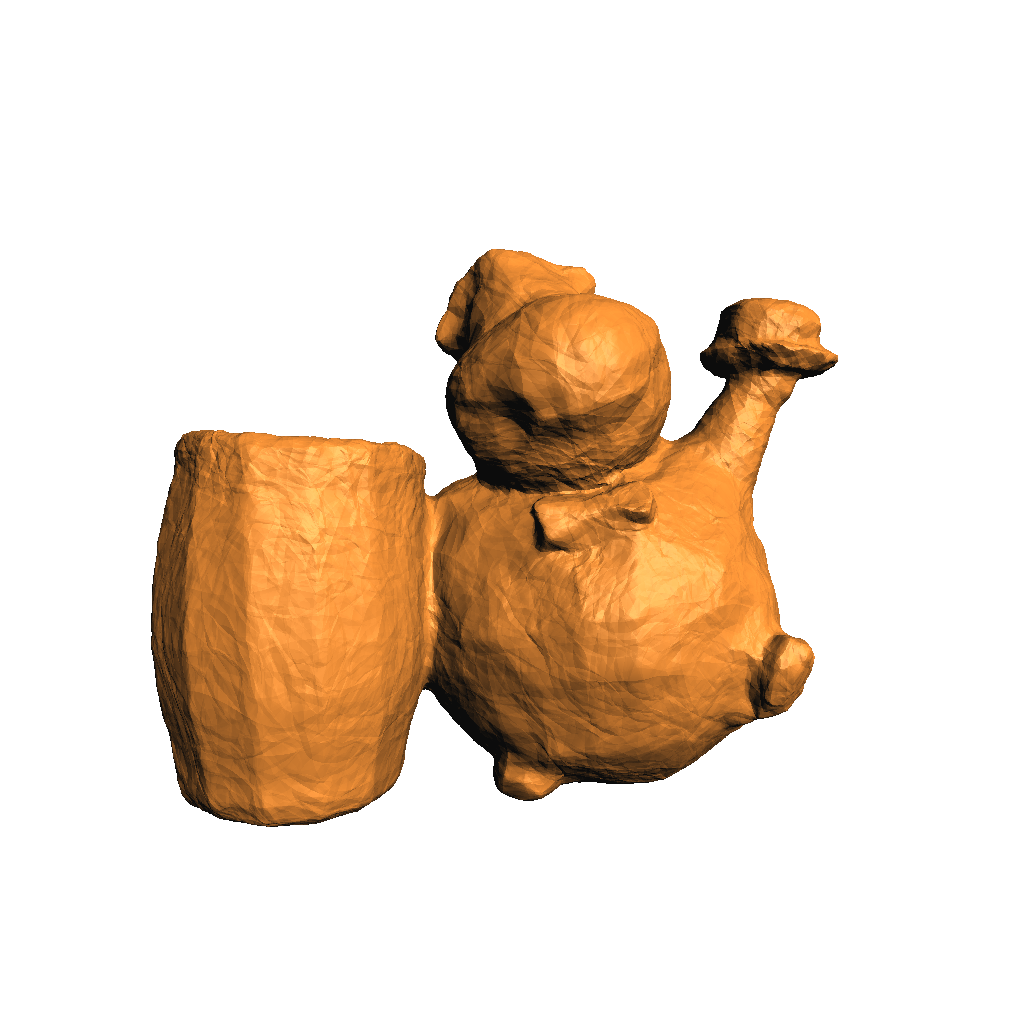}};
            \node at (5.1,-1.7) {\includegraphics[height=45pt,trim={50 50 50 50},clip]{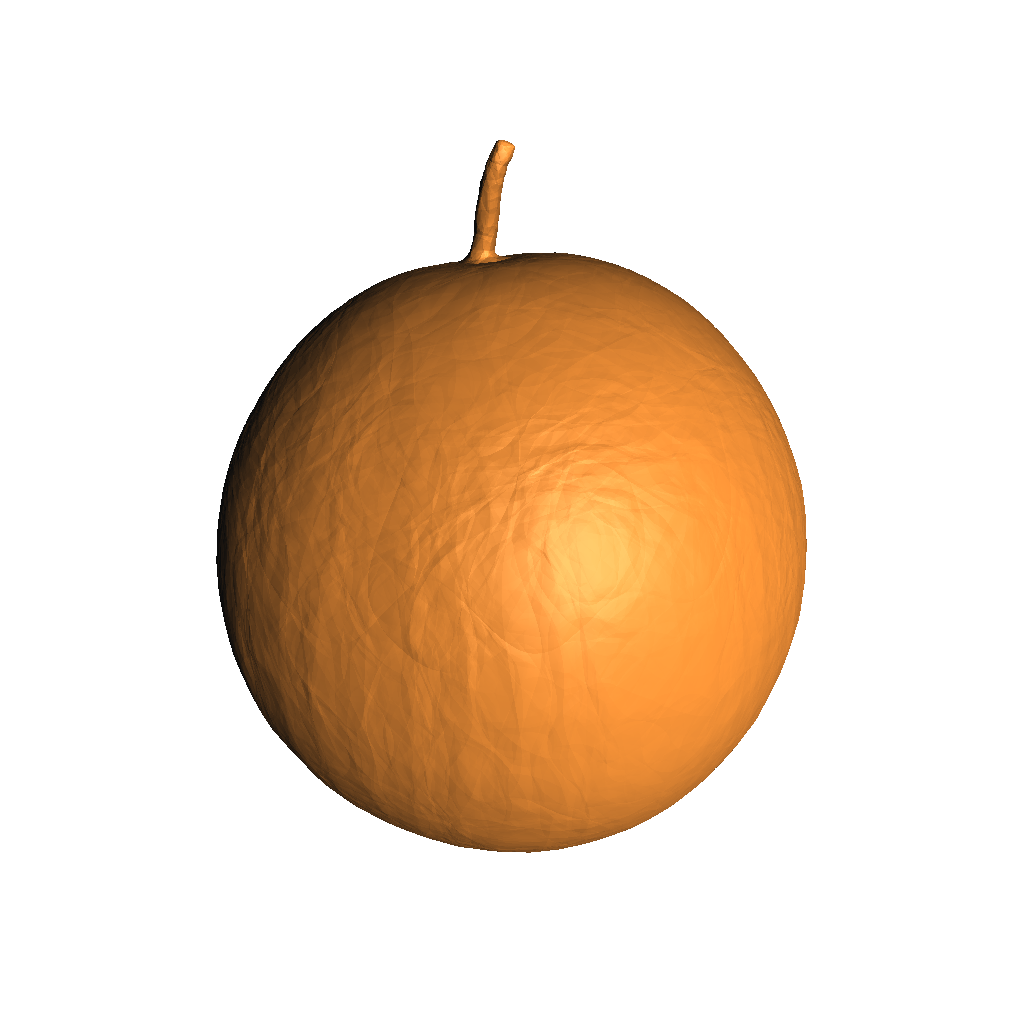}};
        \end{tikzpicture}
        \caption{Without normal mapping}
        \label{fig:normal-mapping-a}
    \end{subfigure}
    \\
    \begin{subfigure}{\linewidth}
        \centering
        \begin{tikzpicture}
            \node at (0,0) {\includegraphics[height=45pt,trim={50 50 50 50},clip]{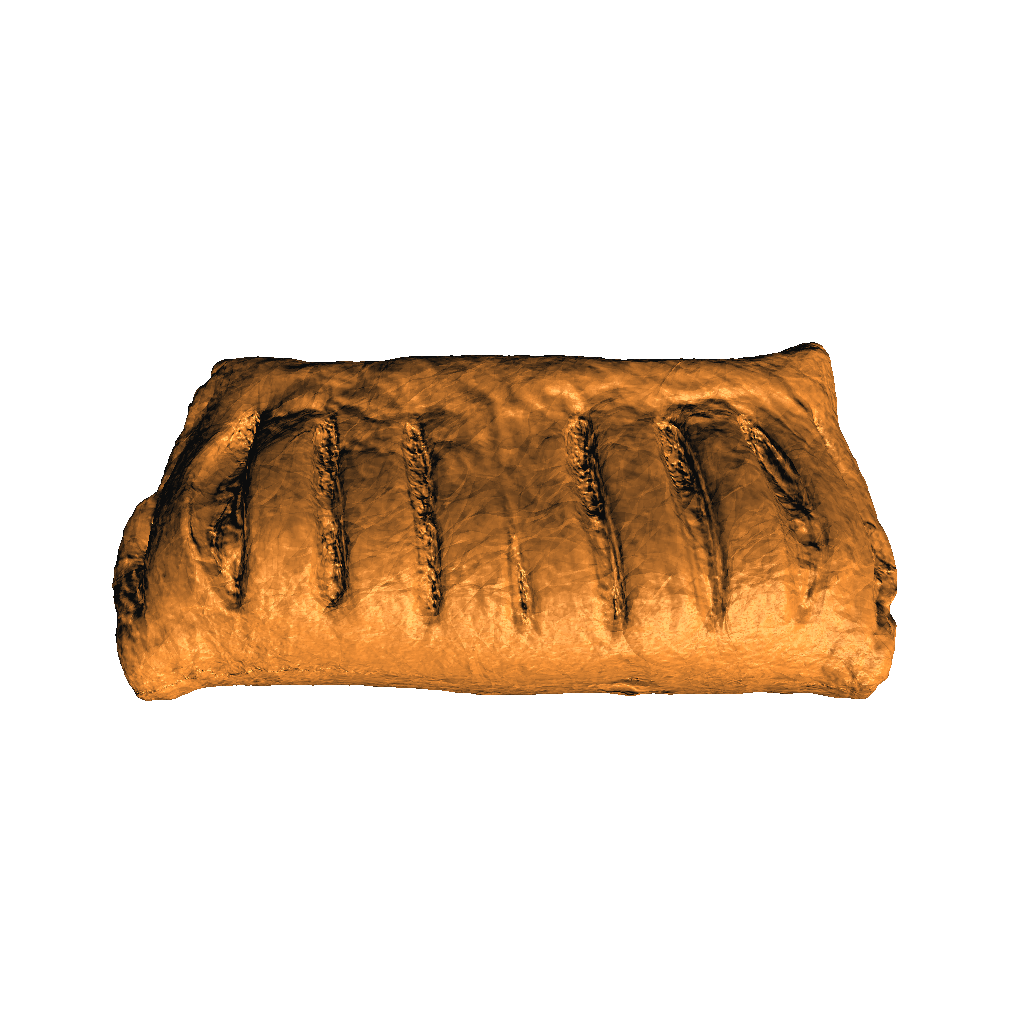}};
            \node at (1.7,0) {\includegraphics[height=45pt,trim={50 50 50 50},clip]{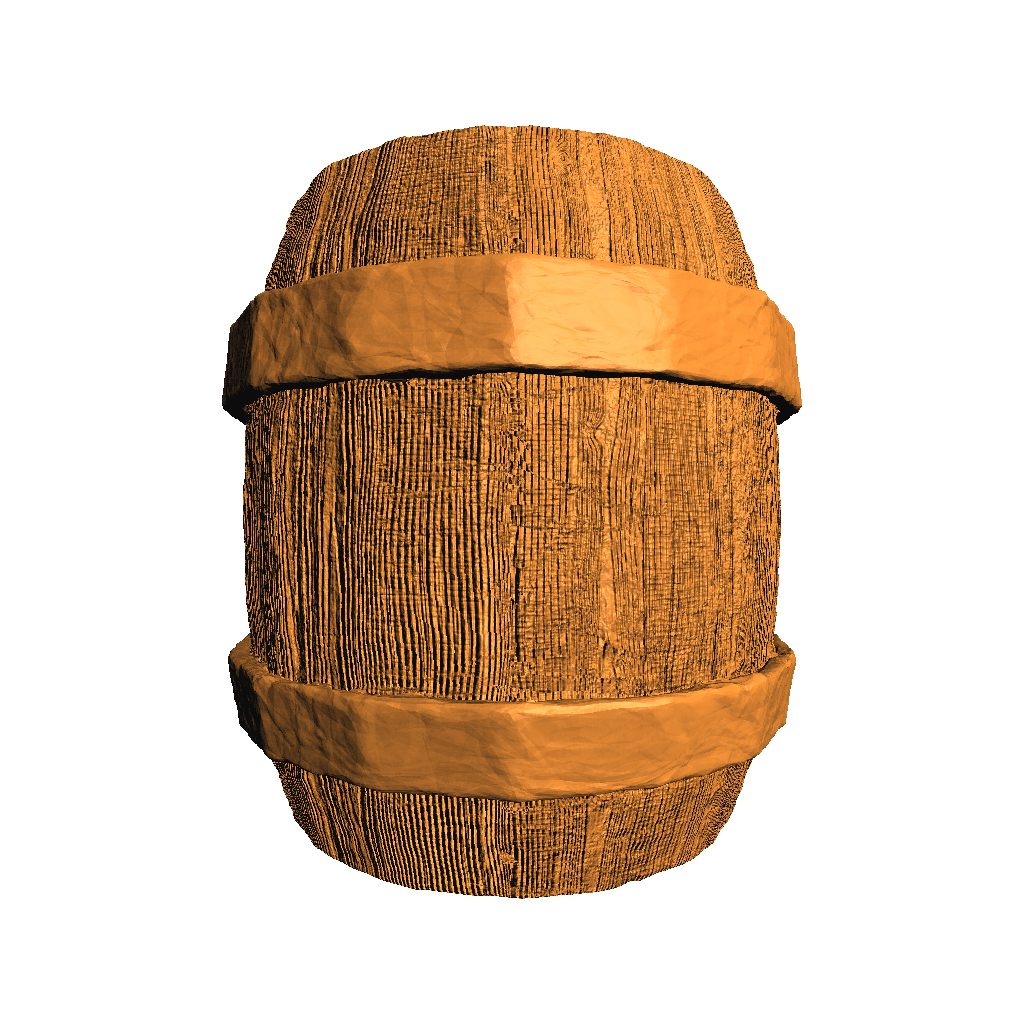}};
            \node at (3.4,0) {\includegraphics[height=45pt,trim={50 50 50 50},clip]{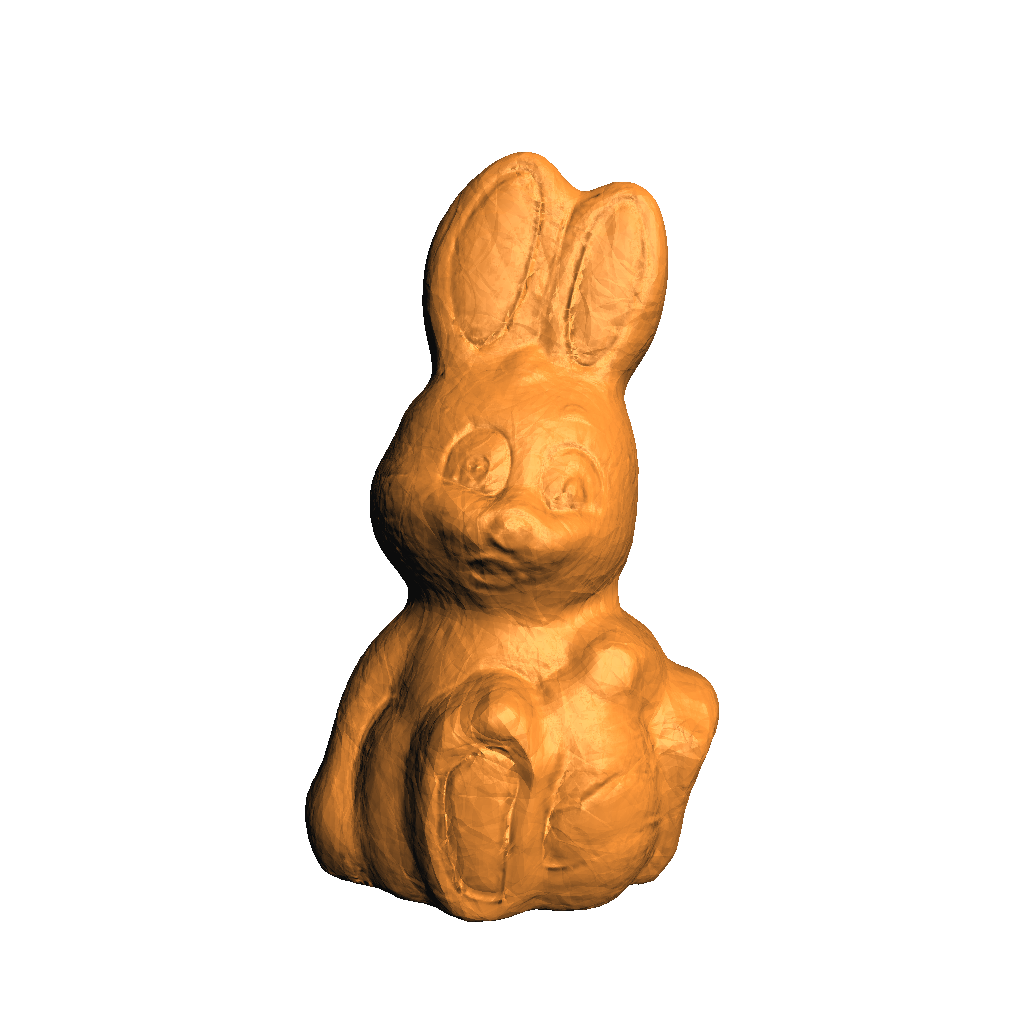}};
            \node at (5.1,0) {\includegraphics[height=45pt,trim={50 50 50 50},clip]{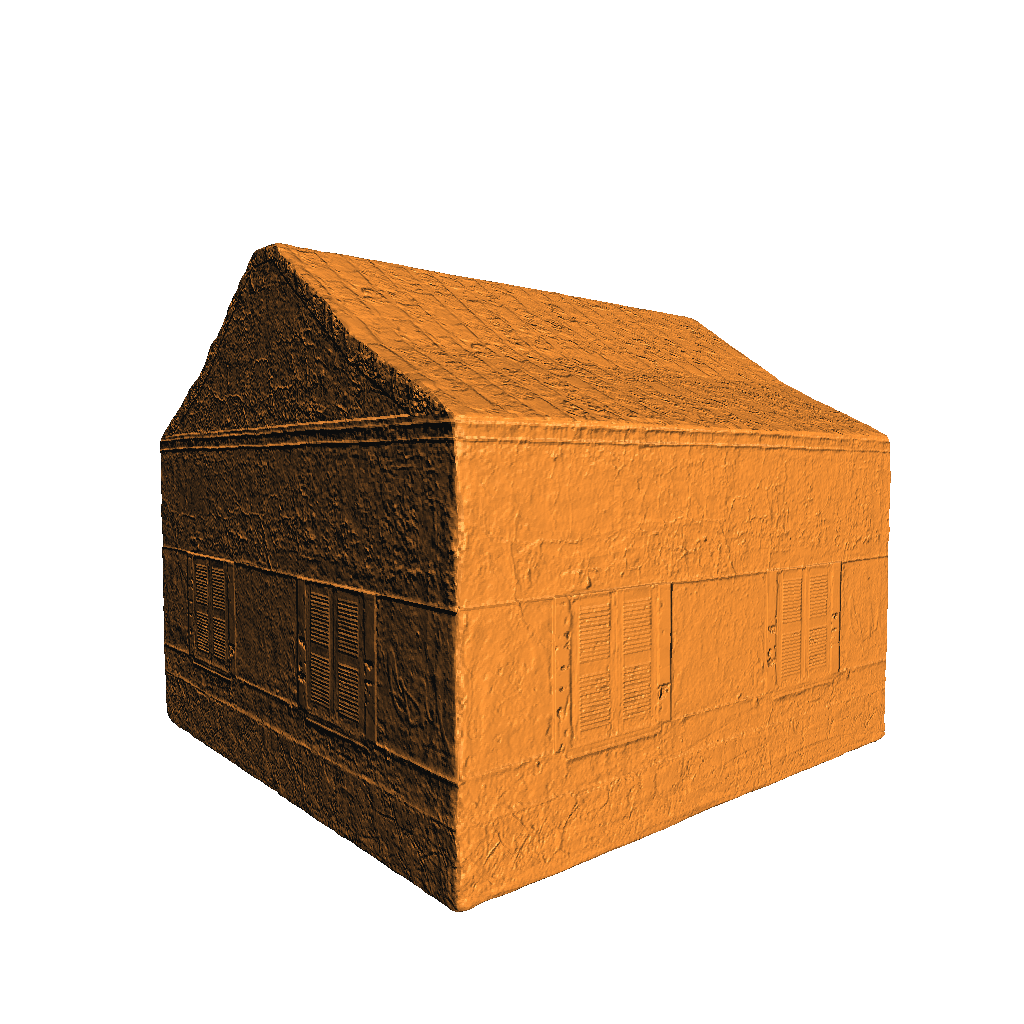}};
            \node at (0,-1.7) {\includegraphics[height=45pt,trim={50 50 50 50},clip]{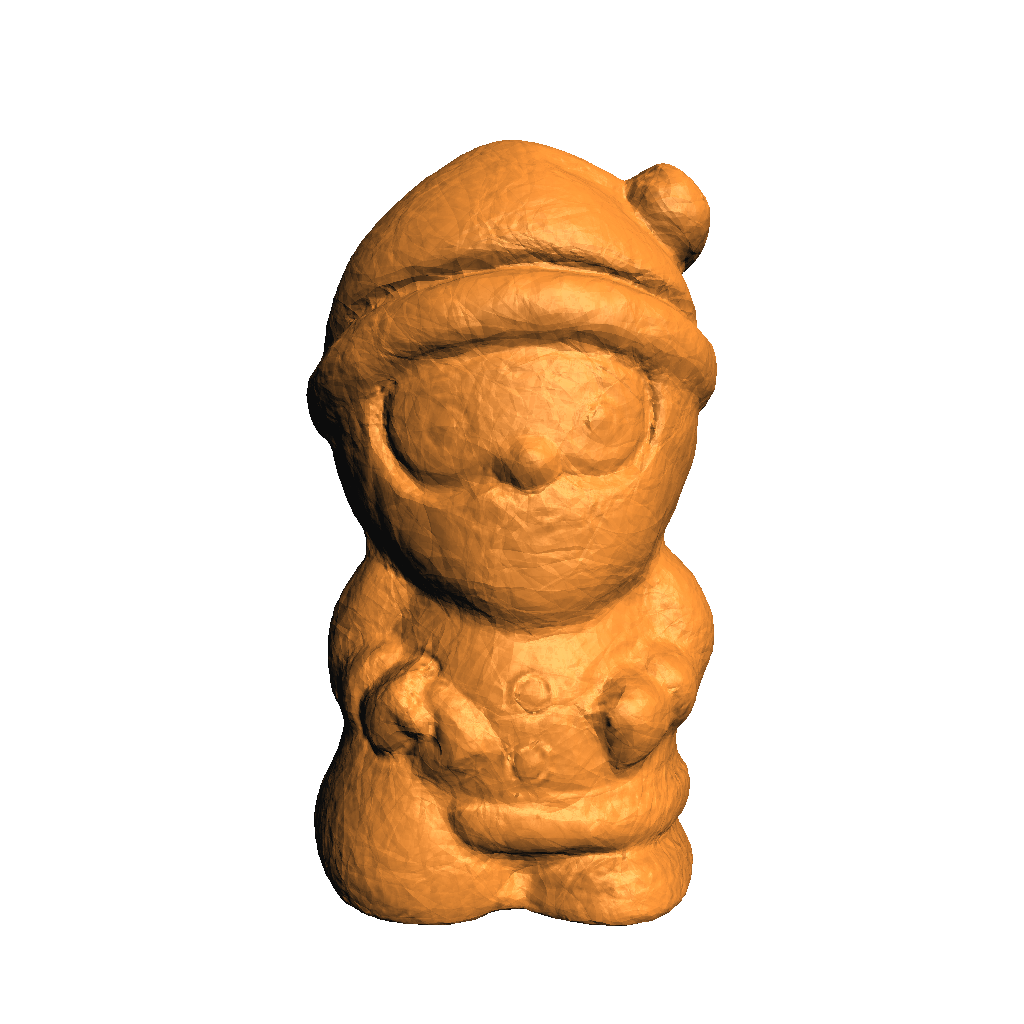}};
            \node at (1.7,-1.7) {\includegraphics[height=45pt,trim={50 50 50 50},clip]{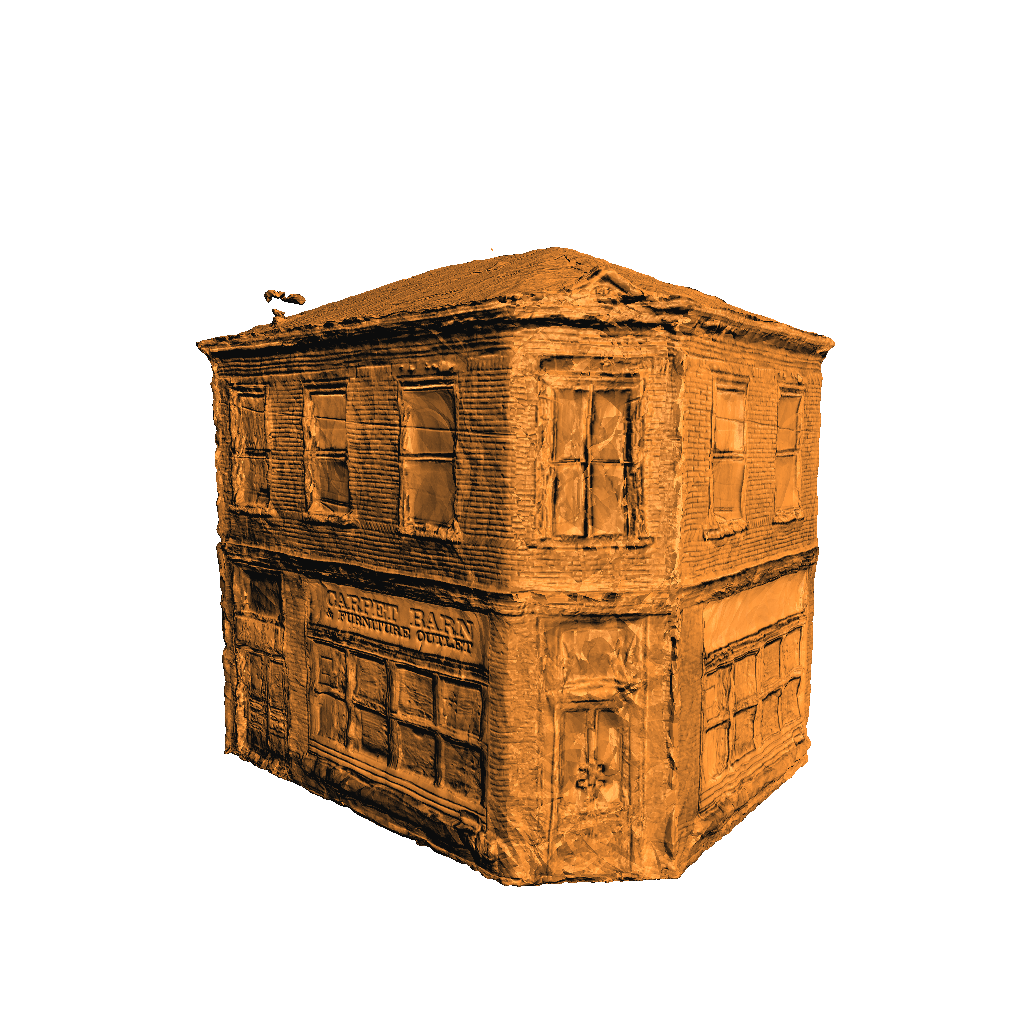}};
            \node at (3.4,-1.7) {\includegraphics[height=45pt,trim={50 50 50 50},clip]{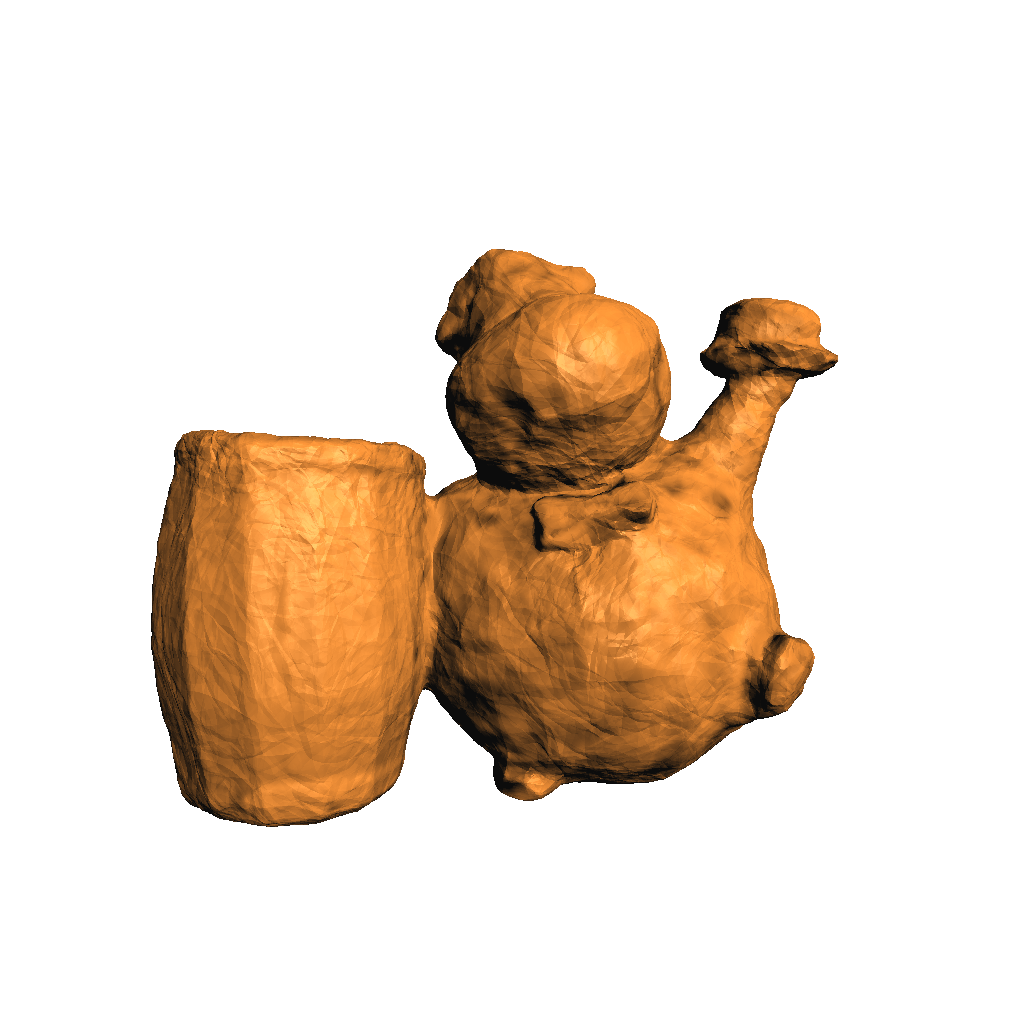}};
            \node at (5.1,-1.7) {\includegraphics[height=45pt,trim={50 50 50 50},clip]{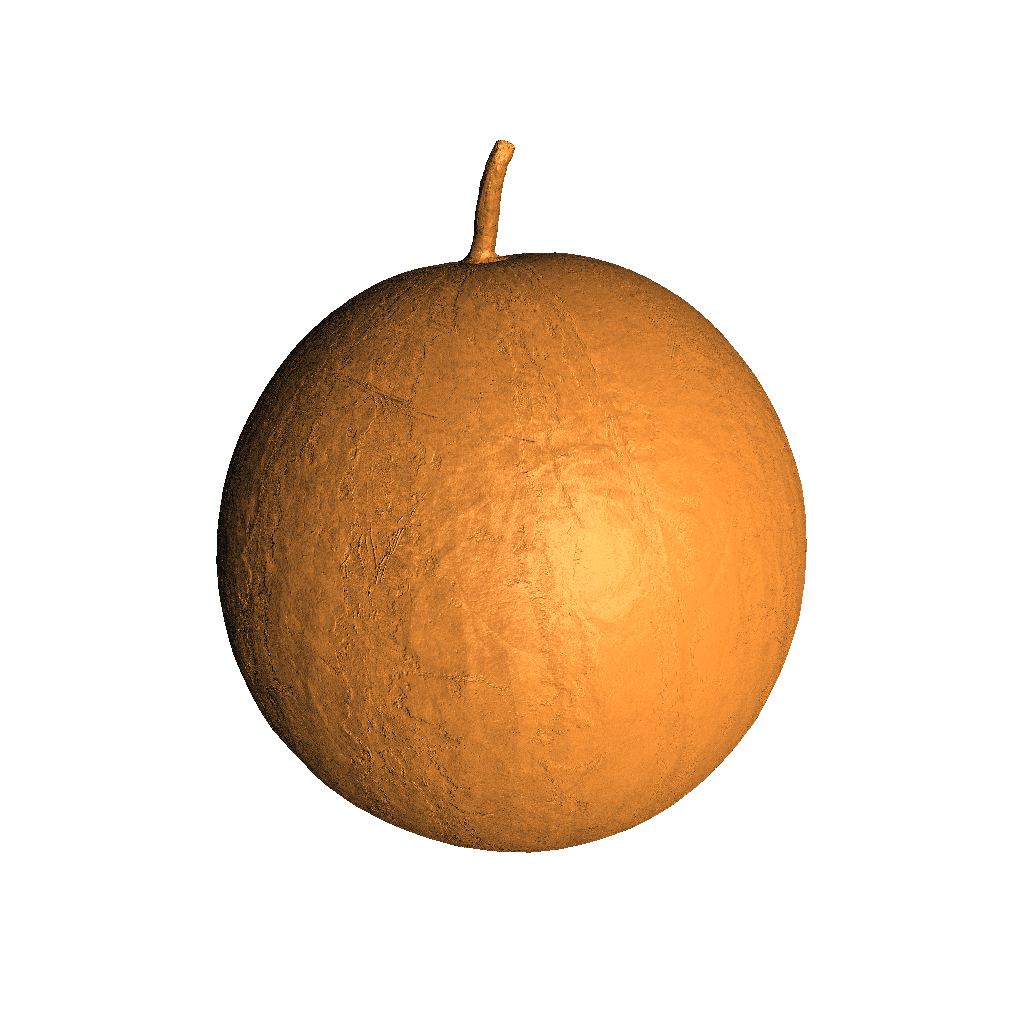}};
        \end{tikzpicture}
        \caption{With normal mapping}
        \label{fig:normal-mapping-b}
    \end{subfigure}
    \caption{Application of normal maps to neural implicit surfaces.}
    \label{fig:normal-mapping}
\end{figure}

%% file: figures/texture_editing.tex
\begin{figure}
    \centering
    \begin{subfigure}{0.49\linewidth}
        \centering
        \begin{tikzpicture}
            \node at (0,0) {\includegraphics[height=49pt]{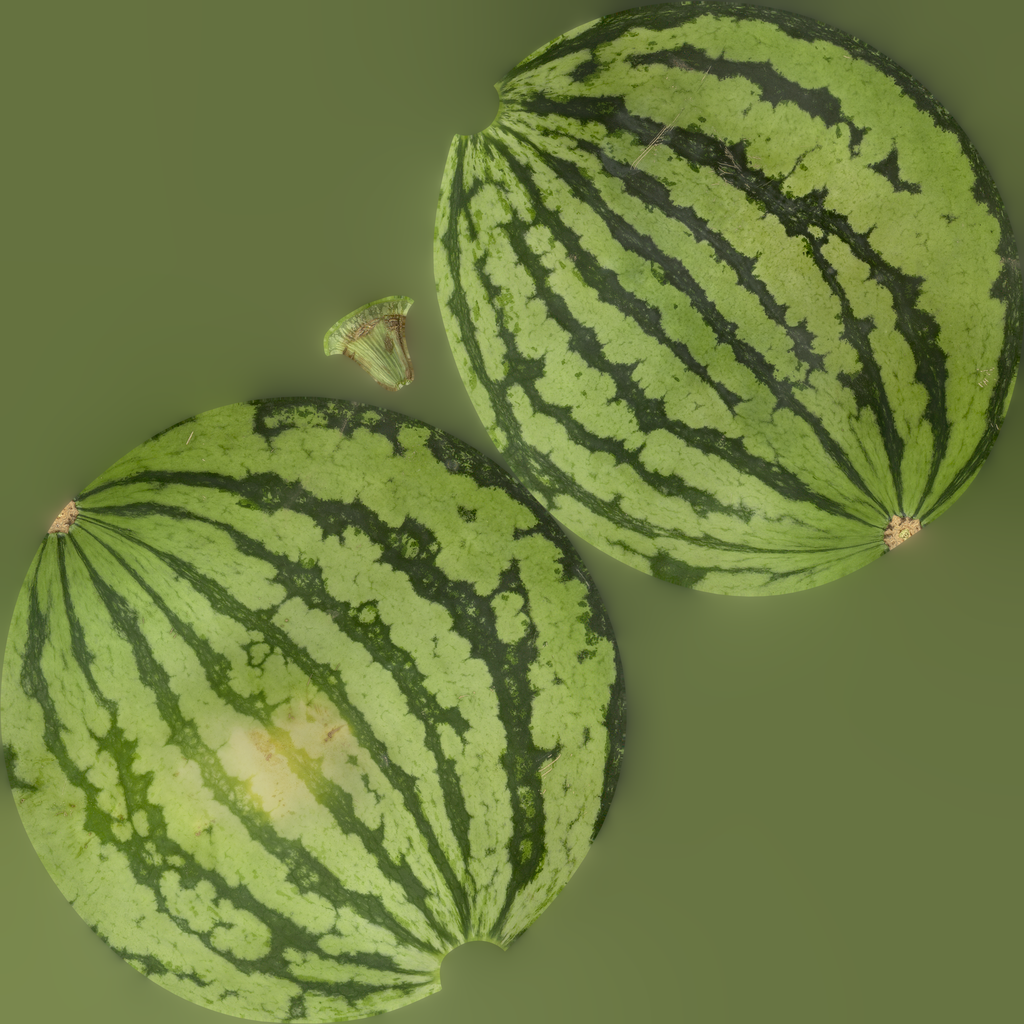}};
            \node at (1.8,0) {\includegraphics[height=49pt,trim={80 80 80 80},clip]{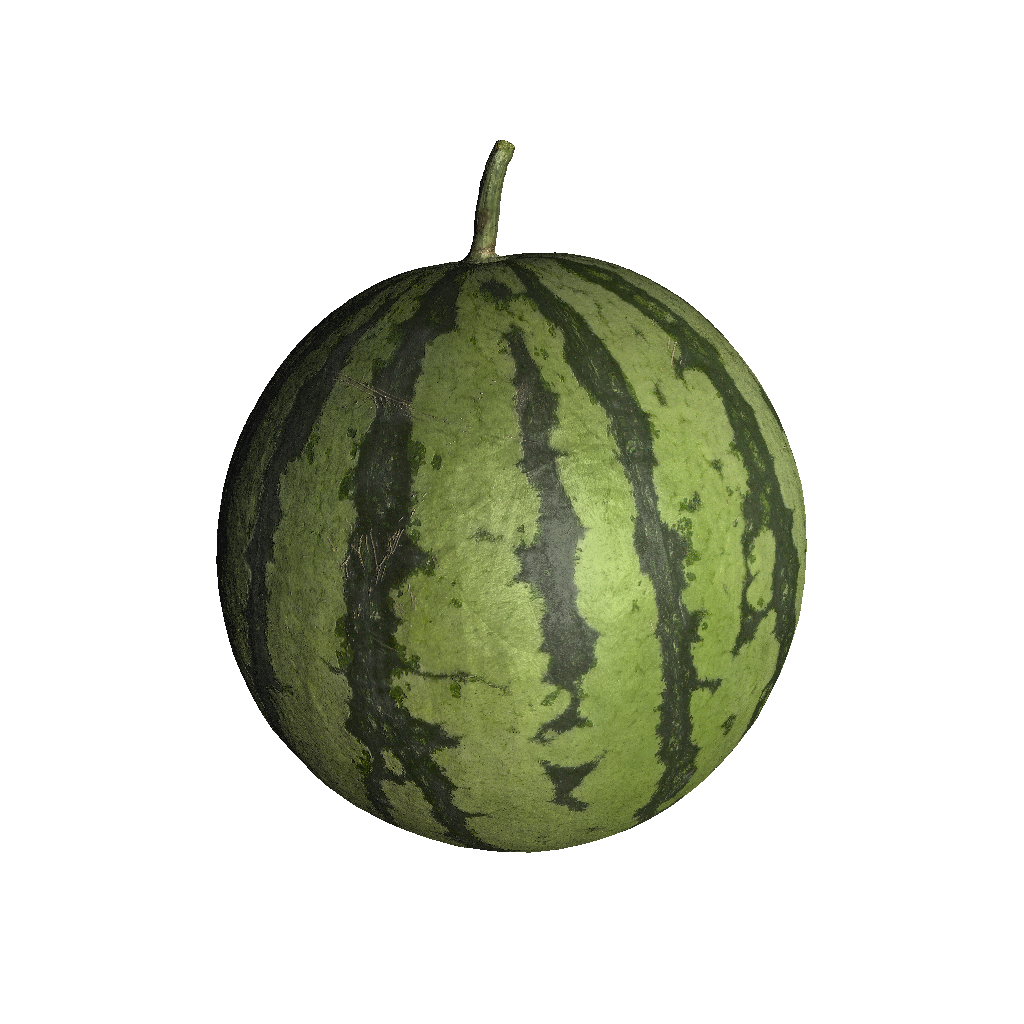}};
            \node at (0,-1.8) {\includegraphics[height=49pt]{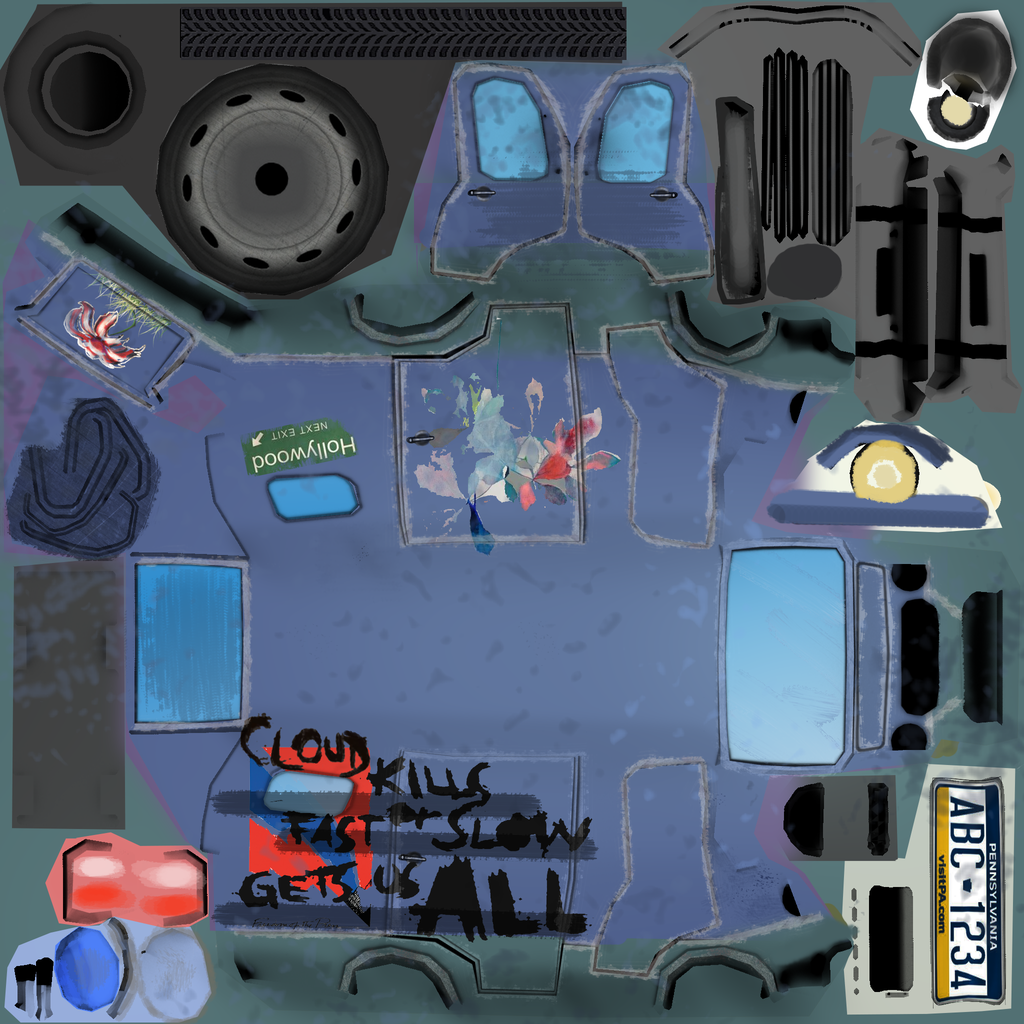}};
            \node at (1.8,-1.8) {\includegraphics[height=49pt,trim={80 80 80 80},clip]{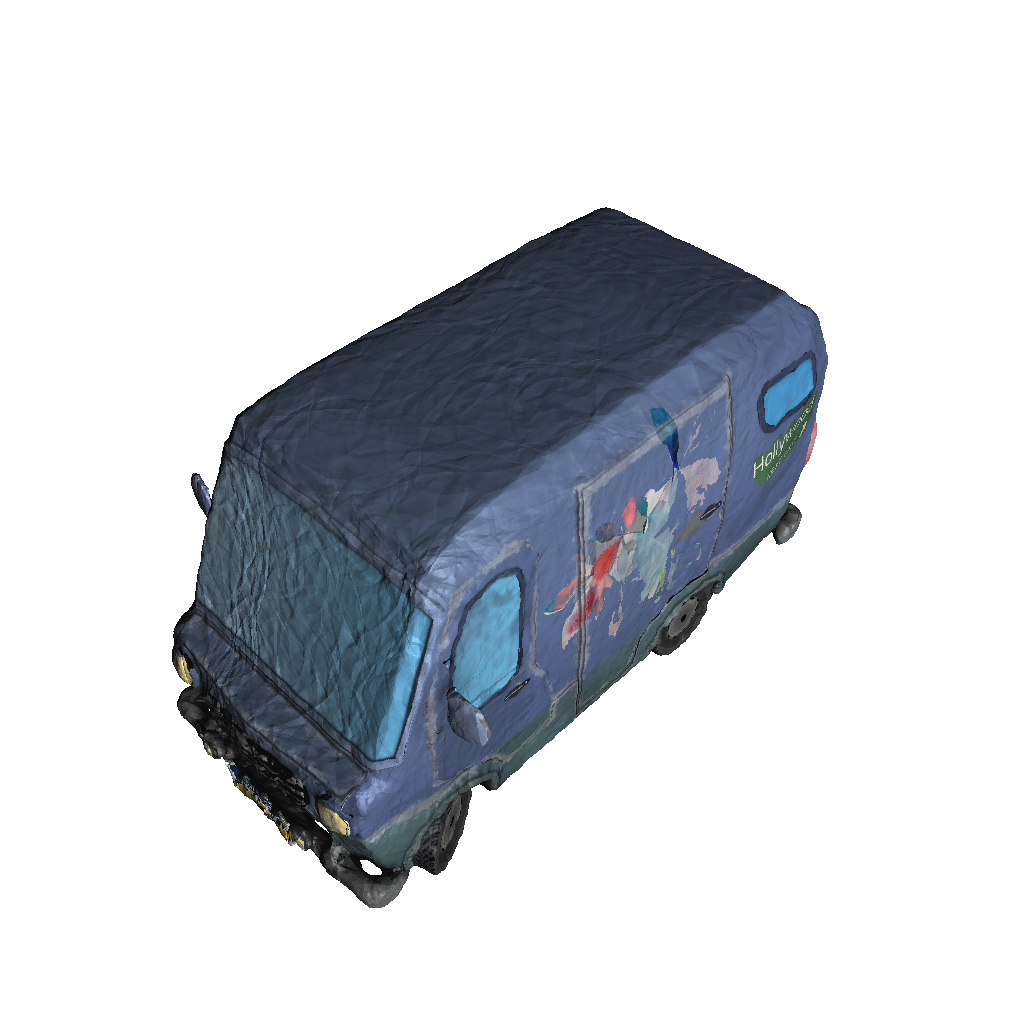}};
        \end{tikzpicture}
        \caption{Original diffuse maps}
        \label{fig:texture-editing-a}
    \end{subfigure}
    \hfil
    \begin{subfigure}{0.49\linewidth}
        \centering
        \begin{tikzpicture}
            \node at (0,0) {\includegraphics[height=49pt]{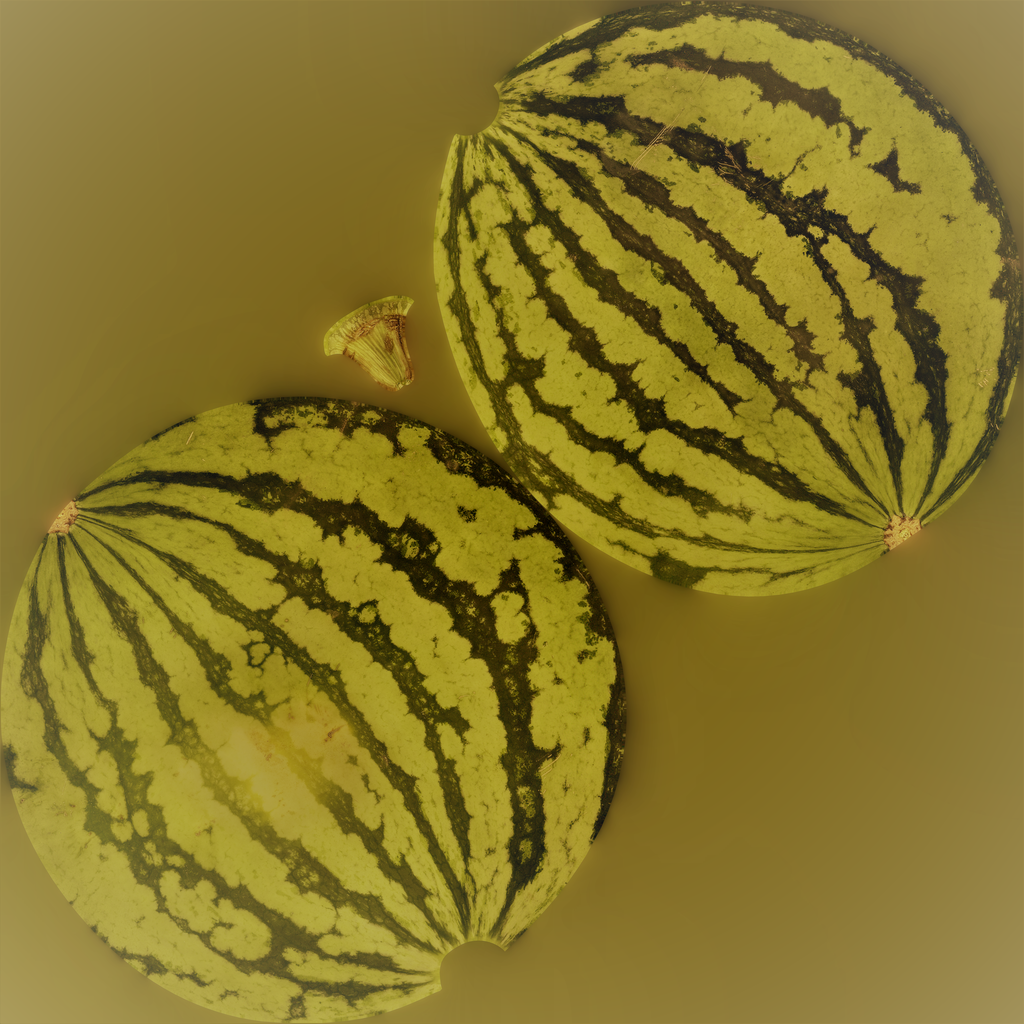}};
            \node at (1.8,0) {\includegraphics[height=49pt,trim={80 80 80 80},clip]{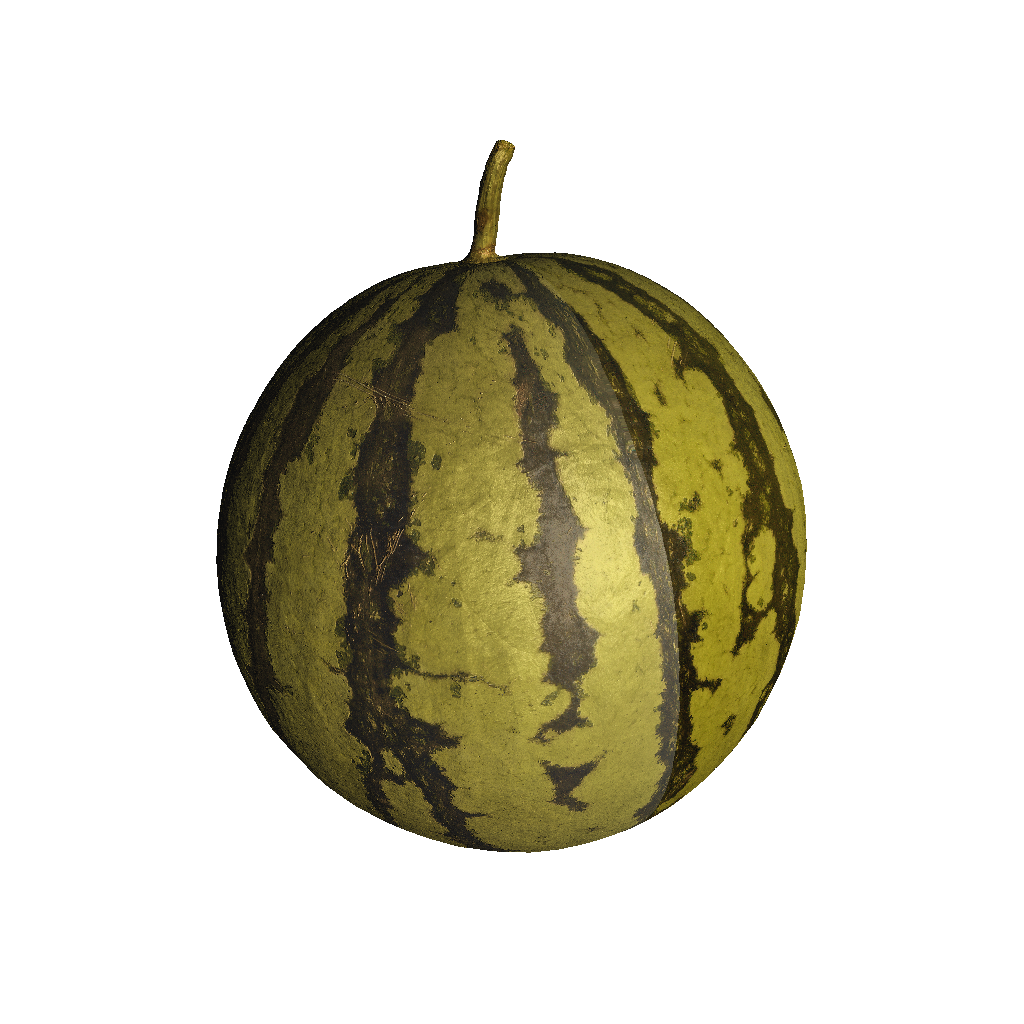}};
            \node at (0,-1.8) {\includegraphics[height=49pt]{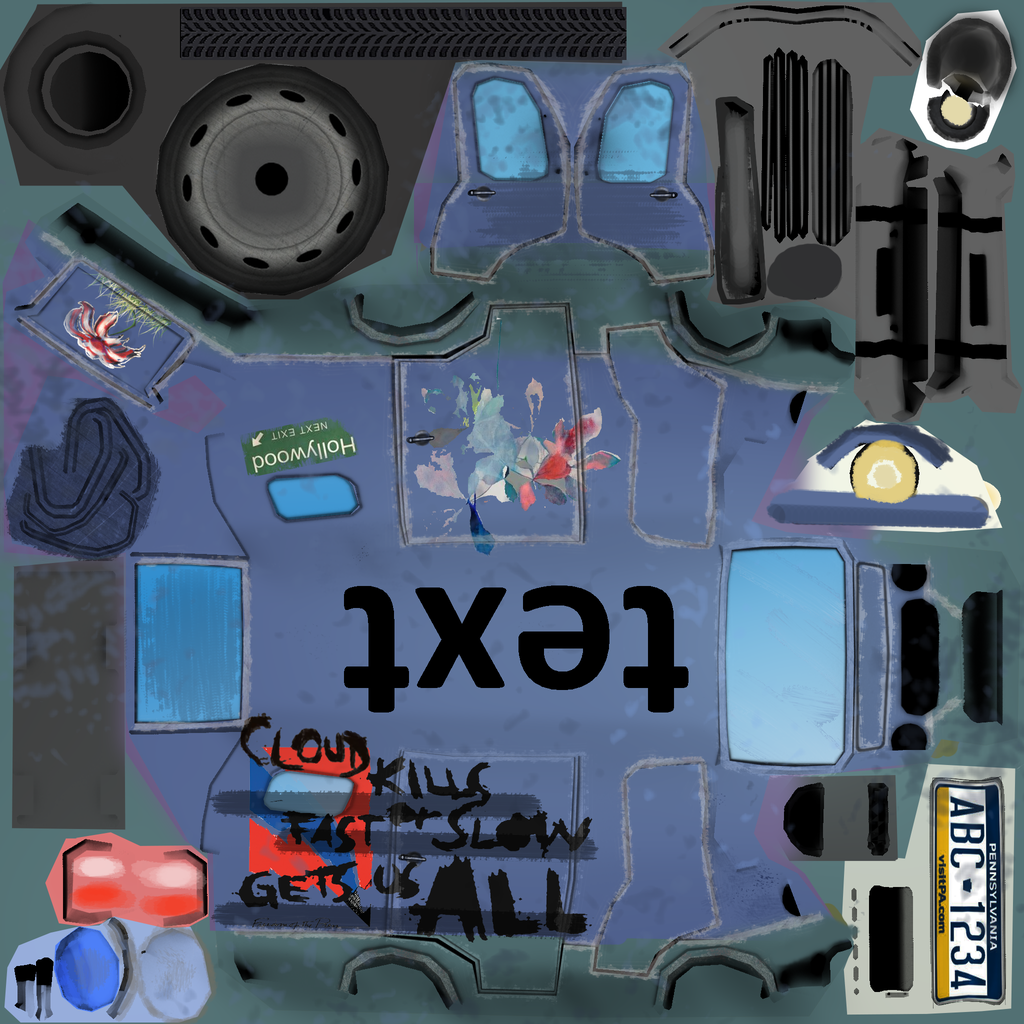}};
            \node at (1.8,-1.8) {\includegraphics[height=49pt,trim={80 80 80 80},clip]{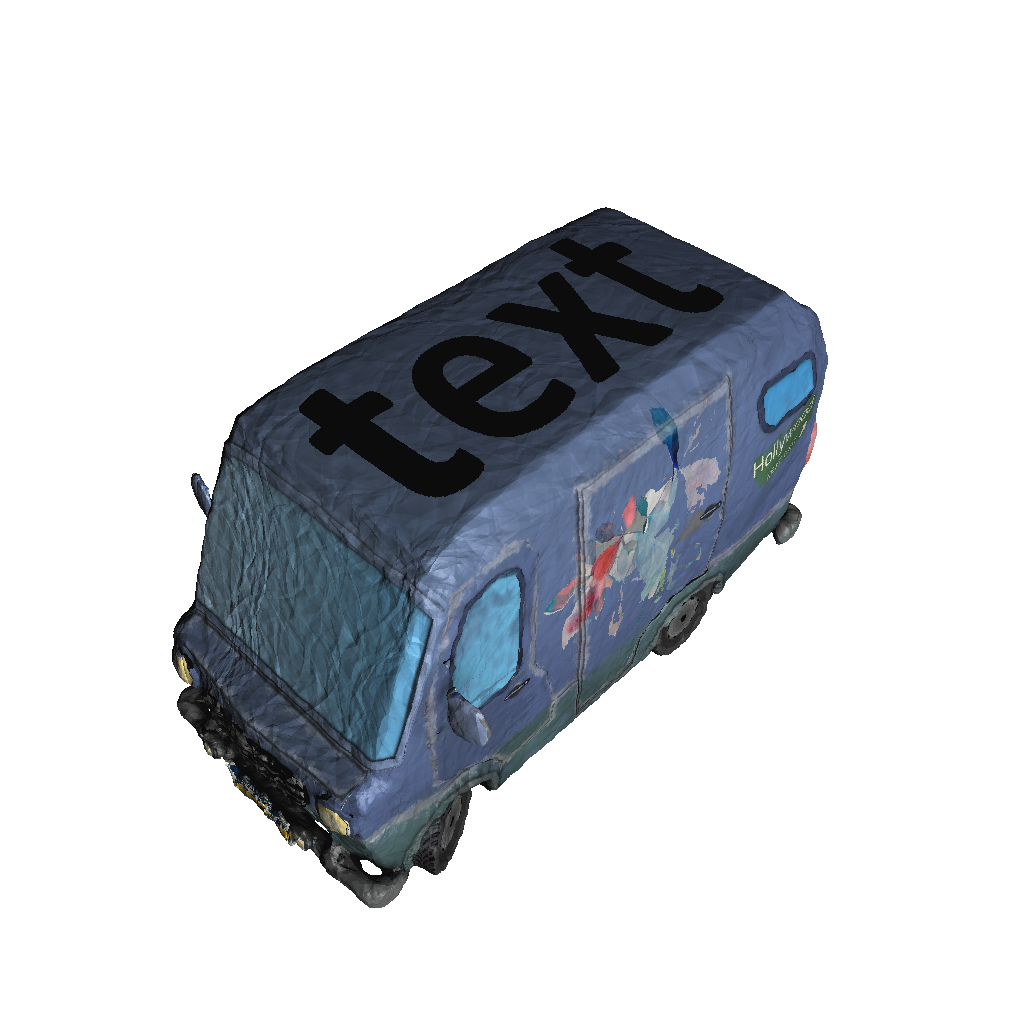}};
        \end{tikzpicture}
        \caption{Edited diffuse maps}
        \label{fig:texture-editing-b}
    \end{subfigure}
    \caption{Editing the diffuse maps applied to neural implicit surfaces, with the rendered results shown on the right side of each diffuse map.}
    \label{fig:texture-editing}
\end{figure}

%% file: tables/distortion_metric.tex
\begin{table}
\caption{Average $\delta_\text{mean}$, $\delta_\text{max}$, and $\delta_\text{std}$ of the learned and GT parameterizations in Figure~\ref{fig:uv-comparision}.}
\label{tab:distortion-metric}
\scriptsize
\centering
\begin{tabular}{ lccc }
\toprule
& $\delta_\text{mean}$ & $\delta_\text{max}$ & $\delta_\text{std}$ \\
\midrule
\texttt{point2UV} & $13.63$ & $6.80\times10^4$ & $284.89$ \\
Ours & $9.60$ & $9.96\times10^3$ & $60.13$ \\
GT & $3.58$ & $6.14\times10^3$ & $17.84$ \\
\bottomrule
\end{tabular}
\end{table}

%% file: figures/model_comparison.tex
\begin{figure}
    \centering
    \begin{subfigure}{0.24\linewidth}
        \centering
        \begin{tikzpicture}
            \node at (0,0) {\includegraphics[height=45pt,trim={16 16 16 16},clip]{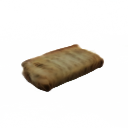}};
            \node at (0,-1.7) {\includegraphics[height=45pt,trim={16 16 16 16},clip]{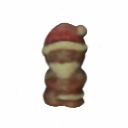}};
            \node at (0,-3.4) {\includegraphics[height=45pt,trim={16 16 16 16},clip]{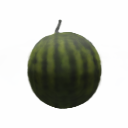}};
        \end{tikzpicture}
        \caption{SRN}
        \label{fig:model-comparison-a}
    \end{subfigure}
    \hfil
    \begin{subfigure}{0.24\linewidth}
        \centering
        \begin{tikzpicture}
            \node at (0,0) {\includegraphics[height=45pt,trim={16 16 16 16},clip]{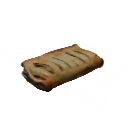}};
            \node at (0,-1.7) {\includegraphics[height=45pt,trim={16 16 16 16},clip]{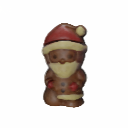}};
            \node at (0,-3.4) {\includegraphics[height=45pt,trim={16 16 16 16},clip]{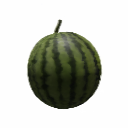}};
        \end{tikzpicture}
        \caption{NeRF}
        \label{fig:model-comparison-b}
    \end{subfigure}
    \hfil
    \begin{subfigure}{0.24\linewidth}
        \centering
        \begin{tikzpicture}
            \node at (0,0) {\includegraphics[height=45pt,trim={128 128 128 128},clip]{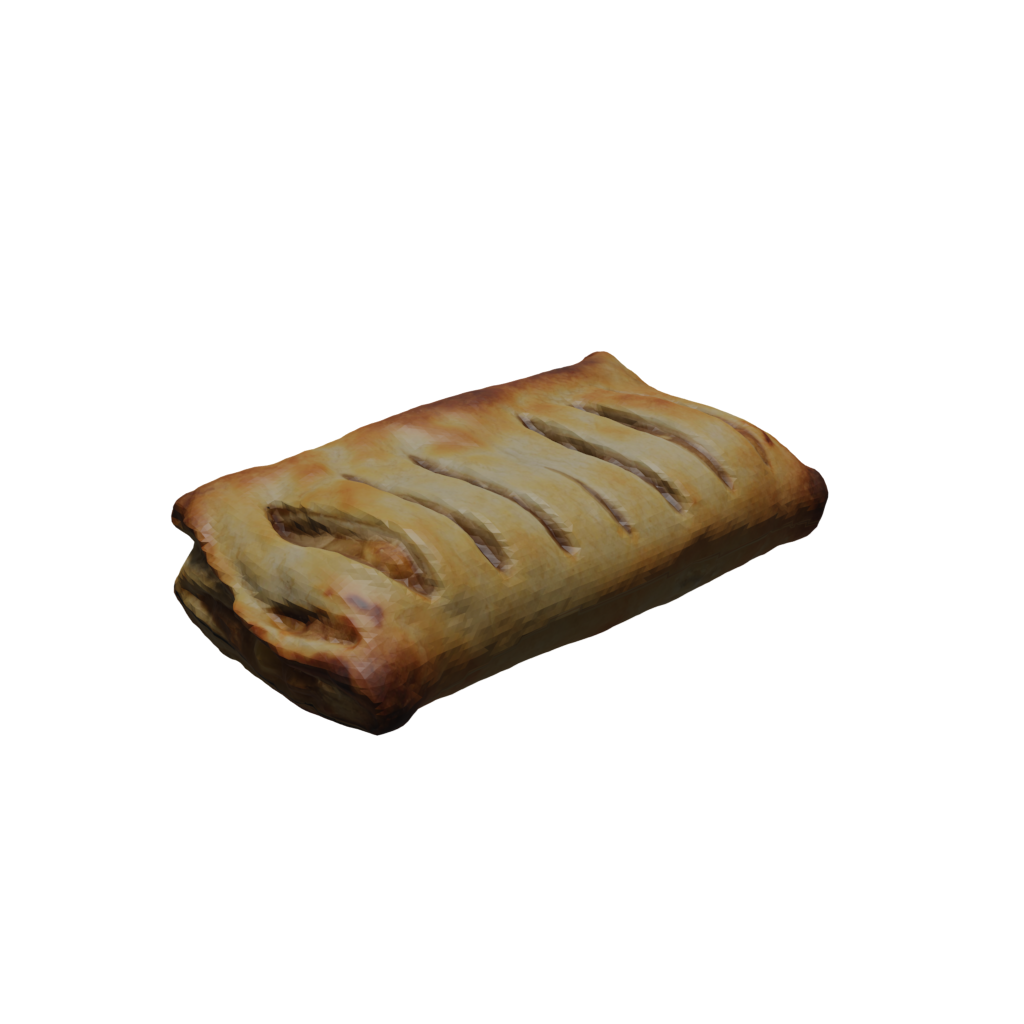}};
            \node at (0,-1.7) {\includegraphics[height=45pt,trim={128 128 128 128},clip]{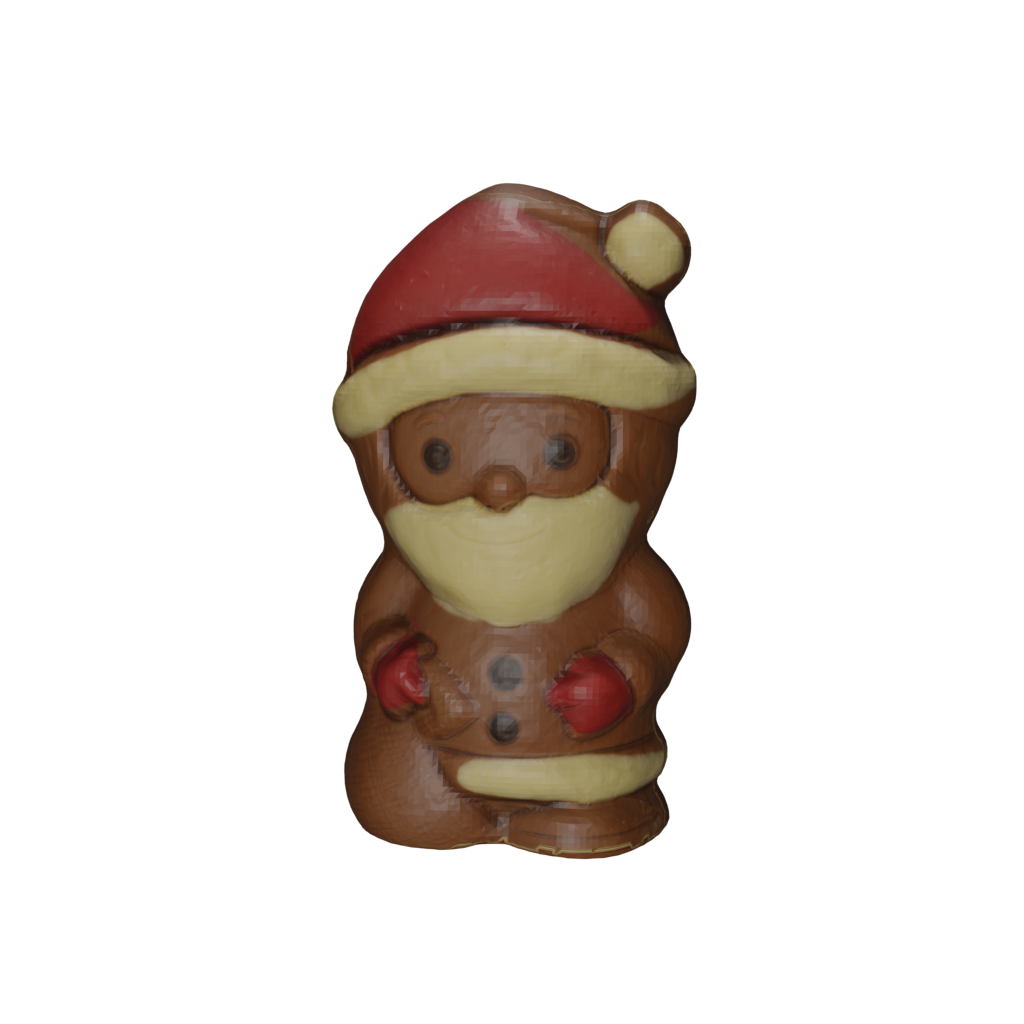}};
            \node at (0,-3.4) {\includegraphics[height=45pt,trim={128 128 128 128},clip]{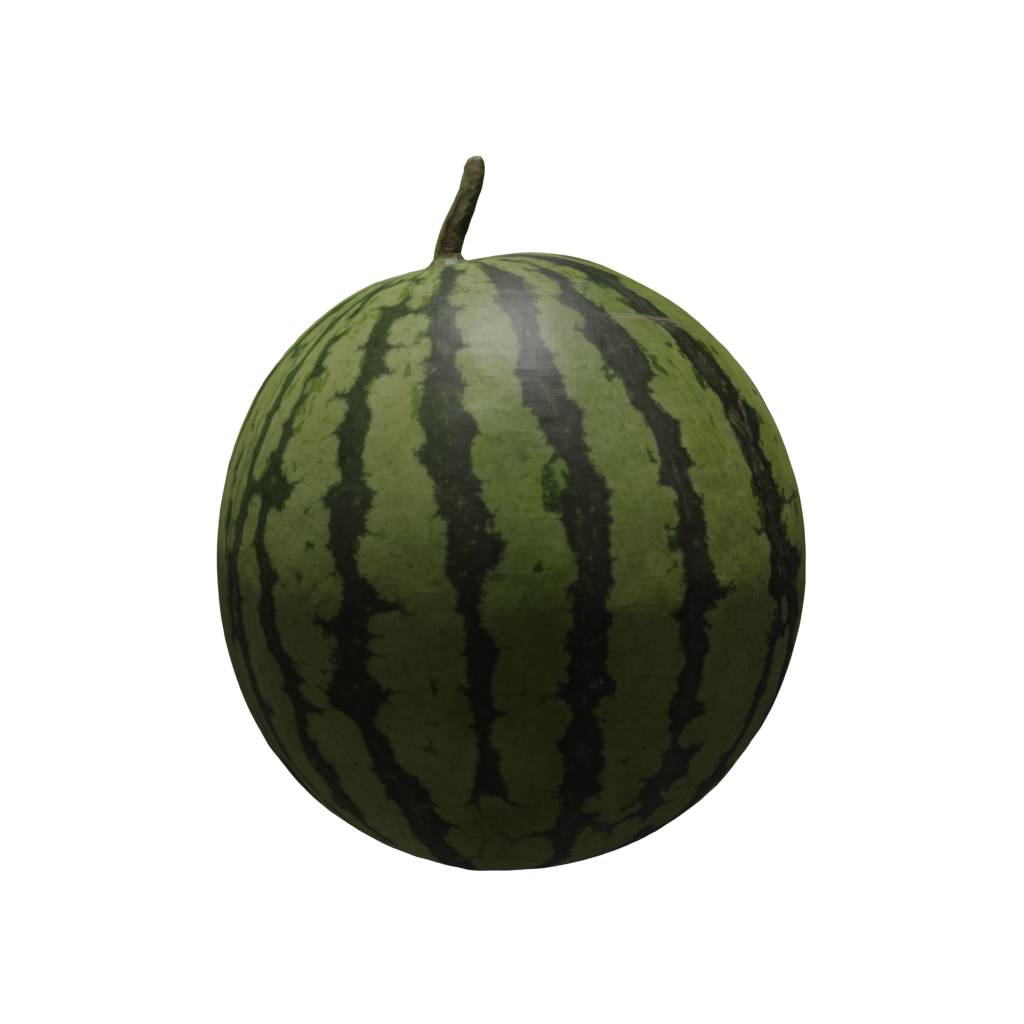}};
        \end{tikzpicture}
        \caption{Ours}
        \label{fig:model-comparison-c}
    \end{subfigure}
    \hfil
    \begin{subfigure}{0.24\linewidth}
        \centering
        \begin{tikzpicture}
            \node at (0,0) {\includegraphics[height=45pt,trim={128 128 128 128},clip]{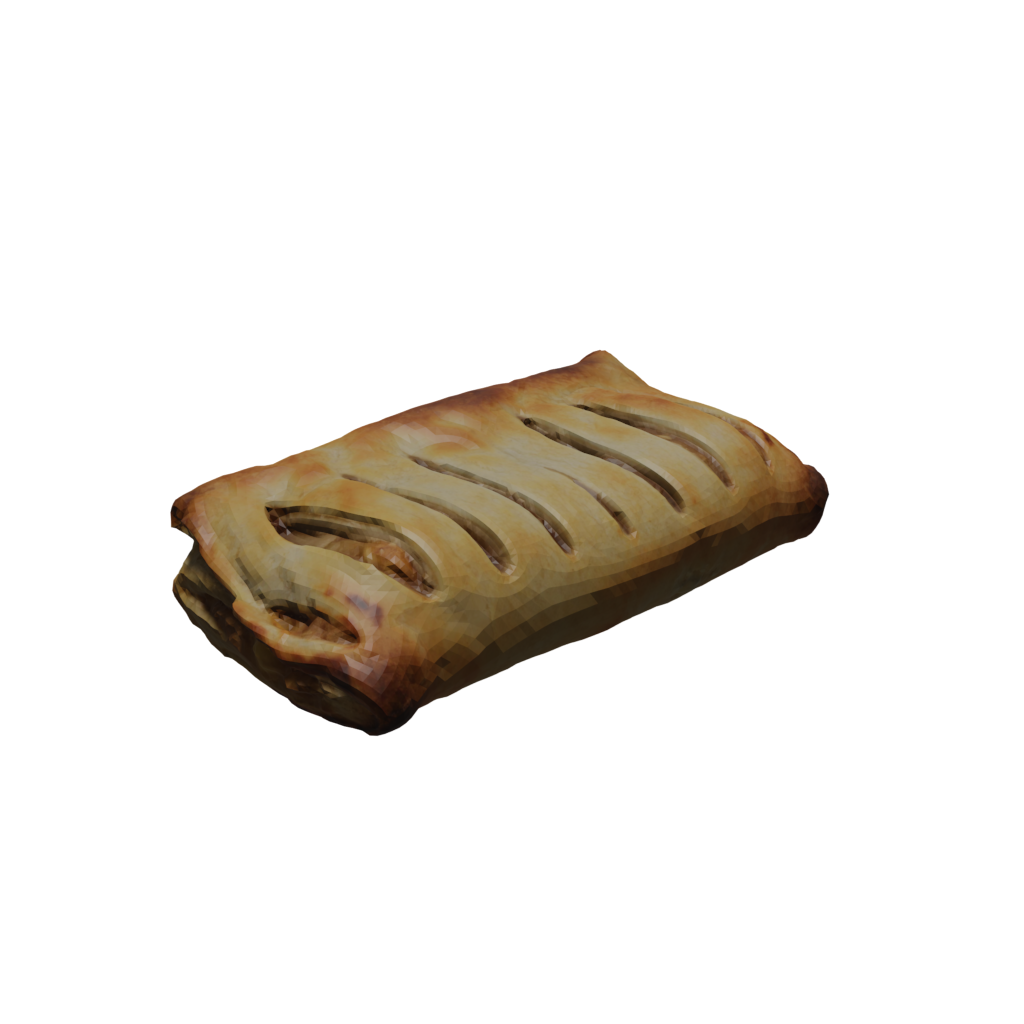}};
            \node at (0,-1.7) {\includegraphics[height=45pt,trim={128 128 128 128},clip]{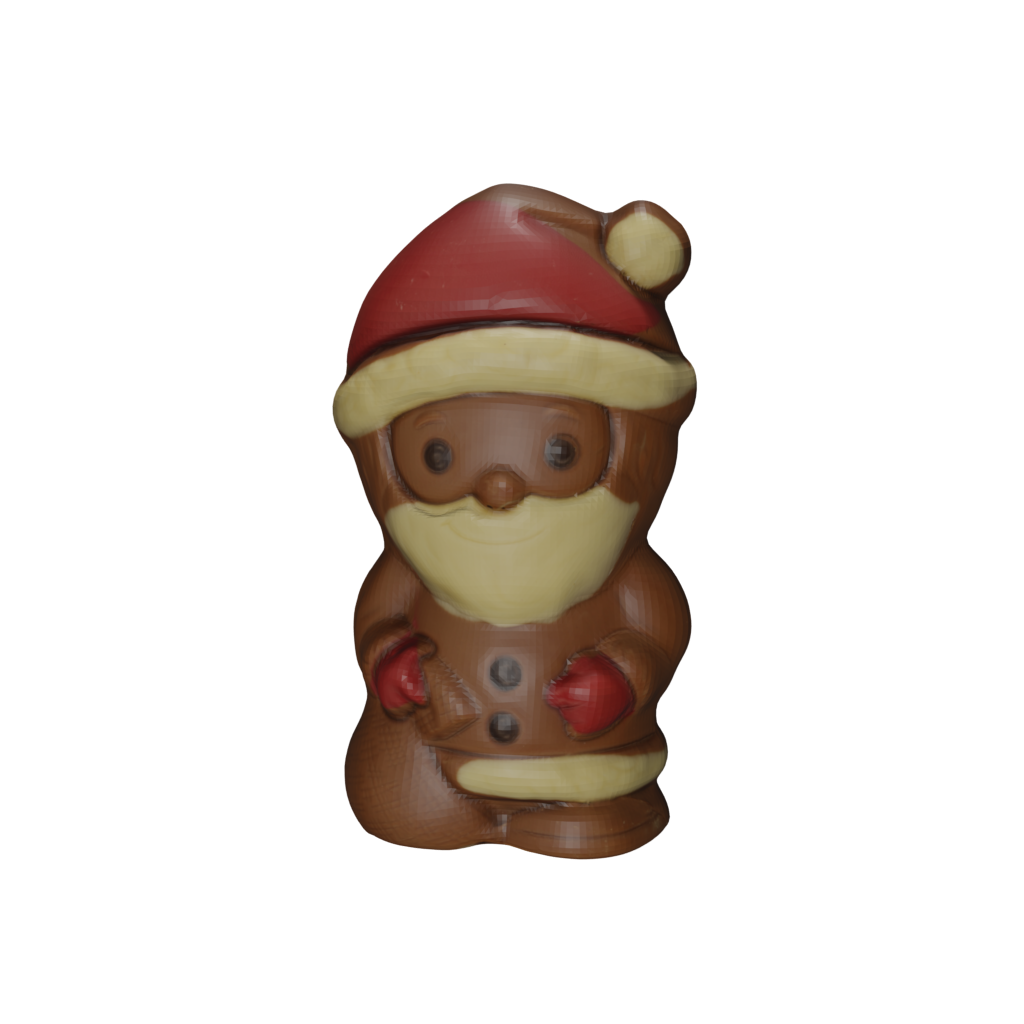}};
            \node at (0,-3.4) {\includegraphics[height=45pt,trim={128 128 128 128},clip]{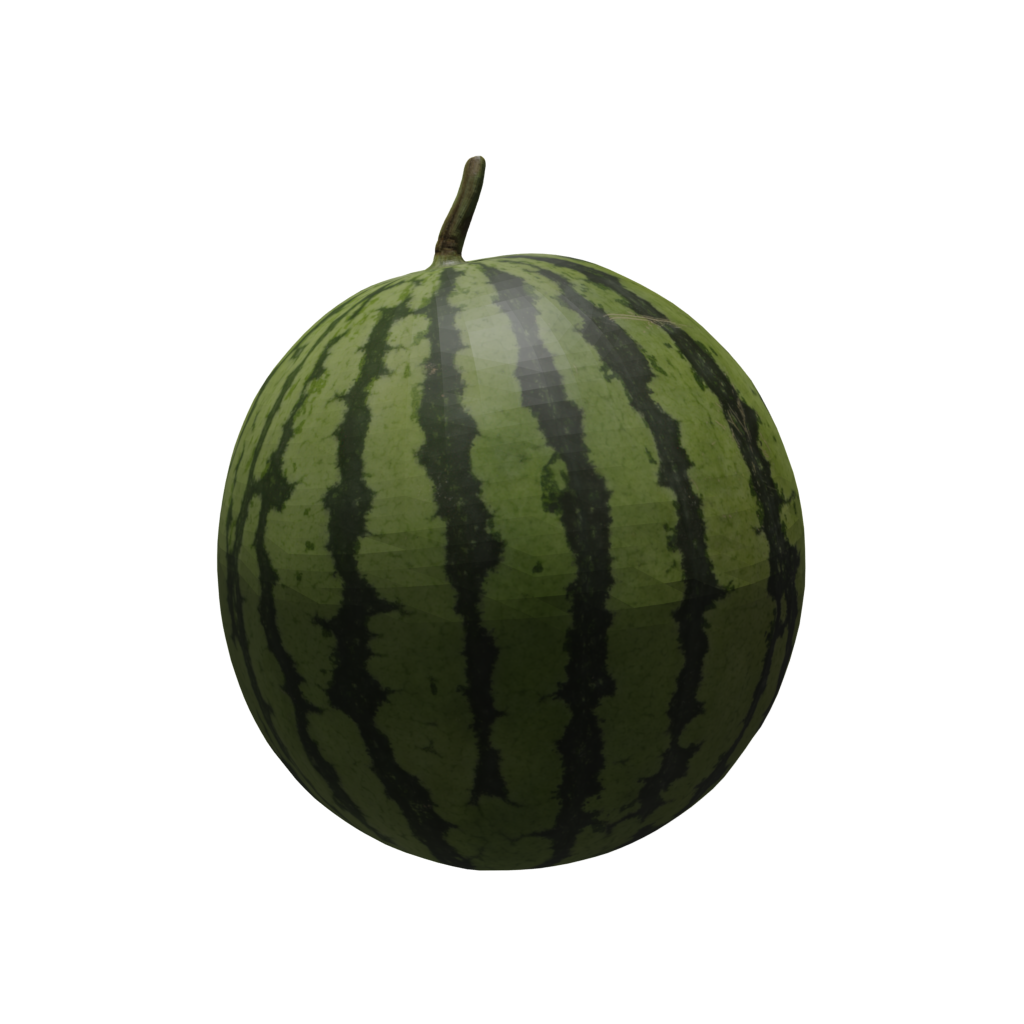}};
        \end{tikzpicture}
        \caption{GT}
        \label{fig:model-comparison-d}
    \end{subfigure}
    \caption{Comparison between the texture mapping results by our method and the synthesized views by neural appearance models.}
    \label{fig:model-comparison}
\end{figure}

%% file: tables/model_comparison_quantitative.tex
\begin{table}[t!]
\caption{Average MSE, PSNR, and SSIM of the synthesized views and our rendered results in Figure~\ref{fig:model-comparison}.}
\label{tab:model-comparison-quantitative}
\scriptsize
\centering
\begin{tabular}{ lccc }
\toprule
& MSE & PSNR & SSIM \\
\midrule
SRN & $40.10$ & $32.18$ & $0.947$ \\
NeRF & $17.95$ & $35.63$ & $0.976$ \\
Ours & $2.95$ & $43.85$ & $0.996$ \\
\bottomrule
\end{tabular}
\end{table}

%% file: figures/model_pruning.tex
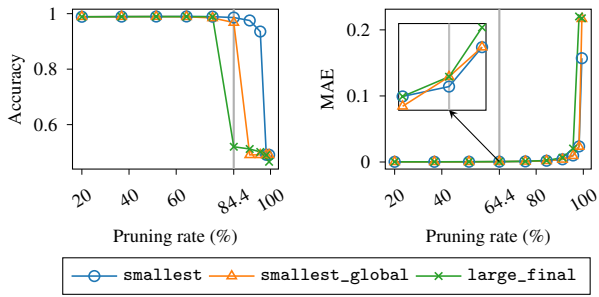
\begin{figure}
    \centering
    \input{images/model-pruning/pruning_comparison}
    \ref*{pruning_strategies}\\
    \caption{Test results when pruning the model using different strategies, with the $x$-axes presenting the pruning rates (in percentage) and $y$-axes being the test accuracy of \texttt{point2component} (left) and the test MAE of \texttt{point2UV} (right).}
    \label{fig:model-pruning}
\end{figure}

%% file: images/model-pruning/pruning_comparison.tex
\definecolor{color0}{rgb}{0.12156862745098,0.466666666666667,0.705882352941177}
\definecolor{color1}{rgb}{1,0.498039215686275,0.0549019607843137}
\definecolor{color2}{rgb}{0.172549019607843,0.627450980392157,0.172549019607843}

\begin{tikzpicture}[trim axis group left, trim axis group right]
\hspace{2mm}
\pgfmathsetlengthmacro\MajorTickLength{
  \pgfkeysvalueof{/pgfplots/major tick length} * 0.5
}

\begin{groupplot}[group style={group size=2 by 1, horizontal sep=40pt},
legend to name={pruning_strategies}, legend columns=-1,
width=0.55\linewidth]
\nextgroupplot[
tick align=outside,
xtick pos=left,
ytick pos=left,
major tick length=\MajorTickLength,
ticklabel style={font=\scriptsize},
x grid style={white!69.0196078431373!black},
xlabel={Pruning rate (\%)},
x label style={font=\scriptsize,anchor=mid},
xmin=16.0305, xmax=103.3595,
xtick style={color=black},
xtick={20,40,60,100},
extra x ticks={84.4},
extra tick style={grid=major,major grid style=thick},
xticklabel style={rotate=30},
y grid style={white!69.0196078431373!black},
ylabel={Accuracy},
y label style={at={(axis description cs:-0.28,0.5)},font=\scriptsize,anchor=mid},
ymin=0.439805, ymax=1.016095,
ytick style={color=black}
]
\addplot [semithick, color0, mark=o, mark size=2, mark options={solid}]
table {%
20 0.9888
36.85 0.9894
51.52 0.9897
64.37 0.9899
75.42 0.9892
84.42 0.9859
91.17 0.9754
95.65 0.9355
98.2 0.4912
99.39 0.4912
};
\addplot [semithick, color1, mark=triangle, mark size=2, mark options={solid}]
table {%
20 0.989
36.85 0.9892
51.52 0.9894
64.37 0.9889
75.42 0.9846
84.42 0.9679
91.17 0.491
95.65 0.491
98.2 0.491
99.39 0.491
};
\addplot [semithick, color2, mark=x, mark size=2, mark options={solid}]
table {%
20 0.9887
36.85 0.9891
51.52 0.9893
64.37 0.9893
75.42 0.9888
84.42 0.5202
91.17 0.5122
95.65 0.5015
98.2 0.4905
99.39 0.466
};

\nextgroupplot[
tick align=outside,
xtick pos=left,
ytick pos=left,
major tick length=\MajorTickLength,
ticklabel style={font=\scriptsize},
x grid style={white!69.0196078431373!black},
xlabel={Pruning rate (\%)},
x label style={font=\scriptsize,anchor=mid},
xmin=16.0305, xmax=103.3595,
xtick style={color=black},
xtick={20,40,80,100},
extra x ticks={64.4},
extra tick style={grid=major,major grid style=thick},
xticklabel style={rotate=30},
y grid style={white!69.0196078431373!black},
ylabel={MAE},
y label style={at={(axis description cs:-0.28,0.5)},font=\scriptsize,anchor=mid},
ymin=-0.0107, ymax=0.2313,
ytick style={color=black}
]
\addplot [semithick, color0, mark=o, mark size=2, mark options={solid}]
table {%
20 0.0003
36.85 0.0003
51.52 0.0004
64.37 0.0005
75.42 0.0009
84.42 0.0019
91.17 0.0041
95.65 0.0099
98.2 0.0236
99.39 0.1571
};
\addplot [semithick, color1, mark=triangle, mark size=2, mark options={solid}]
table {%
20 0.0003
36.85 0.0003
51.52 0.0003
64.37 0.0006
75.42 0.0009
84.42 0.0018
91.17 0.0039
95.65 0.0094
98.2 0.0234
99.39 0.2168
};
\addplot [semithick, color2, mark=x, mark size=2, mark options={solid}]
table {%
20 0.0003
36.85 0.0003
51.52 0.0004
64.37 0.0006
75.42 0.0011
84.42 0.0023
91.17 0.0066
95.65 0.0204
98.2 0.2203
99.39 0.2176
};
\coordinate (A) at (axis cs:64.37,0.0006);

\legend{\texttt{smallest},\texttt{smallest\char`_global},\texttt{large\char`_final}}
\end{groupplot}

\begin{axis}[
at={(0.55\linewidth,0.1\linewidth)},
width=0.35\linewidth,
height=0.35\linewidth,
log basis x={10},
xmin=50.5475744582952, xmax=76.8709169854605,
xmode=log,
extra x ticks={64.4},
extra tick style={grid=major,major grid style=thick},
ticks=none,
ymin=0.00026, ymax=0.00114
]
\addplot [semithick, color0, mark=o, mark size=2, mark options={solid}]
table {%
51.52 0.0004
64.37 0.0005
75.42 0.0009
};
\addplot [semithick, color1, mark=triangle, mark size=2, mark options={solid}]
table {%
51.52 0.0003
64.37 0.0006
75.42 0.0009
};
\addplot [semithick, color2, mark=x, mark size=2, mark options={solid}]
table {%
51.52 0.0004
64.37 0.0006
75.42 0.0011
};
\coordinate (B) at (axis cs:64.37,0.0006);
\end{axis}

\draw[-stealth] (A) -- (B|-0,0.1\linewidth);

\end{tikzpicture}

%% file: figures/model_pruning_visual.tex
\begin{figure}
    \centering
    \begin{subfigure}{\linewidth}
        \centering
        \begin{tikzpicture}
            \node at (0,0) {\includegraphics[height=29pt,trim={250 100 250 100},clip]{images/decomposed-uv/mario_uv_layout_pred.png}};
            \node at (1.2,0) {\includegraphics[height=29pt,trim={250 100 250 100},clip]{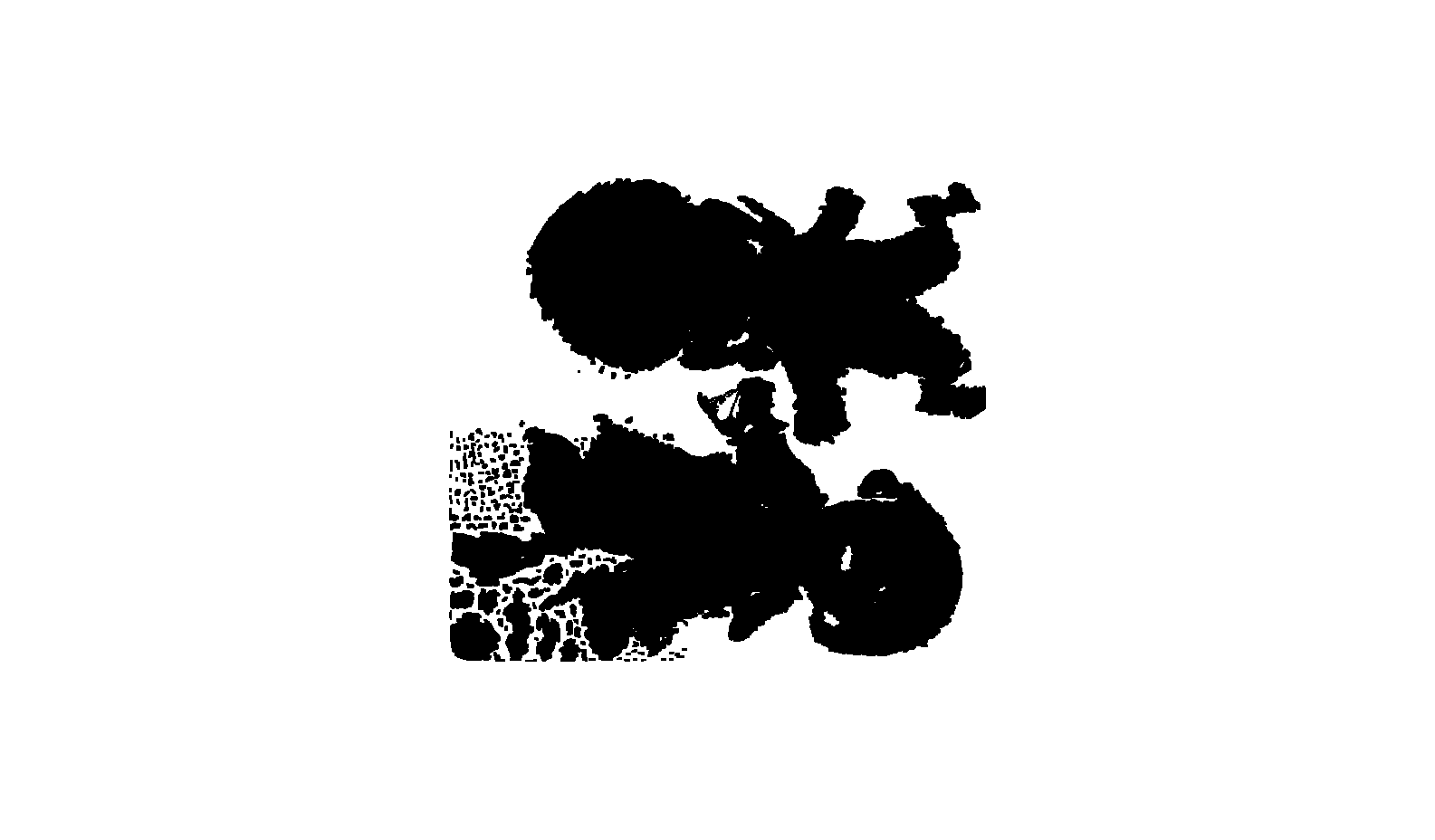}};
            \node at (2.4,0) {\includegraphics[height=29pt,trim={250 100 250 100},clip]{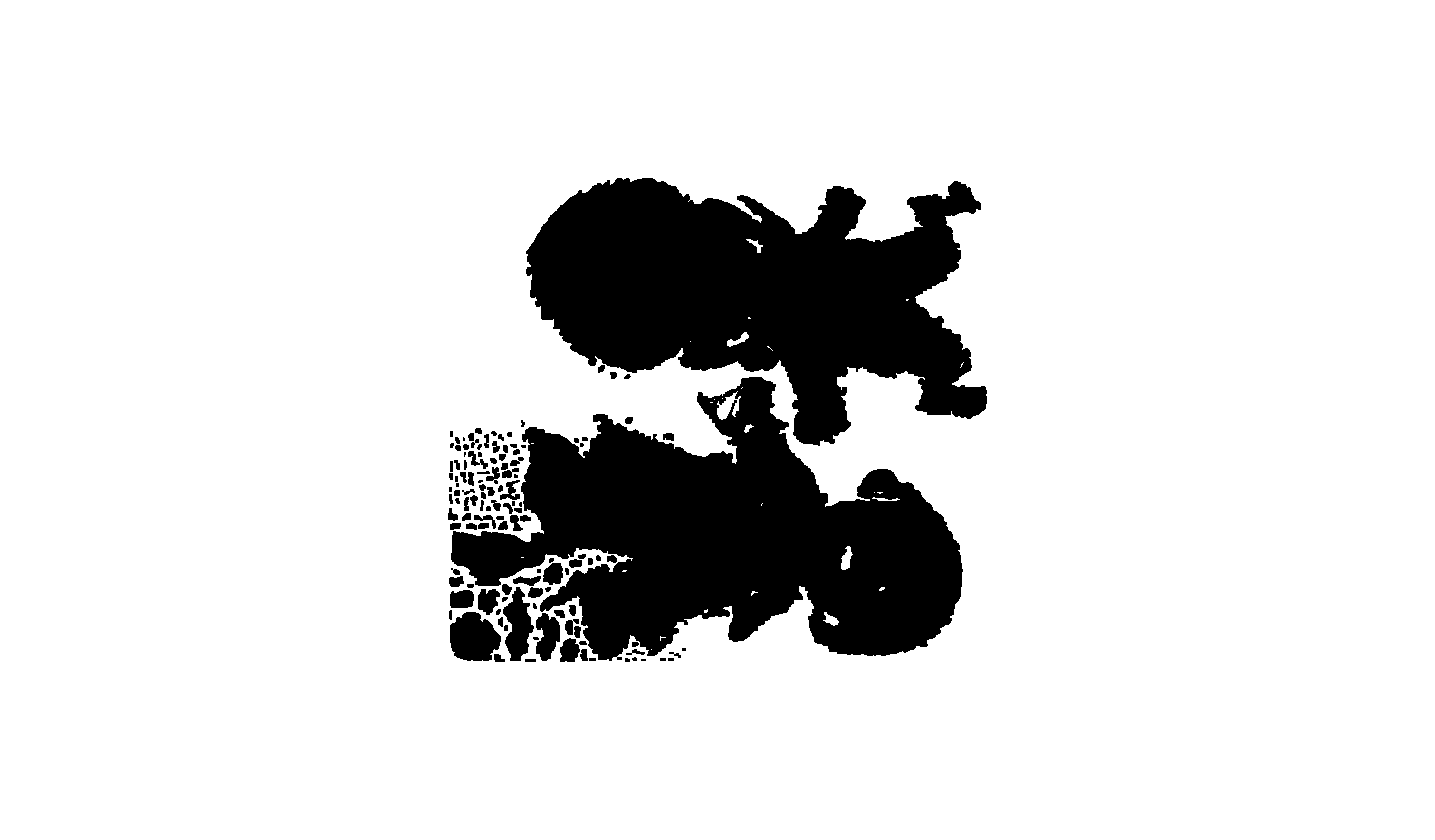}};
            \node at (3.6,0) {\includegraphics[height=29pt,trim={250 100 250 100},clip]{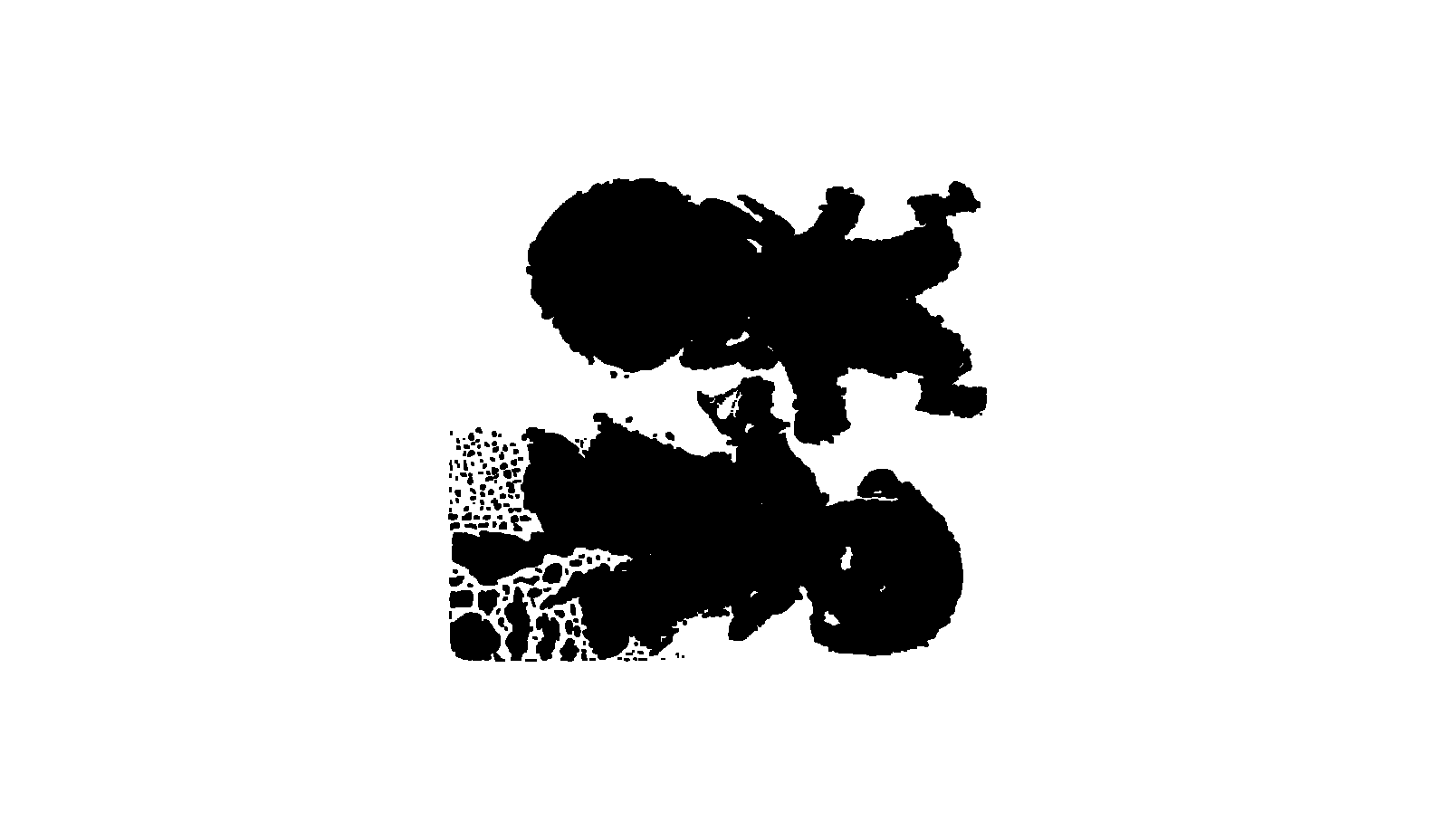}};
            \node at (4.8,0) {\includegraphics[height=29pt,trim={250 100 250 100},clip]{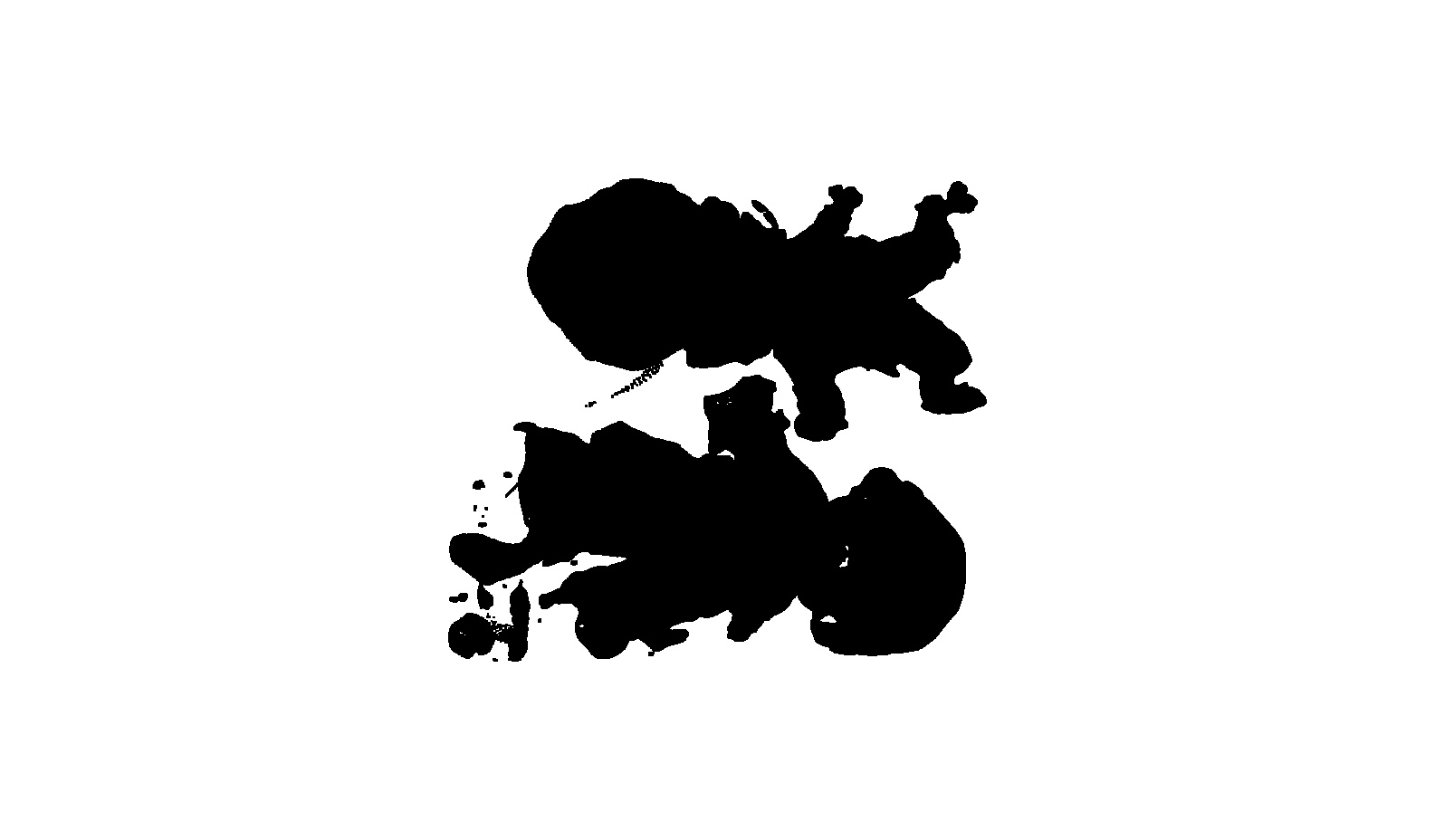}};
            \node at (6,0) {\includegraphics[height=29pt,trim={250 100 250 100},clip]{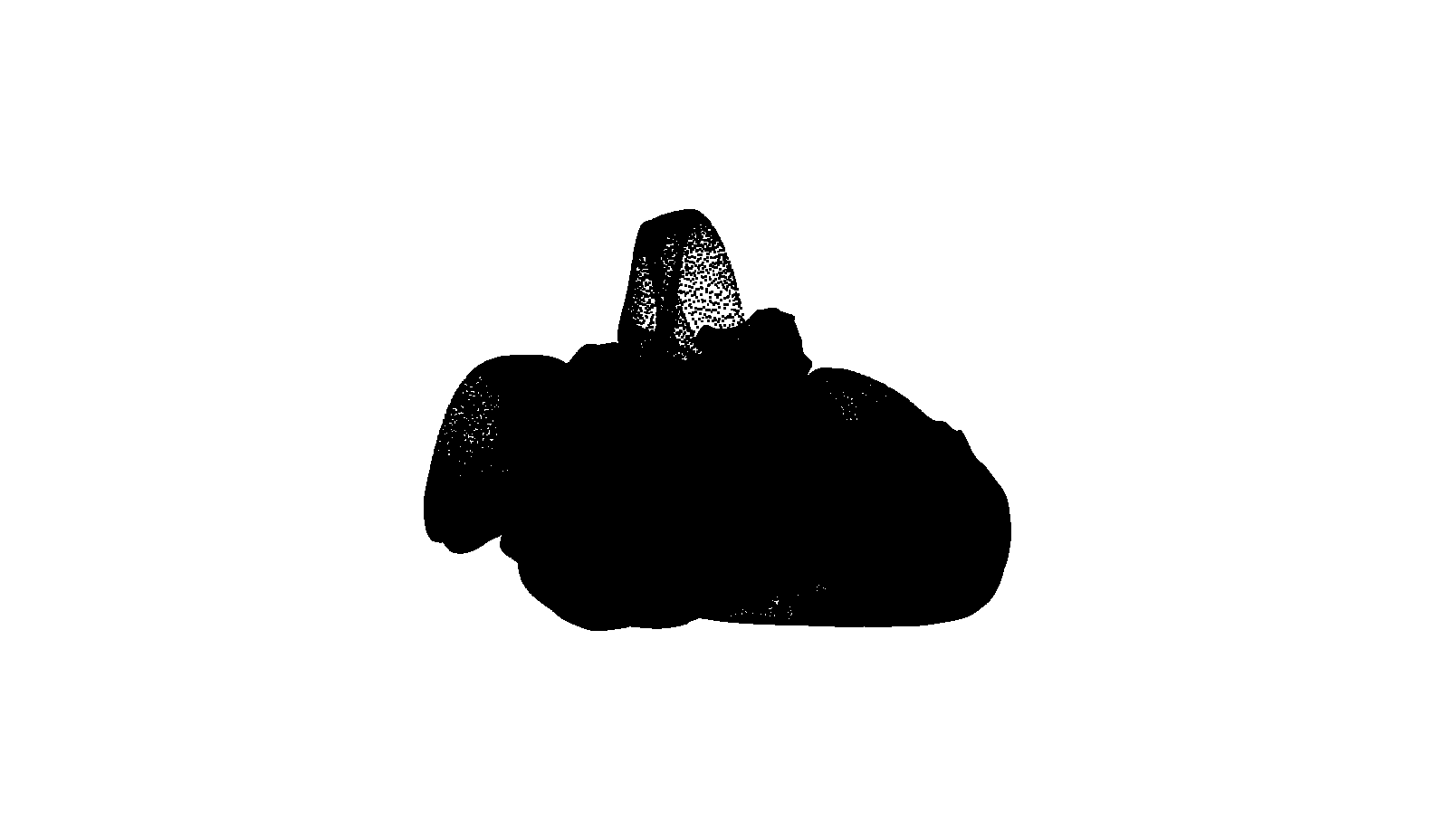}};
            \node at (0,-0.7) {$\SI{0}{\percent}$};
            \node at (1.2,-0.7) {$\SI{36.9}{\percent}$};
            \node at (2.4,-0.7) {$\SI{64.4}{\percent}$};
            \node at (3.6,-0.7) {$\SI{84.4}{\percent}$};
            \node at (4.8,-0.7) {$\SI{95.7}{\percent}$};
            \node at (6,-0.7) {$\SI{99.4}{\percent}$};
        \end{tikzpicture}
        \caption{Pruning \texttt{point2component} only}
        \label{fig:model-pruning-visual-a}
    \end{subfigure}
    \\
    \begin{subfigure}{\linewidth}
        \centering
        \begin{tikzpicture}
            \node at (0,0) {\includegraphics[height=29pt,trim={250 100 250 100},clip]{images/decomposed-uv/mario_uv_layout_pred.png}};
            \node at (1.2,0) {\includegraphics[height=29pt,trim={250 100 250 100},clip]{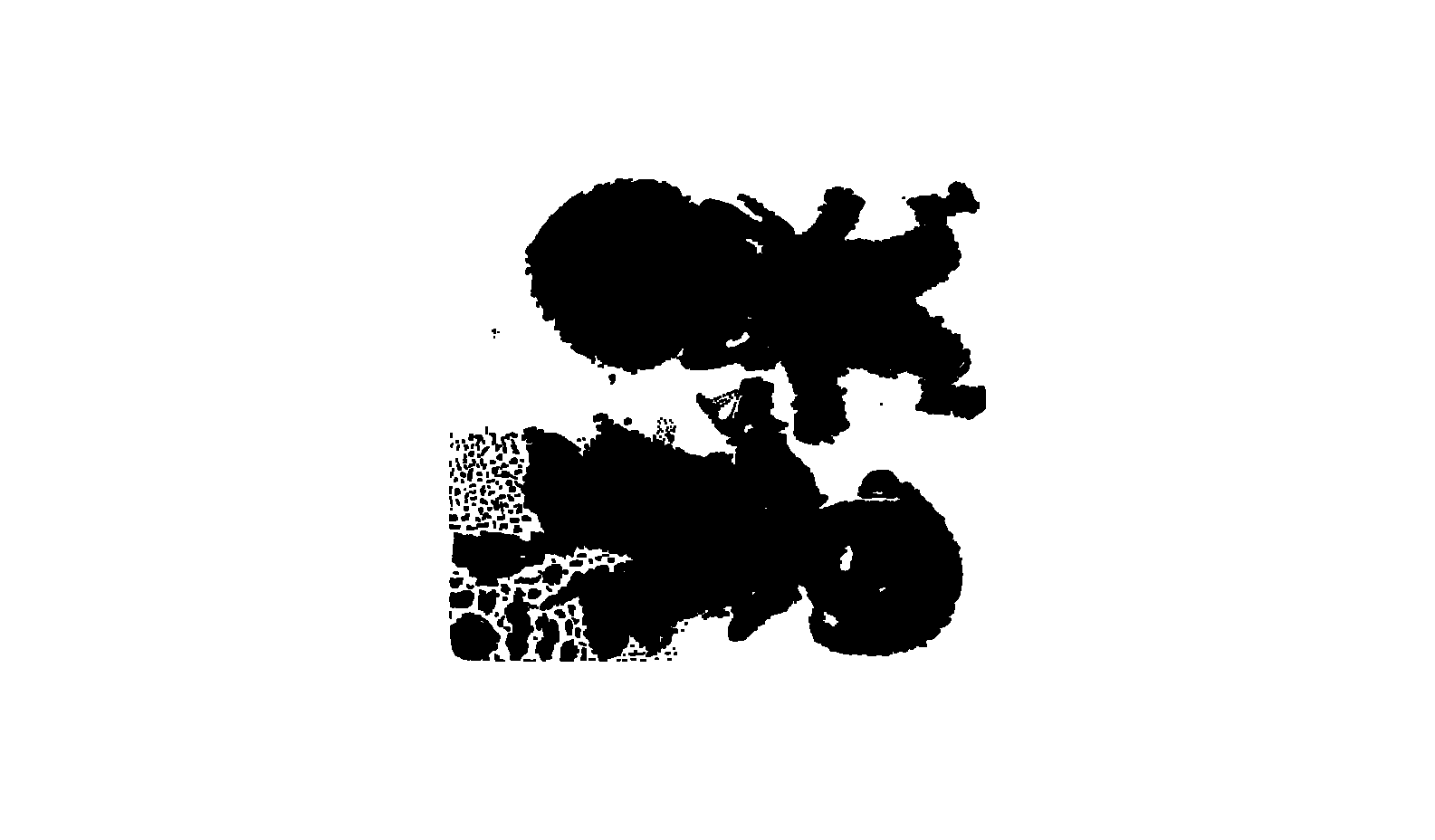}};
            \node at (2.4,0) {\includegraphics[height=29pt,trim={250 100 250 100},clip]{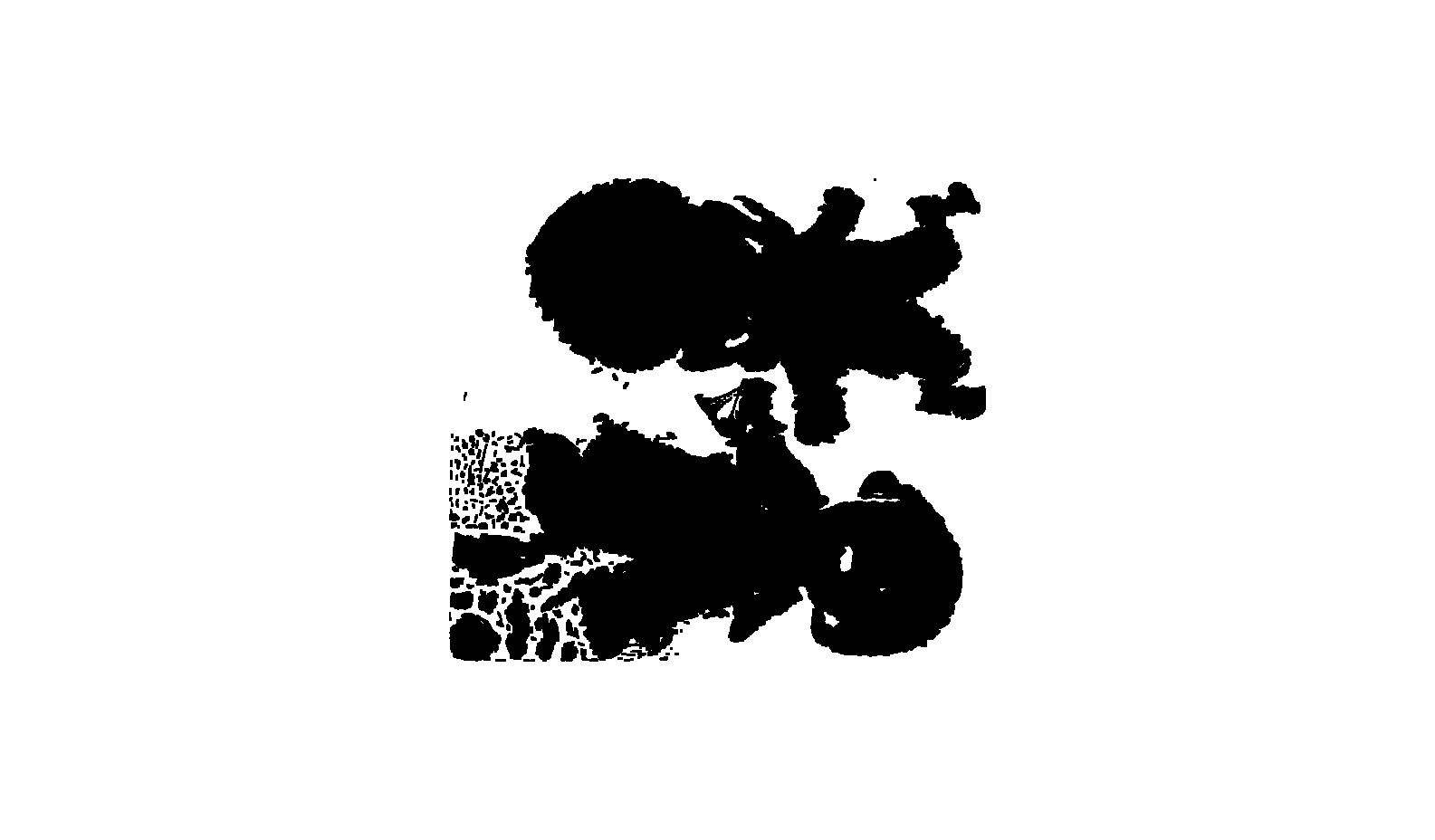}};
            \node at (3.6,0) {\includegraphics[height=29pt,trim={250 100 250 100},clip]{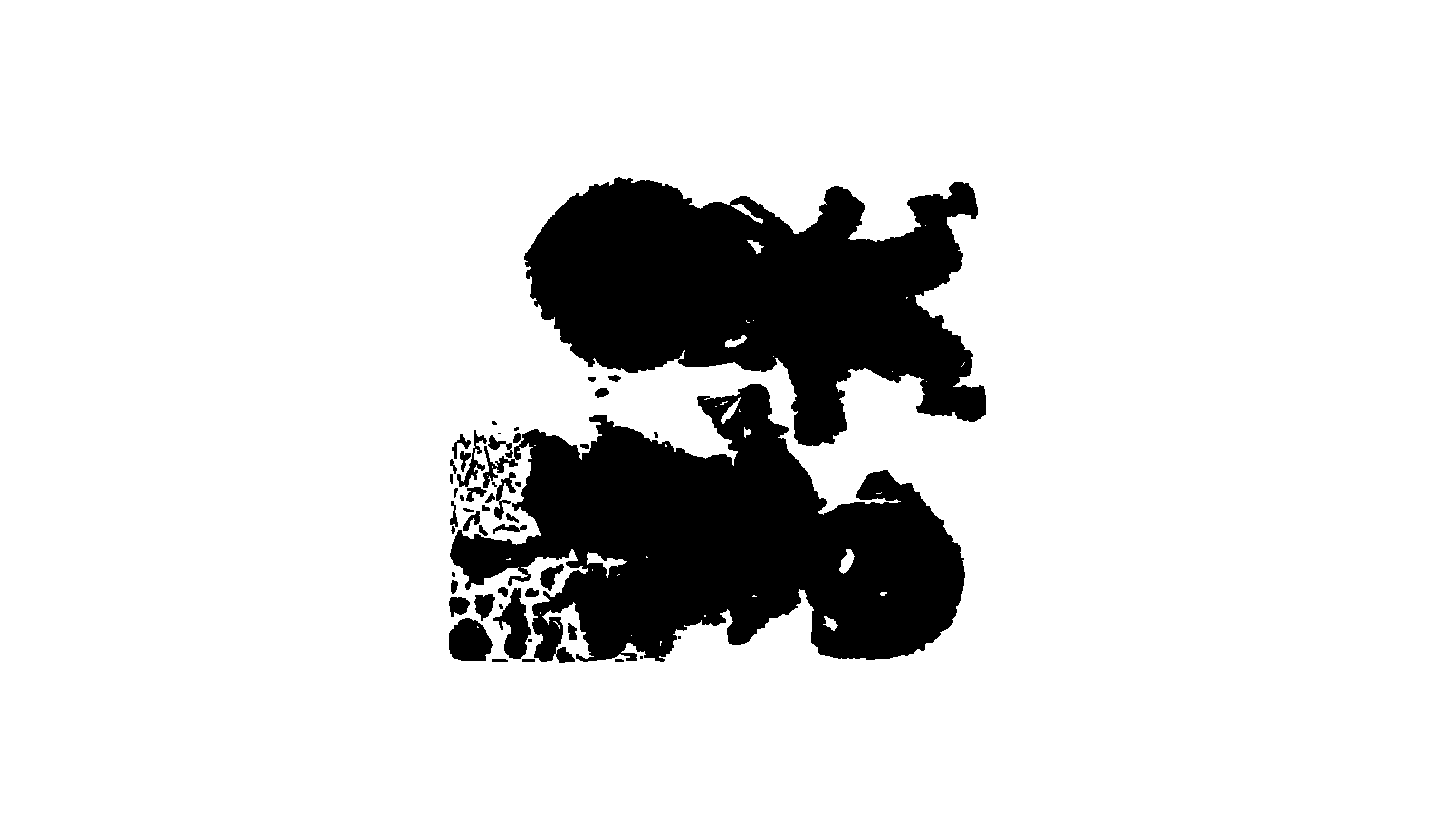}};
            \node at (4.8,0) {\includegraphics[height=29pt,trim={250 100 250 100},clip]{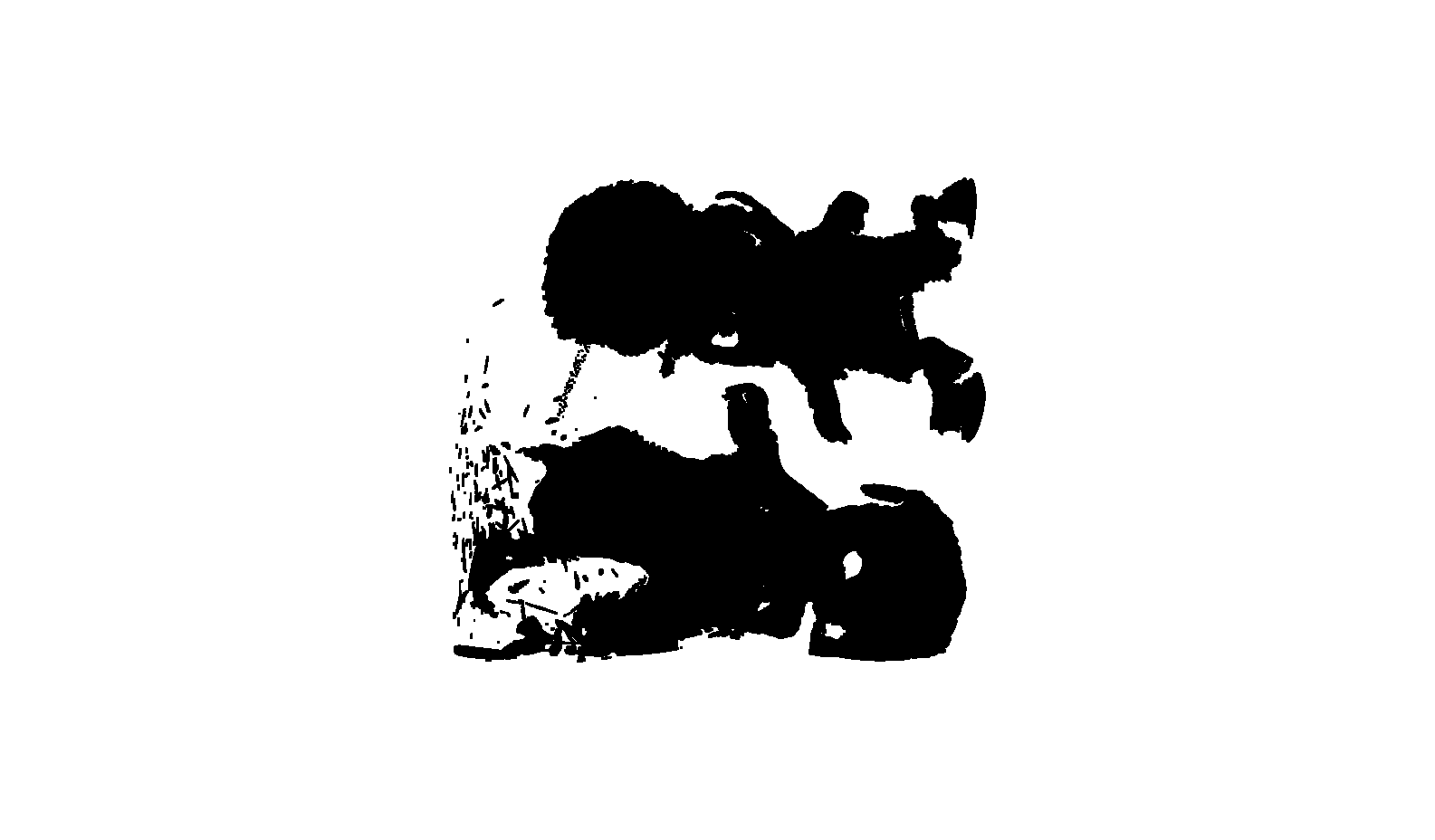}};
            \node at (6,0) {\includegraphics[height=29pt,trim={250 100 250 100},clip]{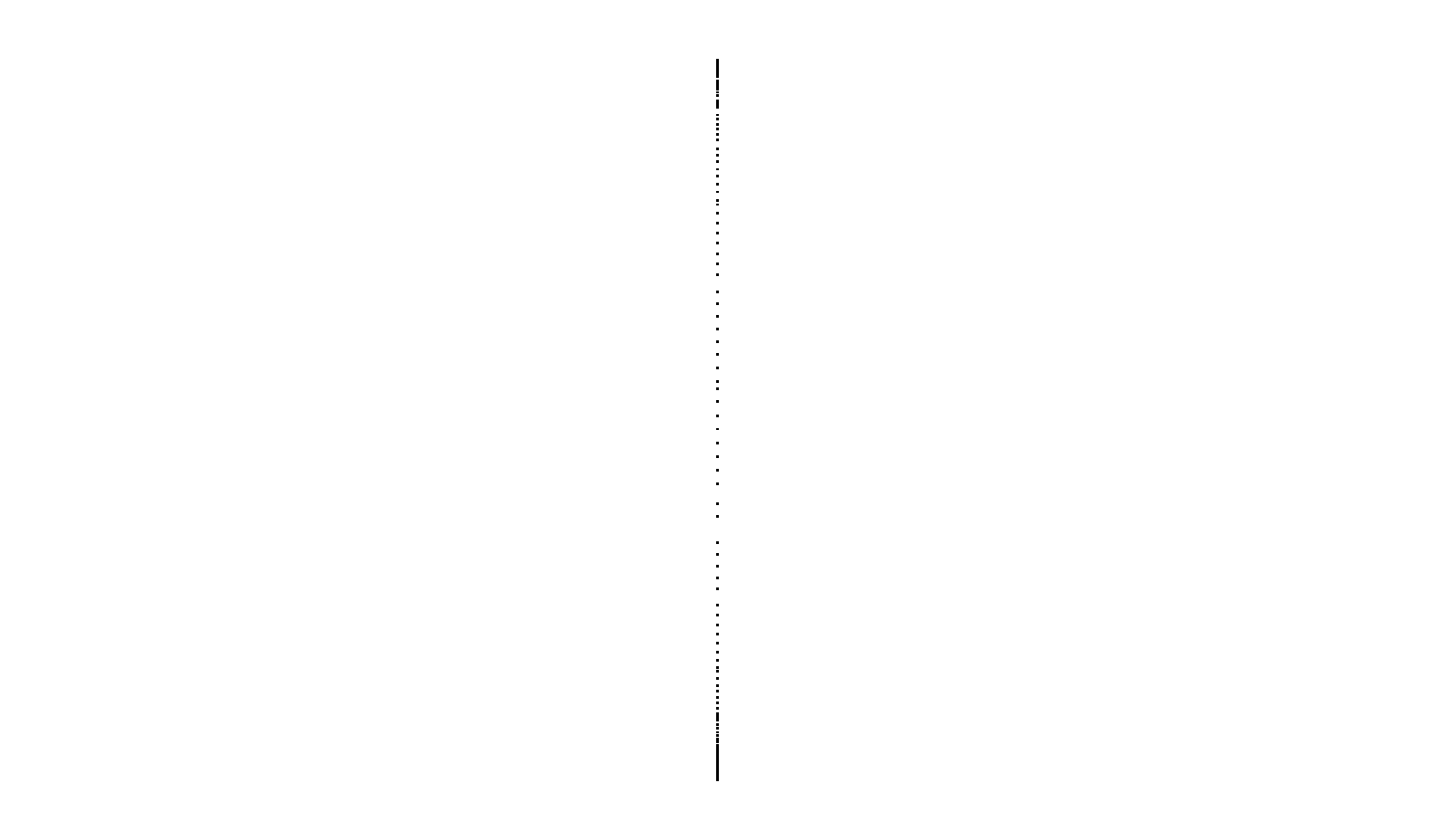}};
            \node at (0,-0.7) {$\SI{0}{\percent}$};
            \node at (1.2,-0.7) {$\SI{36.9}{\percent}$};
            \node at (2.4,-0.7) {$\SI{64.4}{\percent}$};
            \node at (3.6,-0.7) {$\SI{84.4}{\percent}$};
            \node at (4.8,-0.7) {$\SI{95.7}{\percent}$};
            \node at (6,-0.7) {$\SI{99.4}{\percent}$};
        \end{tikzpicture}
        \caption{Pruning \texttt{point2UV} only}
        \label{fig:model-pruning-visual-b}
    \end{subfigure}
    \caption{The predicted UV coordinates (shown as scatter charts) of the Mario object by our model with \texttt{point2component} and \texttt{point2UV} pruned at different rates (shown in percentage below each scatter chart) using the \texttt{smallest} strategy.}
    \label{fig:model-pruning-visual}
\end{figure}

%% file: tables/model_compression.tex
\begin{table}
\caption{Size comparison between the weights, saved in \texttt{.npz} format and measured in kilobytes, of the unpruned model and the pruned model compressed using different coding formats of sparse matrices.}
\label{tab:model-compression}
\scriptsize
\centering
\begin{tabular}{ lccccc }
\toprule
& Unpruned & COO & CSC & CSR & DIA \\
\midrule
\texttt{point2component} & $174.4$ & $54.9$ & $54.5$ & $55.0$ & $58.5$ \\
\texttt{point2UV} & $129.1$ & $71.3$ & $71.4$ & $71.4$ & $73.4$ \\
\midrule
Total & $303.5$ & $126.2$ & $125.9$ & $126.4$ & $131.9$ \\
\bottomrule
\end{tabular}
\end{table}

%% file: conclusion.tex
\section{Conclusion}\label{sec:conclusion}

We proposed to learn neural representations of both an object's 3D surface \emph{and} a surface parameterization suitable for auxiliary appearance data.
Following recent works on overfitted networks, we learn complex multi-chart signal parameterizations as a weight-encoded neural representation. We rely on a novel two-stage network architecture to allow us to capture fine-scale texture domain resolution while respecting the discontinuities in texture atlases. Our model augments existing (neural) implicit surface representations with the benefits of auxiliary texture mapping and editing common to standard 3D digital content creation pipelines. We demonstrated the applicability of our model to appearance-aware neural implicit surface representations, building atop the work of Davies et al.~\cite{davies2020overfit}.

%% file: acknowledgments.tex
\section*{Acknowledgments}

We thank the anonymous reviewers for their valuable comments and suggestions. This work was done during the first author's internship at Huawei Technologies Canada.